\documentclass[11pt,tightenlines,preprintnumbers,superscriptaddress,nofootinbib,fleqn]{revtex4-2}
\pdfoutput=1
\usepackage{fancyhea}
\usepackage{bm}       
\usepackage{graphicx}
\usepackage{hyperref}      
\usepackage{multirow}
\usepackage{slashed}
\usepackage{amsmath,amssymb}
\usepackage{placeins} 
\usepackage{verbatim} 
\usepackage{xcolor} 
\usepackage{xspace}
\usepackage[nottoc]{tocbibind} 

\newcommand{\nc}{\newcommand}  

\def\Acknowledgements{\bigskip  \bigskip \begin{center} \begin{large}
             \bf ACKNOWLEDGEMENTS \end{large}\end{center}}

\def\beq{\begin{equation}}
\def\eeq#1{\label{#1}\end{equation}}
\def\eeqn{\end{equation}}

\newenvironment{Eqnarray}%
   {\arraycolsep 0.14em\begin{eqnarray}}{\end{eqnarray}}
\def\beqa{\begin{Eqnarray}}
\def\eeqa#1{\label{#1}\end{Eqnarray}}
\def\eeqan{\end{Eqnarray}}

\nc{\ra}{\rightarrow}  
\nc{\slsh}{\slash\hspace*{-0.22cm}}
\def\Re{{\cal R \mskip-4mu \lower.1ex \hbox{\it e}\,}}
\def\Im{{\cal I \mskip-5mu \lower.1ex \hbox{\it m}\,}}

\nc{\vev}[1]{ \left\langle {#1} \right\rangle }
\nc{\bra}[1]{ \langle {#1} | }
\nc{\ket}[1]{ | {#1} \rangle }
\nc{\fb}{\,{\rm fb}^{-1}}
\nc{\ev}{{\rm eV}}
\nc{\kev}{{\rm keV}}
\nc{\Mev}{{\rm MeV}}
\nc{\gev}{{\rm GeV}}
\nc{\tev}{{\rm TeV}}
\nc{\mev}{{\rm MeV}}

\def\iab{\mbox{ab$^{-1}$}}
\def\ifb{\mbox{fb$^{-1}$}}
\def\ipb{\mbox{pb$^{-1}$}}

\newcommand{\ttbar}{{\ensuremath{t\bar{t}}}\xspace}

\def\L{{\cal L}}

\def\del{\partial}
\def\Dslash{\not{\hbox{\kern-4pt $D$}}}
\def\dslash{\not{\hbox{\kern-2pt $\del$}}}
\def\pslash{\not{\hbox{\kern-2pt $p$}}}
\def\ETmiss{ \not{\hbox{\kern-4pt $E$}}_T }

\def\msb{{\bar{\ssstyle M \kern -1pt S}}}

\newcommand{\MSbar}{\ensuremath{\overline{\mathrm{MS}}}}

\newcommand{\tsf}{\theta\kern-.20em_{\tilde{f}}}
\newcommand{\tsfp}{\theta\kern-.20em_{\tilde{f}\prime}}
\newcommand{\tsq}{\theta\kern-.15em_{\tilde{q}}}

\newcommand{\VL}{\left( \begin{array}{c}}
\newcommand{\VR}{\end{array} \right)}
\newcommand{\ML}{\left( \begin{array}{cc}}
\newcommand{\MLd}{\left( \begin{array}{ccc}}
\newcommand{\MLv}{\left( \begin{array}{cccc}}
\newcommand{\MR}{\end{array} \right)}

\providecommand{\href}[2]{#2}

\newcommand{\dd}{\mathrm{d}}

\newcommand{\MeV}{\text{MeV}}
\newcommand{\GeV}{\text{GeV}}
\newcommand{\TeV}{\text{TeV}}

\newcommand{\bss}{\begin{tiny}}
\newcommand{\ess}{\end{tiny}}

\newcommand{\abinv}{\ensuremath{\mathrm{ab}^{-1}}}
\newcommand{\fbinv}{\ensuremath{\mathrm{fb}^{-1}}}

\newcommand\snowmass{\begin{center}\rule[-0.2in]{\hsize}{0.01in}\\\rule{\hsize}{0.01in}\\
\vskip 0.1in Report of Energy Frontier Topical Group 3  submitted to the US Community Study\\ 
on the Future of Particle Physics (Snowmass 2021)
\rule{\hsize}{0.01in}\\\rule[+0.2in]{\hsize}{0.01in} \end{center}}

\begin{document}

\raggedbottom

\pagenumbering{roman}

\parindent=0pt
\parskip=8pt
\setlength{\evensidemargin}{0pt}
\setlength{\oddsidemargin}{0pt}
\setlength{\marginparsep}{0.0in}
\setlength{\marginparwidth}{0.0in}
\marginparpush=0pt


\pagenumbering{arabic}

\renewcommand{\arraystretch}{1.25}
\addtolength{\arraycolsep}{-3pt}




\pagestyle{empty}
\snowmass
\centerline{\Large\bf Top quark physics and heavy flavor production}

\begin{center}\begin{boldmath}




\begin{center}

\begin{large} {\bf Conveners:} R.~Schwienhorst$^{1}$ and
D.~Wackeroth$^{2}$
\end{large}

K.~Agashe$^{3}$, S.~Airen$^{3}$, S.~Alioli$^{4}$, J.~Aparisi$^{5}$, G.~Bevilacqua$^{6}$, H.~Y.~Bi$^{7}$, R.~Brock$^{1}$, A.~Gutierrez Camacho$^{5}$, F.~Febres Cordero$^{8}$, J.~de Blas$^{9}$, R.~Demina$^{10}$, Y.~Du$^{11}$, G.~Durieux$^{12}$, J.~Fein$^{1}$, R.~Franceschini$^{13}$, J.~Fuster$^{5}$, M.~V.~Garzelli$^{14}$, A.~Gavardi$^{4}$, J.~Gombas$^{1}$, C.~Grojean$^{15, 16}$, J.~Gu$^{17}$, M.~Guzzi$^{18}$, H.~B.~Hartanto$^{19}$, A.~H.~Hoang$^{20}$, J.~Holguin$^{21}$, A.~Irles$^{5}$, N.~Kidonakis$^{18}$, D.~Kim$^{22}$, M.~Kraus$^{8}$, C.~Lepenik$^{20}$, T.~M.~Liss$^{23}$, L.~Mantani$^{24}$, S.~Mantry$^{25}$, V.~Mateu$^{26}$, D.~Melini$^{27}$, J.~Michel$^{28}$, V.~Miralles$^{29}$, M.~Miralles Lopez$^{5}$, M.~Moreno Llacer$^{5}$, S.-O.~Moch$^{14}$, P.~Nadolsky$^{30}$, T.~Neumann$^{31}$, M.~Narain$^{32}$, J.~Nasufi$^{7}$, K.~Nowak$^{33}$,  A.~Pathak$^{34}$, M.~Peskin$^{35}$, R.~Poncelet$^{24}$, M.~Procura$^{20}$, L.~Reina$^{8}$, G.~Rodrigo$^{5}$, D.~Sathyan$^{3}$, S.~Sawford$^{1}$, R.~Schwienhorst$^{1}$, F.~Simon$^{36}$, M.~Spira$^{37}$, I.~Stewart$^{28}$, S.~Tairafune$^{38}$, J.~Tian$^{39}$, A.~Tricoli$^{31}$, P.~Uwer$^{16}$, E.~Vryonidou$^{34}$, K.~Vo\ss$^{40}$, M.~Vos$^{5}$, D.~Wackeroth$^{2}$,
M.~Worek$^{7}$, K.~Xie$^{41}$, H.~Yamamoto$^{5}$, R.~Yonamine$^{38}$, Z.~Yu$^{1}$, C.-P.~Yuan$^{1}$, A.~F.~\.Zarnecki$^{33}$

$^{1}${\sl Michigan State University, East Lansing, MI, USA}\\
$^{2}${\sl University at Buffalo, Buffalo, NY, USA} \\
$^{3}${\sl University of Maryland, College Park, MD, USA} \\
$^{4}${\sl INFN, Milano-Bicocca, Milan, Italy} \\
$^{5}${\sl IFIC (UV/CSIC), Valencia, Spain} \\
$^{6}${\sl ELKH-DE Particle Physics Research Group, Debrecen, Hungary} \\
$^{7}${\sl RWTH Aachen University, Aachen, Germany} \\
$^{8}${\sl Florida State University, Tallahassee, FL, USA} \\
$^{9}${\sl Universidad de Granada, Granada, Spain} \\
$^{10}${\sl University of Rochester, Rochester, NY, USA} \\
$^{11}${\sl Chinese Academy of Sciences, Beijing, China} \\
$^{12}${\sl CERN, Geneva, Switzerland} \\
$^{13}${\sl Universit{\`a} degli Studi and INFN Roma Tre, Rome, Italy} \\
$^{14}${\sl Universit{\"a}t Hamburg, Hamburg, Germany} \\ 
$^{15}${\sl Deutsches Elektronen-Synchrotron DESY, Hamburg, Germany} \\
$^{16}${\sl Humboldt-Universit{\"a}t zu Berlin, Berlin, Germany} \\
$^{17}${\sl Fudan University, Shanghai, China} \\
$^{18}${\sl Kennesaw State University, Kennesaw, GA, USA} \\
$^{19}${\sl  University of Cambridge, Cambridge, UK} \\
$^{20}${\sl  University of Vienna, Vienna, Austria} \\
$^{21}${\sl Ecole polytechnique, Paris, France} \\  
$^{22}${\sl Texas A\&M University, College Station, TX, USA} \\ 
$^{23}${\sl City College of New York, New York City, NY, USA} \\ 
$^{24}${\sl University of Cambridge, Cambridge, UK} \\
$^{25}${\sl University of North Georgia, Dahlonega, GA, USA} \\ 
$^{26}${\sl Universidad de Salamanca, Salamanca, Spain} \\
$^{27}${\sl Technion, Israel Institute of Technology, Haifa, Israel} \\
$^{28}${\sl Massachusetts Institute of Technology, Cambridge, MA, USA}\\
$^{29}${\sl INFN, Sezione di Roma, Rome, Italy}\\
$^{30}${\sl Southern Methodist University, Dallas, TX, USA} \\
$^{31}${\sl Brookhaven National Laboratory, Upton, NY, USA} \\
$^{32}${\sl Brown University, Providence, RI, USA}\\
$^{33}${\sl University of Warsaw, Warsaw, Poland}\\
$^{34}${\sl University of Manchester, Manchester, UK} \\
$^{35}${\sl SLAC National Accelerator Laboratory, Menlo Park, CA, USA } \\
$^{36}${\sl Max-Planck-Institut f{\"u}r Physik, Munich, Germany} \\
$^{37}${\sl PSI, Villigen, Switzerland} \\
$^{38}${\sl Tohoku University, Sendai, Japan} \\
$^{39}${\sl CEPP, University of Tokyo, Japan} \\
$^{40}${\sl Universit{\"a}t Siegen, Siegen, Germany} \\
$^{41}${\sl University of Pittsburgh, Pittsburgh, PA, USA} \\
\end{center}



\end{boldmath}

Version: 2.4; Date: \today
\end{center}

\begin{center} {\bf Abstract} \end{center}

This report summarizes the work of the Energy Frontier Topical Group on EW Physics: Heavy flavor and top quark physics (EF03) of the 2021 Community Summer Study (Snowmass). It aims to highlight the physics potential of top-quark studies and heavy-flavor production processes (bottom and charm) at the HL-LHC and possible future hadron and lepton colliders and running scenarios.

\newpage

\begin{center} \begin{Large} {\bf Executive Summary}
\end{Large} \end{center}

Top quarks play a central role in the exploration of the energy frontier, second only to the Higgs boson. 
The broad program of top-quark measurements at the LHC and related advanced theoretical calculations will continue into the future, at the HL-LHC and possibly at future lepton and hadron colliders. At lepton colliders, the top-quark program will only start in earnest once the CM energy reaches the top-pair production threshold. 

Studies of the top quark are directly connected to the important questions at the energy frontier. Of particular importance is the top-quark mass, which is a key ingredient in EW precision and QCD calculations. Mass measurements at hadron colliders are limited by theory modeling uncertainties. A precision of better than 500~MeV is required for the HL-LHC.
Achieving this precision requires significant work on the theory side, to understand and calibrate the top-quark mass in different schemes, higher-order QCD and EW calculations, and improvements in parton shower Monte Carlo generators. A significant reduction of PDF uncertainties is required, which can be achieved, in part, through studies of heavy flavor production in the forward region.
The best precision will only be achieved at future lepton colliders running at the top threshold (340~GeV), where uncertainties of 50~MeV or better should be achievable. Circular lepton colliders are able to measure the top-quark mass more precisely than linear colliders because of they can constrain the strong coupling constant, which contributes a large uncertainty on the top-quark mass. 

Top-quark production processes probe all aspects of top-quark couplings to the SM bosons and top-quark final states are sensitive to BSM particles such as SUSY top squarks. The HL-LHC will provide the event samples required to study many of these processes with percent-level precision, which necessitates (N)NNLO and higher-order QCD calculations, and the inclusion of (N)NLO EW corrections. 
Lepton colliders running at CM energies above 500~GeV (ILC and CLIC) are able to constrain the top-quark couplings to bosons beyond the precision of the HL-LHC. Additional constraints for global EFT fits come from
heavy flavor production (bottom and charm quarks) at lepton colliders.

Searches for BSM physics in top-quark final states focus on the third-generation coupling of BSM particles, indirectly through EFT fits or searches for flavor-changing neutral currents, and directly through searches for SUSY and other new particles. Linear colliders will expand the reach of FCNC searches of the top-$Z$ interaction.
Contact interaction and searches for compositeness are examples of BSM physics that top-quark production is sensitive to at TeV energies and above, these are probed at CLIC, FCC-hh as well as muon colliders.

Significant theoretical effort is required to exploit the full potential of future colliders, as pointed out throughout this document. Some of the biggest challenges are:
\begin{itemize}
    \item Calibration of the top quark MC mass to a well-defined scheme in perturbation theory with a precision comparable to the experimental uncertainty.
    \item Computing cross-sections, inclusively and differentially at higher orders in perturbation theory, going to N$^{3}$LO in QCD for top pair production plus resummation, going to NNLO in QCD for associated production processes, and including EW higher order corrections.
    \item Reducing the PDF uncertainties, which are already now the largest theory uncertainties for several processes, most importantly top-pair production. This requires close interconnections between theory and experiment and new differential measurements of top production processes.
    \item Improving the modeling of the full event at the LHC and future hadron and lepton colliders and reducing parton shower uncertainties.
\end{itemize}

\pagestyle{myheadings}
\newpage

\tableofcontents

\section{Introduction}
\label{sec:TOPHF-introduction}

\begin{figure}[ht]
\centering
\includegraphics[width=0.6 \linewidth]{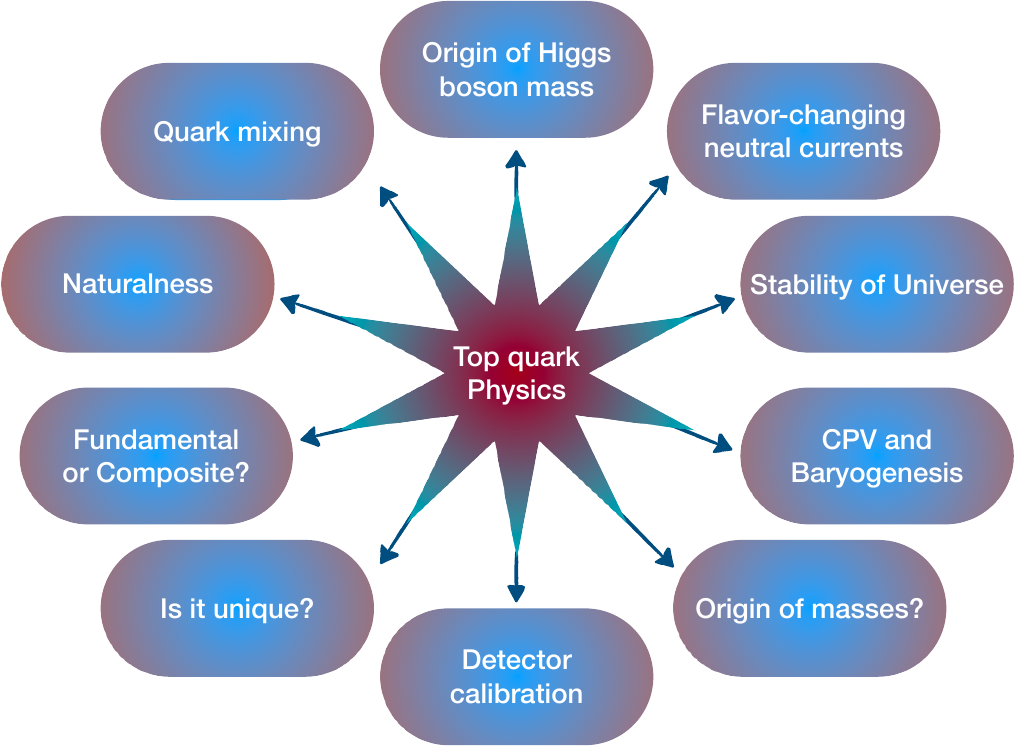}
\caption{Illustration of the different aspects of the top-quark physics program.
}
\label{fig:topoverview}
\end{figure}

Since the discovery of the top quark 
at the Fermilab Tevatron proton-antiproton collider in 1995~\cite{CDF:1995wbb,D0:1995jca}, the properties, productions
and decays of the to-date heaviest fundamental particle have been
under close experimental scrutiny. Given its large mass ($m_t$), the
top quark plays a special role in the EW (EW) sector of the
Standard Model (SM), with a Yukawa coupling ($y_t$) of order unity
($y_t=\sqrt{2} m_t/v \approx 1$ and $v$ is the vacuum expectation
value of the Higgs field), introducing large quadratic corrections to
the Higgs-boson mass, and affecting the stability of the EW
vacuum~\cite{Degrassi:2012ry}. The top-quark sector is therefore
especially suitable for precision EW tests and to search for possible beyond-the-SM (BSM)
physics. Being so heavy also makes for a unique phenomenology: the top
quark dominantly decays to a $W$ boson and bottom quark with a
lifetime approximately ten times shorter than the time needed for the
formation of hadrons ($1/\Lambda_{QCD} \approx 10^{-24}$s) or
bound states, and several orders of magnitude smaller than the spin decorrelation time ($m_t/\Lambda_{QCD}^2 \approx 10^{-21}$s). As a
consequence, the top quark decays before it hadronizes and the
information about its spin state is preserved in distributions of
top-quark decay products. The observation of many millions of top-quark events
at the LHC together with the development of innovative analysis
techniques allowed to perform measurements at a high level of
precision and to explore a wide variety of top quark observables in
new kinematic regimes and event topologies. Also on the theory side,
immense effort and ingenuity went into improving predictions for top
quark observables and in devising direct and indirect searches
strategy for BSM physics. A diagram of physics questions that can be addressed with top quarks is shown in Figure~\ref{fig:topoverview}. For recent reviews on top-quark physics at
hadron colliders see, e.g.,\cite{Zyla:2020zbs,Hoang:2020iah,Giammanco:2017xyn}. As the heaviest SM particle, the top quark is also a key to BSM searches at the LHC and future colliders. Its large mass and decay products enable the precision calibration of detectors at hadron collider and the different top-quark production processes contribute important backgrounds in almost all analyses.

Given the uniqueness and importance of the top-quark sector in the SM
and in many BSM scenarios it is of the utmost importance to take full
advantage of the LHC as a top-quark factory to perform precision
studies and BSM searches in top-quark production processes and
decays. Other heavy-quark production processes involving bottom (the other 3rd generation quark) and charm quarks 
also offer interesting opportunities for precision tests of the SM and searches for BSM physics, and examples are presented in this report. 

This report aims to highlight the physics potential of top-quark studies and heavy-flavor production processes (bottom and charm) at the HL-LHC and possible future hadron and lepton colliders and running scenarios. The following specific, overarching questions were useful in guiding the discussion: 
\begin{itemize}
\item    
What is the ultimate precision that can be reached for the measurement of a well-defined  top-quark mass? What is the impact on global EW precision fits (see EF04 report)? What is needed to reduce the theoretical uncertainties in order to push for the highest possible precision, especially in $e^+ e^- \to t\bar t$ threshold cross sections at the required level? This is addressed in Section~\ref{sec:TOPHF-mtop} and the EW report~\cite{EWreport}.
\item
What is the potential for discovery and studies of rare top-quark production processes, such as multiple top production and top in association with other heavy SM particles, and what is their impact on a global EFT fit and direct BSM searches? This is addressed in Sections~\ref{sec:TOPHF-XS-LHC-ttX} and~\ref{sec:TOPHF-4top-production} and in the Higgs report~\cite{Higgsreport}, the EW report~\cite{EWreport} and the BSM report~\cite{Bose:2022obr}.
\item
What can be learned from measurements of top quark production and top quark properties other than the top quark mass and couplings such as spin correlations, asymmetries, polarization in new kinematic regimes, and what is the achievable/required precision? This is addressed in Section~\ref{sec:TOPHF-EW-couplings}.
The relevant EFT operators and the global fit are discussed in the EW report~\cite{EWreport}.
\item
What is the potential of multi-differential cross sections in top quark production processes to simultaneously extract $\alpha_s, m_t$, and the gluon parton distribution function (PDF)? This is addressed in Sections~\ref{sec:TOPHF-XS}, ~\ref{sec:TOPHF-mtop-pole} and~\ref{sec:TOPHF-HF-PDF}. See also the QCD report~\cite{QCDreport}.
\item
What is the potential of heavy-quark production cross sections (also in association with EW gauge bosons) to probe heavy-quark PDFs, and could this impact the achievable precision of Higgs production in association with heavy quarks? This is addressed in Section~\ref{sec:TOPHF-XS-LHC-ttX} and the QCD report~\cite{QCDreport}.
\item
What can be learned from precision measurements of heavy-quark production ($c\bar c,b\bar b,t\bar t$) at lepton colliders, and are systematic uncertainties from theory under control, especially higher-order EW corrections? An example of this is the running of the bottom quark mass, see Section~\ref{sec:TOPHF-bmass}, see also the QCD report~\cite{QCDreport}.
\item
There will be improvements in theory and analysis techniques after the LHC is finished taking data. How can these be used to improve our understanding of the top quark and its interactions in the foreseeable absence of top physics data? This will require persistence of HL-LHC data and the ability to re-analyze it as tools improve. This impacts most of the top-quark measurements described here, and particularly the top quark pole mass measurements (Section~\ref{sec:TOPHF-mtop-pole}).
\item
Is it possible to realistically project systematic uncertainties affecting precision measurements of top quark observables and heavy flavor production to be able to compare different collider options. Can we learn from the lessons of Tevatron to LHC or Snowmass 2013 compared to LHC data with 150 fb$^{-1}$? This is addressed in Section~\ref{sec:TOPHF-mtop-direct}.
\end{itemize}

In Section~\ref{sec:TOPHF-conclusions} we summarize our findings and highlight the need for significant theory effort to exploit the full potential of future colliders.


  

\section{Top-quark mass}
\label{sec:TOPHF-mtop}

The top-quark mass $m_t$ is one of the most important parameters of the SM and relevant as an input for precise predictions and for the understanding of SM properties such as the stability of the spontaneously broken vacuum state.  As such, precise $m_t$ measurements are just as important as their interpretation in terms of a theoretically well-defined mass. For instance, the crucial role of $m_t$ in EW precision tests of the SM as both an SM input parameter and an EW precision observable (EWPO) is highlighted in the EW report~\cite{EWreport}. Loop corrections to the mass of the $W$~boson from top quarks result in a $W$~mass uncertainty of about 1~MeV for a top-quark mass uncertainty of 100~MeV~\cite{Awramik:2003rn}. The precision EW fits therefore set the scale for the required top-quark mass uncertainty~\cite{deBlas:2021wap,Haller:2018nnx}. A $W$~boson mass uncertainty of about 5~MeV (see Table~1 in Ref.~\cite{EWreport}) therefore necessitates the top-quark mass to be measured with an 500~MeV. Similarly, a $W$~boson mass uncertainty of 0.5~MeV at a lepton collider (see Table~2 in Ref.~\cite{EWreport}) necessitates a top-quark mass uncertainty below 50~MeV. This evaluation assumes the top-quark mass is measured in a well-defined scheme, otherwise additional uncertainties need to be included, see Section~\ref{sec:TOPHF-mtop-theory}.

The evaluation of EWPOs is usually performed either in the on-shell or $\overline{\rm MS}$ scheme and the prospects for measurements of the top-quark mass in these schemes at current and future colliders and their relation to other mass definitions is therefore important. This section starts with a discussion of the $m_t$ theory aspects, followed by the status and prospects for $m_t$ measurements, from the decay (direct or Monte Carlo (MC) mass) and production cross sections in well-defined schemes (e.g., pole mass) at a hadron collider, and from an energy scan around the $t \bar t$ threshold at a lepton collider.

\FloatBarrier
\subsection{Theory aspects and challenges}
\label{sec:TOPHF-mtop-theory}

Since isolated quarks cannot be observed, the top-quark mass is
not directly physical, but a renormalization scheme dependent quantity. This scheme dependence can only be well-defined and controlled for mass sensitive 
observables that are calculable in perturbation theory (at least at the next-to-leading order (NLO) level). Even though the scheme dependence of predictions vanishes in the large order limit,
in practice adopting a proper scheme can affect the quality of the perturbative expansion and the size of the perturbative uncertainties at each order.
In contrast to the strong coupling $\alpha_s(\mu_R)$, for which exclusively the $\overline{\rm MS}$ scheme is used and the issue of scheme-dependence reduces to making an adequate 
choice of the renormalization scale $\mu_R$, for the top-quark mass many different schemes are in use. Some are renormalization scale dependent ($\overline{\rm MS}$, MSR, kinetic, PS, RS)
and some are not (pole, 1S). The difference between the pole mass $m_t^{\rm pole}$ and the other schemes is of order $R\times\alpha_s(R)$ where $R$ is between $1$~GeV and $m_t$. 
For the $\overline{\rm MS}$ mass $\overline m_t(\mu)$ the scale $R$ is equal to $\overline m_t(\mu)$ and for the MSR mass  $m_t^{\rm MSR}(R)$ it is $R$.
The renormalization scale of the MSR mass $m_t^{\rm MSR}(R)$ allows to interpolate between the $\overline{\rm MS}$ mass $\overline m_t(\overline m_t)$ and the pole mass 
allowing for dynamical scale setting with $R$ values below the top-quark mass.~\cite{Hoang:2017suc,Hoang:2017btd}. While currently, for most theoretical NLO and next-to-next-to-next-to-leading order (NNLO) calculations of 
top-quark cross section, the top-quark mass scheme is fixed to a particular scheme (mostly the pole mass), it should eventually become common practice that such calculations allow for a flexible
scheme choice with dynamical scale setting in analogy to the renormalization scale of the strong coupling. 

The QCD corrections between most top-quark mass schemes are known to ${\cal O}(\alpha_s^4)$ precision and allow for fixed-order and renormalization group improved conversions with
uncertainties at the level of $10$-$20$ MeV as far as QCD corrections are concerned. The exception is the pole mass, which due to the on-shell condition has an intrinsic renormalon ambiguity of $110$ to $250$~MeV~\cite{Beneke:2016cbu,Hoang:2017btd}, and also leads to in general larger higher order corrections. Top-quark mass determinations aiming for a precision of a few hundred MeV or below have to avoid the pole mass scheme.  The QCD corrections in the scheme conversions are available in the software packages Rundec~\cite{Herren:2017osy} and REvolver~\cite{Hoang:2021fhn}, where the latter includes renormalization group improvement for scales below $m_t$ including a treatment of bottom and charm thresholds. 
The EW corrections at NLO and NNLO as well as mixed QCD-EW NLO~\cite{Hempfling:1994ar} and NNLO~\cite{Jegerlehner:2012kn,Kniehl:2014yia} are known for the relation of the top quark $\overline{\rm MS}$ and the pole mass. While the EW corrections are in general smaller than
the QCD corrections, there are large Yukawa tadpole corrections that are comparable to the NLO QCD corrections and for which different schemes have been suggested~\cite{Fleischer:1980ub,Bohm:1986rj,Dittmaier:2022maf}. In current top-quark mass determinations EW corrections are not yet included systematically. For uncertainties of $1$~GeV or less, EW corrections should, however, be included. 

The currently most precise top-quark mass determinations at the LHC, typically yielding uncertainties of $500$-$600$~\cite{ATLAS:2018fwq,CMS:2015lbj} for individual measurements, and even reaching the level of below 400 MeV~\cite{CMS:2022kcl} for an individual analysis, are obtained from the direct reconstruction of the top decay products (jets and leptons) from which top decay distributions are constructed that exhibit highly top-quark mass sensitive kinematic structures (peaks, shoulders and endpoints) (see also the discussion in Section~\ref{sec:TOPHF-mtop-direct}). These top decay sensitive observables are constructed to be largely independent of the top production mechanism and only weakly sensitive to uncertainties related to the underlying event or multi-particle interactions.
Since for these observables analytic perturbative QCD calculations are not available due to their complex exclusive character, and since hadronization corrections are large, these ‘direct’ measurements exclusively rely on multipurpose MC event generators, so the measured mass is the MC top-quark mass parameter $m_t^{\rm MC}$. 
There are theoretical arguments showing that $m_t^{\rm MC}$ should be numerically close to the the pole mass or the MSR mass $m_t^{\rm MSR}(1\,\mbox{GeV})$ within around $500$~MeV~\cite{Nason:2017cxd,Hoang:2020iah}, but limitations concerning the perturbative precision of the parton showers used in current MC event generators, which depending on the observable can have NLL precision or less, do not yet allow for a more precise statement. Due to the parton shower cutoff it is, however, guaranteed that the MC top-quark mass parameter $m_t^{\rm MC}$ does not suffer from the renormalon ambiguity inherent to the pole mass~\cite{Hoang:2008xm}. For parton showers that have next-to-leading-logarithmic (NLL) precision in the soft-collinear region, the relation between $m_t^{\rm MC}$ and well-defined top-quark mass schemes depends on the implementation of the parton shower and its cutoff and can be calculated at NLO at least for event-shape type mass sensitive observables~\cite{Hoang:2018zrp}. The results indicate that the difference between $m_t^{\rm MC}$ and $m_t^{\rm pole}$ and $m_t^{\rm MSR}(1\,\mbox{GeV})$ is in the range of $500$ and $200$~MeV, respectively. For the 2-jettiness distribution in the peak region in $e^+e^-$ annihilation, concrete calibration fits between particle level NNLL+NLO calculations and the PYTHIA8.205 MC confirm these estimates~\cite{Butenschoen:2016lpz}. Issues such as the universality of such results among different MCs and observables and the impact of hadronization effects, such that they can be applied reliably to the interpretation of the direct measurements, remain important questions to be studied to pin down the relation of $m_t^{\rm MC}$ to well-defined top-quark mass schemes at the level that matches the experimental uncertainties that are projected to approach $170$~MeV at the HL-LHC~\cite{CMS-PAS-FTR-16-006}.

Top-quark mass measurements in a well-defined mass scheme rely on top-quark mass sensitive cross sections that can be calculated systematically in perturbation theory. Since such calculations at the LHC are currently only possible for more inclusive observables, this in general comes with the price of a reduced mass sensitivity, a stronger sensitivity to the top production mechanism or a higher sensitivity to underlying event or multi-particle interactions. In addition this approach frequently involves some unfolding to the calculated parton level cross section and theoretical uncertainties arising from the finite truncation order of the perturbation series. 
A well-studied method of this kind relies on the total inclusive top-antitop production cross section based on NNLO QCD calculations~\cite{Barnreuther:2012wtj,Czakon:2012zr,Czakon:2012pz,Czakon:2013goa}, and the pole $m_t^{\rm pole}$ as well as the $\overline{\rm MS}$ mass  $\overline m_t(\overline m_t)$ have been determined (see Section~\ref{sec:TOPHF-mtop-pole}). Due to a strong sensitivity to the value of $\alpha_s$ and the gluon PDF, the uncertainties of the total cross section method are at the level of 2 GeV. Due to the strong correlation of $\alpha_s$, the gluon PDF and the top-quark mass~\cite{CMS:2018fks}, a significant improvement of this method relies on a more precise absolute knowledge of the gluon PDF. Reaching uncertainties well below 1~GeV also corrections beyond NNLO have to be accounted for.

Top-antitop production cross sections that are differential offer the advantage of having a stronger sensitivity to the top-quark mass due to kinematic threshold and endpoint kinematics in certain regions of the distributions and a reduced dependence on the PDFs. They are however, more difficult to calculate, and frequently only NLO predictions are used in the experimental analyses in many cases. For top-antitop events with at least one additional jet, the shape of the distribution in the $t\bar t j$ invariant mass is very top-quark mass sensitive, but also sufficiently inclusive to allow for a reliable calculation using fixed-order perturbation theory~\cite{Dittmaier:2007wz,Alioli:2013mxa,Bevilacqua:2015qha,Fuster:2017rev}. The $t\bar t j$ jet distribution has been used for measurements of the pole  $m_t^{\rm pole}$ and the $\overline{\rm MS}$ mass  $\overline m_t(\overline m_t)$ with combined uncertainties of 1~GeV and 2~GeV, respectively from LHC 8 TeV data~\cite{ATLAS:2019guf}. The larger uncertainty using the $\overline{\rm MS}$ mass indicates that a renormalon-free mass definition with a renormalization scale below $m_t$ may be more suitable. The calculation of NNLO QCD corrections and a reduction of experimental systematic uncertainties may lead to further improvements of the method. 

The $t\bar t$ invariant mass distribution is highly top-quark mass sensitive in the threshold region close to twice the top-quark mass. Fixed-order NLO QCD predictions have been used in Ref.~\cite{CMS:2019esx} to determine the pole mass with an uncertainty of $0.8$~GeV showing the high potential of this method. NNLO calculations with QCD resummation and EW corrections for the differential $t\bar t$ cross section~\cite{Czakon:2015owf,Czakon:2018nun,Czakon:2019txp} as well as NNLO QCD corrections to the on-shell top quark decay~\cite{Gao:2012ja,Brucherseifer:2013iv} are available. Fixed-order calculations are, however, not sufficient for a reliable top-quark mass determination due to Coulomb, retardation and off-shell effects that are important in the threshold region for color-singlet as well as color-octet $t\bar t$ configurations. Results accounting for Coulomb effects are available~\cite{Hagiwara:2008df,Kiyo:2008bv,Ju:2020otc}, but no complete theoretical treatment is available. 

A number of additional methods have been suggested (see also the discussion in Section~\ref{sec:TOPHF-mtop-new}) and partly studied experimentally to determine the top-quark mass aiming for at least NLO precision concerning the scheme dependence. This includes charged lepton momentum observables kinematically sensitive to the top-quark mass~\cite{Kawabata:2014wob,CMS:2017znf,ATLAS:2017dhr}, the energy distribution of $b$-jets~\cite{Agashe:2016bok}, the $\gamma\gamma$ invariant mass close to $2m_t$ that is sensitive to QCD Coulomb effects~\cite{Kawabata:2016aya}. 
Novel methods based on resummed and factorized QCD calculations at hadron level using Soft-Collinear Effective Theory (SCET) applicable for boosted top quarks which become available with higher statistics at the HL-LHC using top jet grooming~\cite{Hoang:2017kmk,ATLAS:2021urs} and based on energy-energy correlations~\cite{Holguin:2022epo} have been suggested.  These examples of alternative methods offer promising and valuable alternative top-quark mass determinations. 

Overall, top-quark mass determinations in well-defined mass schemes are still less precise than the direct measurements at hadron colliders, but there are prospects that their precision increases
with further theoretical work. One also has to always recall
that the direct measurements suffer from the interpretation problem of the MC top-quark mass parameter $m_t^{\rm MC}$
which adds an additional uncertainty (at the level of 0.5 GeV)~\cite{Hoang:2020iah} on top of uncertainties quoted in the analyses,
when the values of $m_t^{\rm MC}$ are employed for making high-precision theoretical predictions. 

Considering future lepton colliders, in principle most of the top-quark mass measurement methods used at the LHC can also be applied there. Due to the substantially cleaner hadronic environment aspects related to hadronic initial state radiation such as underlying event and multi-particle interactions are absent (and replaced by photonic backgrounds), so that in general higher precision can be reached~\cite{Vos:2016til,CLICdp:2018esa}. The dependence on the precise knowledge of the PDFs at hadron colliders is replaced by the dependence on the beam’s luminosity spectrum.  For the direct top-quark mass measurement the interpretation problem of $m_t^{\rm MC}$ remains essentially unchanged, but can be addressed more cleanly compared to the LHC due to the simpler structure of hadronization corrections~\cite{Hoang:2018zrp}. Because the $t\bar t$ pairs are produced predominantly in a color-single state and the low background level of identifying $t\bar t$ states the threshold scan method~\cite{Fadin:1987wz,Fadin:1990wx,Strassler:1990nw,Bigi:1986jk} emerges as an extremely powerful method, where the rising total inclusive $t\bar t+X$ cross section at c.m.\ energies around the threshold close to $2m_t$ has very high top-quark mass sensitivity combined with an almost ideal computational environment, where perturbative QCD and EW calculations can be employed to high precision and hadronization effects only play a minor role. Overall, measurements of the top-quark mass in well-defined schemes with low renormalization scales ($R\sim m_t\alpha_s\sim 20$~GeV) with uncertainties at the level of $50$~MeV are possible. Precise knowledge on the luminosity spectrum is crucial and a major source of uncertainty of the method. Theoretical calculations of the total cross section are based on non-relativistic effective theories of QCD (NRQCD)~\cite{Bodwin:1992ye}, where the Coulomb bound state dynamics is accounted for exactly using non-relativistic fixed-order (pNRQCD)~\cite{Pineda:1997bj,Brambilla:1999xf} or renormalization group improved (vNRQCD)~\cite{Luke:1999kz,Manohar:1999xd,Hoang:2002yy} expansions. Theoretical QCD calculation have been achieved at NNLO~\cite{Hoang:2000yr} and NNNLO~\cite{Beneke:2015kwa,Beneke:2016kkb} in the fixed order approach and at NNLL order in the renormalization group improved approach~\cite{Hoang:2013uda} yielding uncertainties for the cross section normalization at the few percent level. Electroweak, finite-lifetime and off-shell corrections~\cite{Kuhn:2005it,Hoang:2006pd,Beneke:2010mp,Penin:2011gg,Ruiz-Femenia:2014ava,Beneke:2015lwa} and the impact of phase space cuts for inclusive measurements~\cite{Hoang:2010gu,Beneke:2017rdn} are known to high precision as well. At this time only the Whizard event generator~\cite{Bach:2017ggt} can provide simulations that account for the bound state Coulomb effects in $t\bar t$ threshold production at NLL for the total cross section, and there is very limited knowledge on differential distributions~\cite{Hoang:1999zc,Peter:1997rk} due to the complexity of low-energy ultrasoft gluon radiation. 
This mean that some aspects related to the projections for top-quark mass measurements at the top-antitop threshold at $e^+e^-$ colliders discussed in Section~\ref{sec:TOPHF-mtop-ee} are still not yet explored based on full simulations.
The task to improve the precision of top-quark mass threshold measurements substantially below the level $50$~MeV constitutes a very challenging theoretical as well as experimental effort.
Is was also shown in Ref.~\cite{Boronat:2019cgt} that $e^+e^-\to t\bar t+\gamma$ events with an identified photon can be used to carry out radiative return measurements of the $t\bar t$ threshold at higher c.m.\ energies. Theoretically the radiative return method has in principle a precision very similar to the threshold scan, but is limited by statistics.

\FloatBarrier
\subsection{Experimental aspects}
\label{sec:TOPHF-mtop-exp}

This section discusses experimental measurements of the top-quark mass and the different measurement techniques: 
\begin{itemize}
    \item Direct top-quark mass measurements at the LHC are discussed in Section~\ref{sec:TOPHF-mtop-direct}. The measurements are limited by systematic uncertainties, in particular the jet energy calibration and the modeling of top-quark pair events. While the ultimate sensitivity is reached through combinations, the measurements to be combined need to have different sensitivities to systematic uncertainties to be able to gain in the combination. While these measurements, which measure the MC top-quark mass parameter $m_t^{\rm MC}$, will likely continue to have the smallest experimental uncertainties at the LHC, the high precision partly involves adaptations of the MC generators used for the analysis that may also affect the interpretation of $m_t^{\rm MC}$. The interpretation of $m_t^{\rm MC}$ in terms of a theoretically well-defined top-quark mass scheme adds additional uncertainties that cannot be ignored, but also cannot be quantified right now in a completely rigorous manner (see Section~\ref{sec:TOPHF-mtop-theory} for a discussion). 
    \item 
    Top-quark mass measurements in a well-defined scheme based on cross section calculations at the LHC are discussed in Section~\ref{sec:TOPHF-mtop-pole}. Since these rely on a comparison of unfolded differential distributions with higher-order theory calculations, the theory uncertainties are currently large. These top-quark mass measurements are mostly carried out in the pole mass scheme. For this method in principle all sources of uncertainties are accounted for and precision can be increased when uncertainties in the theory calculations and the unfolding procedure are reduced. 
    \item Threshold scans at $e^+e^-$ colliders are discussed in Section~\ref{sec:TOPHF-mtop-ee}. The precision of these is also limited by systematic uncertainties, though circular colliders can reach smaller statistical uncertainties than linear colliders.
    \item Measurements of the $\overline{\rm MS}$ mass from radiative events at $e^+e^-$ colliders in the continuum are also discussed in Section~\ref{sec:TOPHF-mtop-ee}.
\end{itemize}

\FloatBarrier
\subsubsection{Direct or MC mass: Measurements from top-quark decays at hadron colliders}
\label{sec:TOPHF-mtop-direct}

Direct measurements of the top-quark mass from top-quark decays is achieved primarily through reconstruction of the invariant mass of the decay products, and supplemented by independent kinematic distributions that are sensitive to the top-quark mass. The measurement precision is limited by systematic uncertainties in production and decay modeling. The most precise measurements come from $t\bar{t}$ events where at least one $W$-boson from the decay of the top quarks decays leptonically. In these events, the ambiguity regarding the momentum of the neutrino from the $W$~boson decay is offset by the precision of the measurement of the charged-lepton momentum. The precision of these measurements is at the 0.5\% level or better and is dominated by systematic effects. The mass is extracted from the detector-level kinematic distribution by comparing to MC generator predictions ($m_t^{\rm MC}$).

To account for background contribution, parton miss-assignment, and other effects, the most common top-quark mass measurement techniques use a template fit to the invariant mass, or other kinematic variable, distributions. The fit is done at the detector level, fitting MC simulation templates~\cite{Campbell:2022qmc} to the data. Experimental systematic effects, such as jet-energy scales for light and heavy quarks, have been constrained using multi-dimensional templates. A summary of top-quark mass measurements and measurement techniques is given in~\cite{PDG:2022}.

At the Tevatron, the combination of direct measurements by the CDF and D0 collaborations gives a top-quark mass of $174.30 \pm 0.35 \text{(stat.)} \pm 0.54 \text{(syst.)}$~GeV ~\cite{CDF:2016vzt}. Systematic uncertainties are large, mainly due to the jet energy calibration and the modeling of the $\ttbar$ system. Measurements by ATLAS and CMS at 7 and 8~TeV are based on much larger data sets than the Tevatron and are dominated by systematic uncertainties. The ATLAS measurement of the combined 7 and 8~TeV measurements in the dilepton and lepton+jets final states is $172.69 \pm 0.25 \text{(stat.)} \pm 0.41 \text{(syst.)}$~GeV~\cite{ATLAS:2018fwq}.
The CMS measurement of the top-quark mass, combining 7 and 8~TeV data is $172.44 \pm 0.13\, \text{(stat.)} \pm 0.47\,\text{(syst.)}$~GeV.\cite{CMS:2015lbj}. 
The most recent, and most precise, single measurement of the top-quark mass comes from the CMS collaboration, using 36 fb$^{-1}$ of $pp$ collisions at $\sqrt{s}=$13 TeV~\cite{CMS:2022kcl}. The analysis uses $t\bar{t}$ events decaying into a single electron or muon and at least four jets. A profile likelihood technique is used to constrain, and thereby reduce, systematic uncertainties. The analysis employs a kinematic fit to the $t\bar{t}$ hypothesis, and uses only the best-fit assignment (minimum $\chi^2$) of observed objects to partons. Additional kinematic variables are used in the profile likelihood fit to constrain the uncertainties due to the modeling and reconstruction of simulated events. These uncertainties are incorporated as nuisance parameters in the final maximum likelihood fit. The final result, $m_t =171.77 \pm 0.04 \text{(stat.)} \pm 0.38 \text{(syst.)}$~GeV~\cite{CMS:2022kcl}, corresponds to a remarkable 0.22\% precision. This is already approaching the latest projections of the precision at the HL-LHC.

Combinations of results in different final states and between ATLAS and CMS are not available yet. These direct measurements at the LHC are all dominated by systematic uncertainties. They are finding a region of phase space that is less sensitive to systematic uncertainties (ATLAS lepton+jets at 8~TeV) and using data to constrain the systematic uncertainties (CMS at 8 and 13~TeV).

The projections for Run~3 at the LHC with an expected integrated luminosity of 300~fb$^{-1}$ and for the HL-LHC with an expected integrated luminosity of 3000~fb$^{-1}$ are based on the 7+8~TeV CMS lepton+jets measurement with an uncertainty of 0.49~GeV~\cite{CMS:2015lbj} and also including projections for other measurement methods~\cite{ATLAS:2022hsp,CMS-PAS-FTR-16-006}. These other methods are based on single-top events or based on the characteristics of the $b$ meson from the top-quark decay provide additional information but are nevertheless limited in their systematic uncertainties.

Since the top-quark mass measurements described above are dominated by jet energy scale, promising methods rely on reconstructing the final-state leptons, mesons or baryons rather than the full jet. A study by ATLAS using J/$\psi$ mesons identified in top quark decays projects an uncertainty of $\pm 0.14 \text{(stat.)} \pm 0.48 \text{(syst.)}$~GeV~\cite{ATLAS:2018hvc}. Another new idea is to obtain the top-quark mass from the peak in the energy distribution of the $b$ quark from the top quark decay and to analyze the decay length of the $B$ meson rather than reconstructing jets~\cite{Agashe:2204.02928}, see Section~\ref{sec:TOPHF-mtop-len}. 
Resolving the difference between direct top-quark mass measurements (from the decay products, i.e. the MC mass) and a theoretically well-defined top-quark mass (see Section~\ref{sec:TOPHF-mtop-theory}) is possible by using large-momentum, boosted top-quark jets. These boosted jets make it possible to use analytic resummation rather than having to rely on parton shower simulations, see Section~\ref{sec:TOPHF-mtop-soft}. 

\begin{figure}[ht]
\centering
\includegraphics[width=0.6 \linewidth
]{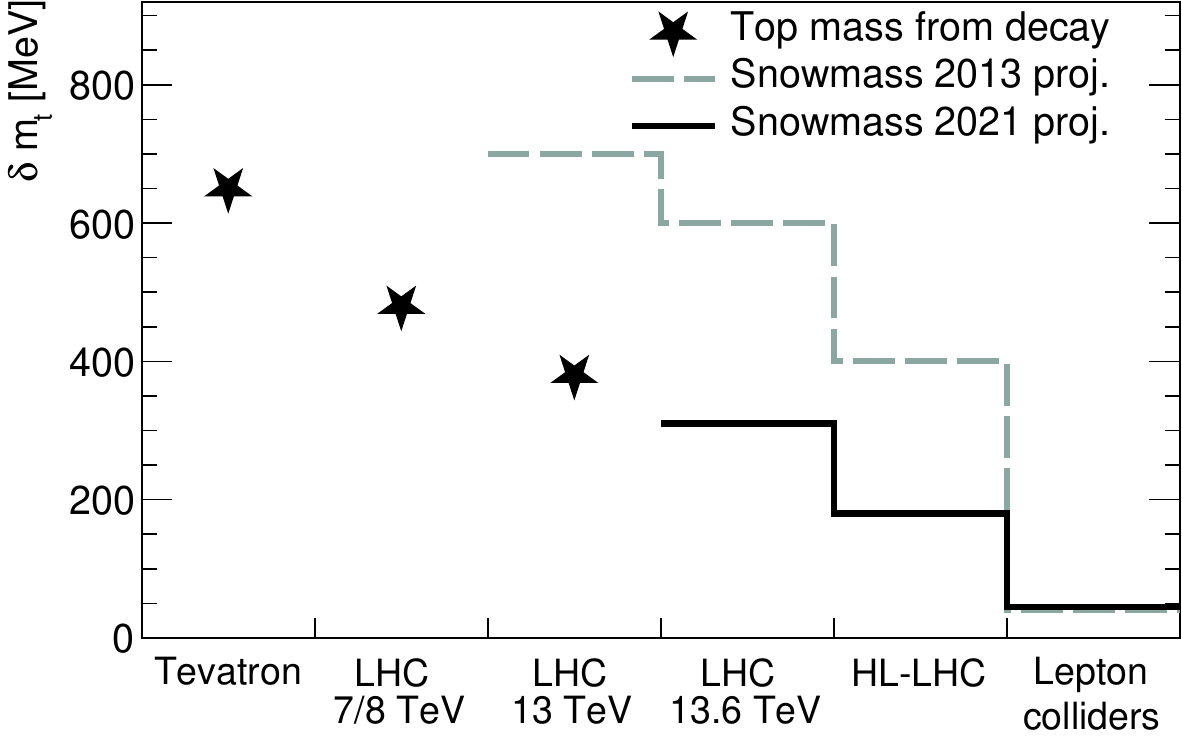}
\caption{Comparison of top-quark mass measurements from top-quark decays (MC mass or direct measurement) between measurements at the Tevatron and the LHC, and projections from Snowmass 2013 for future LHC and HL-LHC sensitivity and for a top threshold scan at a future lepton collider, and projections from Snowmass 2021. The interpretation uncertainty for the mass from decay is not included (see Figure~\ref{fig:mtop_msbar}).
}
\label{fig:mtop_direct}
\end{figure}

The measurements of the top-quark MC mass as well as the projections are summarized in Table~\ref{tab:TOPHF-MCmtop}. They are compared to the projections from Snowmass 2013 in Figure~\ref{fig:mtop_direct}. The projections from Snowmass 2013 were made in 2012, years before the current Tevatron results (2016) and LHC Run~1 results (2015 and 2018) were published. The predictions for systematic uncertainty reduction expected in 2013 were very conservative. The measurements made a lot of improvements on the understanding of the detectors and the modeling of $\ttbar$ events in the intervening years. The current projections could be seen as similarly conservative as discussed above. As discussed above, these do not include the new ideas presented in Section~\ref{sec:TOPHF-mtop-new} or combinations of different measurements that reduce the systematic uncertainty dependence further. 
However, as discussed in Section~\ref{sec:TOPHF-mtop-theory}, there is the additional ambiguity in relating these MC top-quark mass measurements with a well-defined top-quark mass scheme and the improved modelling may affect the interpretation in a non-trivial manner.

\begin{table}[ht]
\begin{center}
\begin{tabular}{l|c|c|c|c|c|c|c}
$\delta m_t^{MC}$ [MeV] & Tevatron & \multicolumn{5}{|c|}{LHC}  & HL-LHC  \\ \hline
& & \multicolumn{2}{|c|}{Run 1} & \multicolumn{2}{|c|}{Run 2} & Run 3 & \\ \cline{3-7}
& & ATLAS &  CMS & ATLAS & CMS& & \\ \cline{3-6}
$\sqrt{s}$ [TeV] & 1.96 & 7,8 & 7,8& 13 & 13& 13.6 & 14 \\
$\L [\fbinv]$    & 9.7 & 5, 20& 5, 20 & 36 & 36 & 300 & 3,000  \\ \hline
Statistical uncert. & 350 & 250 & 130 & 400 & 40&  40 & 20 \\
Systematic uncert. & 540 & 410& 470 & 670 & 380 & 300& 170 \\ \hline
Total uncert. & 650 & 480 & 480 & 780  & 380&  310 & 170  \\ \hline
\end{tabular}
\caption{Current (Tevatron~\cite{CDF:2016vzt}, LHC Run 1: ATLAS~\cite{ATLAS:2018fwq} and CMS~\cite{CMS:2015lbj} and Run 2: ATLAS~\cite{ATLAS:2019ezb} and CMS~\cite{CMS:2022kcl}) and anticipated (LHC Run 3 and HL-LHC (lepton+jets channel)~\cite{CMS-PAS-FTR-16-006})
statistical and systematic uncertainties in the measurement of the MC mass, $m_t^{MC}$ (measurement from decay), at hadron colliders. No estimates exist for the FCC-hh. Numbers are rounded to two significant digits. Interpreting this mass from decay in a well-defined scheme requires an additional uncertainty due to ambiguities in the top-quark mass definition (see Section~\ref{sec:TOPHF-mtop-theory}). }\label{tab:TOPHF-MCmtop}
\end{center}
\end{table}

\FloatBarrier
\subsubsection{Indirect or pole mass: mass measurements in well-defined schemes from top-quark production at hadron colliders}
\label{sec:TOPHF-mtop-pole}
Direct top-quark mass measurements have reached the best quoted experimental accuracy (see Section~\ref{sec:TOPHF-mtop-direct}) but the translation to a renormalized mass is not clear (see Section~\ref{sec:TOPHF-mtop-theory}). 
Determining the top-quark mass in a well-defined scheme is achieved based on differential and total cross-section measurements, unfolded to the parton level, and then compared to theoretical predictions. These predictions should be at least NLO and use a top-quark mass renormalized in a well-defined scheme. This approach is frequently called “indirect” or "from production".  At first sight, this way of proceeding seems more rigorous than the MC mass measurements, because the theoretical predictions used for the comparison with the experimental data have a well-defined accuracy in perturbative QCD (currently NLO or NNLO~\cite{Czakon:2011xx,Catani:2019iny}) or, in the most advanced cases, they also include the effects of resummation of large logarithms (e.g. threshold logarithms at NNLL). Recently, approximate N$^3$LO total cross-sections for $t\bar{t}$ production at hadron colliders have also become available~\cite{Kidonakis:2203.03698} (see Section~\ref{sec:TOPHF-XS-LHC-tt-theo}).

The indirect measurements so far have measured the top-quark pole mass, which is reasonable given that the experimental uncertainties are still much larger than the theory uncertainty due to the renormalon ambiguity (see Section~\ref{sec:TOPHF-mtop-theory}). Measurements of the $\overline{\rm MS}$ mass (which avoid this problem) are now starting, and from the experimental perspective, there is little difference between the two schemes, thus this section will focus on the pole mass. But the statements equally apply to the $\overline{\rm MS}$ mass.

When measuring the top-quark mass from unfolded differential distributions, those distributions should be used that, on the one hand, have a shape very sensitive to the top-quark mass value and, on the other hand, do not depend too much on non-perturbative QCD effects (such as, e.g., beam remnant and multiple parton interactions), which requires also careful studies of non-perturbative effects and selection of the most suitable observables. Alternatively, total cross-sections are used.

At present, the quoted uncertainties for these top-quark pole mass measurements are in general larger than those using the direct methods, but there is no additional ambiguity in the theoretical definition of the top-quark mass. A summary of recent measurements at the LHC using the indirect method is provided in the following. 

Several measurements exist from both the ATLAS and the CMS collaborations, which extract the top-quark mass in a well-defined mass renormalization scheme through the comparison of theory predictions for
total and differential cross sections, parameterized in the top-quark mass. Both the $t\bar{t}$ and the $t\bar{t}j$ processes are exploited, where for the former, both the inclusive and differential cross sections are considered, while for the latter only differential distributions are used. In all the experimental analyses, special care is taken not to introduce a large dependence on the value of the top-quark mass $m_t^{\rm{MC}}$ in the unfolding or extrapolation to the parton level. This avoids dependence on the MC generator used to produce the MC event samples which simulate the detector response. Different approaches to this problem exist, as discussed in the following for example analyses.

Owing to the large statistics available in $t\bar{t}$ events, the cleanest channel is the one in which both the $W^+$ and the $W^-$ bosons from the $t$ and $\bar{t}$ subsequently decay leptonically. In the recent ATLAS~\cite{ATLAS:2019hau} and CMS~\cite{CMS:2018fks} $\sqrt{s}=13\,$TeV analyses, only the $e\mu$ channel was used for the top-quark pole mass extraction, since background events (e.g., from $Z$ decay) are highly suppressed in this channel. This strategy was already previously employed in the $\sqrt{s}= 7$ and $8$~TeV analyses by ATLAS~\cite{ATLAS:2014nxi} and CMS~\cite{CMS:2016yys} and the ATLAS+CMS combination of 7+8~TeV measurements~\cite{ATLAS:2022aof}. In these analyses, single top $tW$ production constitutes the main background, especially because in the typical event selection, one or two $b$-tagged jets are required. For the 2 $b$-jet signature, $tW$ events enter as background only if a light quark or gluon jet is misreconstructed as a $b$-jet. The top-quark pole mass was extracted through maximising a Bayesian likelihood as function of $m_t^{\rm{pole}}$ by ATLAS, while a $\chi^2$ fit was used in the CMS analysis and the LHC combination~\cite{ATLAS:2022aof}. Further,
using the same analysis set-up,
the CMS analysis~\cite{CMS:2018fks} also extracted  the top-quark mass in the $\overline{\rm{MS}}$ renormalization scheme. The influence of the $m_t^{\rm{MC}}$ mass was estimated through the variation of the experimental acceptance and the $tW$ background and was found to be small in the ATLAS analysis, while the $m_t^{\rm{MC}}$ value was fitted as a nuisance parameter in the CMS analysis and subsequently fixed to the fitted value in the extraction of $m_t^{\rm{pole}}$ and $m_t^{\overline{\rm{MS}}}$. The theoretical uncertainties in these extractions are larger than the experimental ones, mainly from PDF, $\alpha_S$ and scale variations. Reducing the uncertainties due to PDF and $\alpha_S$ can be accomplished through precision QCD measurements, while reducing the scale uncertainty requires higher-order calculations~\cite{QCDreport}. 

Information on the $t\bar{t}$ process at the differential level provides complementary information that has smaller dependence on PDF and $\alpha_S$ uncertainties. A study of double-differential distributions in the dilepton ($e\mu$) channel by the ATLAS collaboration led to the extraction of the top-quark pole mass using various techniques, like fits using templates from NLO~+~PS simulated event samples, fits to fixed-order NLO QCD predictions and mass determinations using moments of distributions~\cite{ATLAS:2017dhr}. On the other hand, the CMS analysis of triple-differential cross-sections in Ref.~\cite{CMS:2019esx} allowed for a simultaneous determination of PDFs, $\alpha_S$ and the top-quark pole mass, preserving the correlations among these quantities. A subsequent study demonstrated the possibility of applying the same methodology to the determination of the top-quark mass in the $\overline{\rm{MS}}$ and MSR schemes, with similar fit uncertainties as in the pole mass case, as reported in Table~3 of Ref.~\cite{Garzelli:2020fmd}, and reduced theory uncertainties due to the lack of mass renormalon ambiguities in the two short-distance mass renormalization schemes. In particular, this method to determine of $m_t(m_t)$ may lead to reduced uncertainties with respect to all other methods used so far for this quantity, as shown in Fig.~18 of Ref.~\cite{Garzelli:2020fmd}. 

The $t\bar{t}j$ process was considered by both the ATLAS and the CMS collaborations to extract the top-quark mass. The $\rho_S$ distribution was used, which is proportional to the inverse of the invariant mass of the $t\bar{t}j$ system (see Eq.~\ref{eq:TOPHF-mtr}). This was first introduced as a sensitive distribution to the top-quark mass and further it was shown that the sensitivity is enhanced for the $t\bar{t}j$ process with respect to the $t\bar{t}$ one, since the light jet QCD emission is quite sensitive to the mass of the radiating quark~\cite{Alioli:2013mxa}. The ATLAS analyses at $\sqrt{s}=7$~\cite{ATLAS:2015pfy} and $8$~TeV~\cite{ATLAS:2019guf} and the CMS analysis at $\sqrt{s}=7\,$TeV~\cite{CMS:2016khu} use the lepton+jets channel.  The strategy of these analyses is to unfold the experimental data to the parton level and then perform a $\chi^2$ fit with fixed-order theoretical predictions, calculated using as input different top-quark mass values. The mild dependence on the $m_t^{\rm{MC}}$ parameter is checked by looking at the dependence of the fitted unfolding matrix on this parameter.  In the most recent CMS $\sqrt{s}=13\,$TeV~\cite{CMS:2022emx} analysis, the $e^+e^-$, $\mu^+ \mu^-$ and $e^+\mu^-$ dilepton channels were used. This study came to the conclusion that the uncertainty related to the $\chi^2$ fit, PDF and extrapolation (effect of the relevant theoretical uncertainties on the signal acceptance) is larger than the one due to scale variation. 

Considering the most accurate analyses so far, the uncertainties on top-quark mass extraction using the $t\bar{t}j$ $\rho_s$ distribution are larger than those using $t\bar{t}$ double and triple differential cross-section distributions due to the large uncertainty of PDFs at large $x$, considering that the $t\bar{t}j$ process is sensitive to larger $x$ values than the $t\bar{t}$ one. In the future, the possibility of using double-differential distributions also for $t\bar{t}j$ production should definitely be exploited~\cite{Gombas:2203.08064}. To get the most out of them, however, it is mandatory to increase the accuracy of $t\bar{t}j$ theoretical predictions in the comparison with the experimental data, by incorporating NNLO corrections and further higher-order effects. The reduction of large-$x$ PDF uncertainties, as foreseen thanks to new data from the Electron-Ion Collider (EIC)~\cite{Accardi:2012qut}, and simultaneous analyses of LHC and EIC data will be crucial to further explore correlations between $m_t$, PDFs and $\alpha_s$ and pin down the residual uncertainties. 

\begin{figure}[ht]
\centering
\includegraphics[width=0.6 \linewidth
]{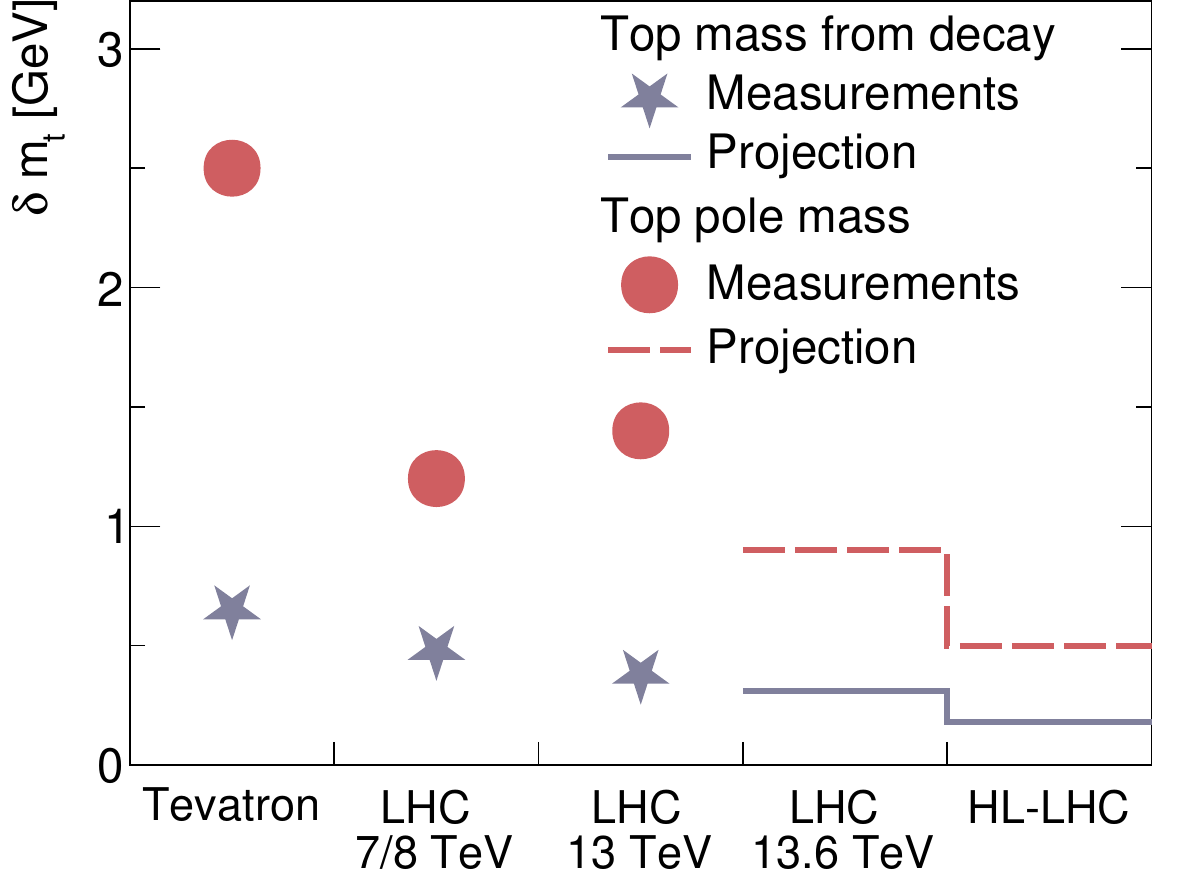}
\caption{Comparison of top-quark pole mass and mass from decay measurements at the Tevatron and the LHC, and projections for future LHC sensitivity.
}
\label{fig:mtop_pole}
\end{figure}

The measurements and projections for the top-quark pole mass are shown in Table~\ref{tab:TOPHF-Polemtop}, and graphically in Figure~\ref{fig:mtop_pole}. The pole mass measurement using the D0 differential cross section is $169.1 \pm 2.5$~GeV~\cite{Czakon:2016teh}. The pole mass measured by ATLAS in Run 1 at the 8~TeV LHC in $ttj$ events is $171.1 \pm 0.4 \text{(stat.)} \pm 0.9 \text{(syst.)} ^{+0.7}_{-0.3}$~(theo.)~GeV~\cite{ATLAS:2019guf}. The best Run I measurement by CMS is based on $\sigma_{t\bar t}$ at NNLO+NLL: $173.8^{+1.7}_{-1.8}$ GeV~\cite{CMS:2016yys}. 
The pole mass measured by CMS in Run 2 in $ttj$ events at the 13~TeV LHC is $172.94 \pm 1.27 \text{(fit)} \pm 0.5$~(scale)~GeV~\cite{CMS:2022emx},
where the fit includes contributions from the PDF uncertainty in the theory prediction. The mass has also been extracted in a fit to $\ttbar$ differential distribution by CMS~\cite{CMS:2019esx}, with an experimental uncertainty of only 0.8~GeV. 
For this measurement, a +1 GeV theory uncertainty is added as an estimate of Coulomb and soft-gluon resummation effects which were not accounted for in the theory prediction (see also Section~\ref{sec:TOPHF-mtop-theory} for a discussion of the status of theory predictions for the $t\bar t$ invariant mass distribution at threshold). Pole mass measurements from $ttj$ events in Run~3 and at the HL-LHC are expected to have experimental uncertainties of 0.5~GeV or better. 

No projections exist for the top-quark pole mass, the following assumptions are made here: Combining ATLAS and CMS measurements, and combining measurements from $t\bar{t}$ and $ttj$ events should reduce the Run~2 experimental uncertainty significantly, by about a factor of two, to 0.8~GeV. The theory predictions in the fit used by ATLAS and CMS are the same, thus this uncertainty is only reduced through additional effort to calculate higher-order terms and to reduce the PDF uncertainties, It is assumed that the Run~3 measurement can accomplish an experimental uncertainty of 0.8~GeV, and that this can be reduced to 0.4~GeV at the HL-LHC. It is assumed that the theory uncertainty for Run~3 is unchanged at 0.5~GeV, and that this is reduced by a factor two for HL-LHC.

The total uncertainty is currently dominated by theoretical uncertainties due to PDFs. The central value for the top mass extraction from $ttj$ events changes by almost a GeV for the recent CMS measurement~\cite{CMS:2022emx} when comparing the ABMP16 and CT18 PDF sets. For the projection, it is assumed here that the theory uncertainty can be reduced through dedicated PDF fits in $t\bar{t}$ events with the large expected samples to 0.5~GeV in Run~3 and to 0.25~GeV at the HL-LHC~\cite{Gombas:2203.08064}, and through higher-order calculations of the $ttj$ process.

\begin{table}[ht]
\begin{center}
\begin{tabular}{l|c|c|c|c|c}
$\delta m_t^{pole}$ [GeV] & Tevatron & LHC Run 1 & LHC Run 2 & LHC Run 3 & HL-LHC  \\ \hline
$\sqrt{s}$ [TeV] & 1.96 & 7/8 & 13 & 13.6 & 14  \\
$\L [\fbinv]$       & 10 & 20 & 140 & 300 & 3,000    \\ \hline
Experimental uncertainty & 2.2 & 1.0 & 1.3 & 0.8 & 0.4 \\
Theoretical uncertainty  & 1.4 & 0.7 & 0.5 & 0.5 & 0.25 \\ \hline
Total uncertainty        & 2.5 & 1.2 & 1.4 & 0.9 & 0.5 \\ \hline
\end{tabular}
\caption{Current (Tevatron~\cite{Czakon:2016teh}, LHC Run 1~\cite{ATLAS:2019guf}, and LHC Run 2) and anticipated (Run 3 and HL-LHC) 
experimental and theoretical uncertainties in the measurement of the top-quark pole mass, $m_t^{pole}$ (indirect measurement) at hadron colliders. Note that the theory uncertainty quoted for the LHC Run~2 does not include PDF uncertainty contributions, those are included in the experimental uncertainty~\cite{CMS:2022emx}. }
\label{tab:TOPHF-Polemtop}
\end{center}
\end{table}

\FloatBarrier
\subsubsection{Top-quark mass measurements at \texorpdfstring{$e^+ e^-$}{e+e-} colliders}
\label{sec:TOPHF-mtop-ee}

Electron-positron colliders offer in principle three main avenues towards measuring the top-quark mass. The most precise one is a scan of the top-quark pair production threshold, which will be discussed in the following. In addition, the top-quark mass can be measured in the continuum above the threshold, either by kinematic reconstruction of the invariant mass of the decay products or via the cross section of radiative top-quark pair events~\cite{Boronat:2019cgt}. While the former suffers from the same interpretation uncertainties as the standard measurements at the LHC, the latter is conceptually the same as a scan of the production threshold, using the detected ISR photon to determine the effective collision energy. These two techniques are particularly relevant in collider scenarios where a top-quark threshold scan will only be performed after extended running at energies above the threshold, such as running scenarios discussed for CLIC and ILC. A direct measurement of the top-quark mass, $m_t^{MC}$, at CLIC at $\sqrt{s}=380$ GeV with 500 $\fbinv$ in the $l+jets$ (all hadronic) channel is estimated to be possible with a statistical uncertainty of 30 MeV (40 MeV)~\cite{CLICdp:2018esa}. In Ref.~\cite{Boronat:2019cgt} it is estimated that the $\overline {\rm MS}$ mass can be measured in radiative events at ILC and CLIC with a total uncertainty of 360(150) MeV ($\sqrt{s}=500$ GeV and ${\cal L}=500(4000)\fbinv$) and 150(110) MeV ($\sqrt{s}=380$ GeV and ${\cal L}=500(1000)\fbinv$, respectively. This estimate includes statistical, experimental systematic and theory uncertainties.    
Already prior to the experimental observation of the top quark, the potential for a determination of its mass and other properties via the measurement of the cross section in the threshold region compared to theory predictions was recognized \cite{Fadin:1987wz,Fadin:1988fn,Strassler:1990nw}.

\begin{table}[t]
\begin{center}
\begin{tabular}{l|c|c|c}
$\delta m_t^{{\rm PS}}$ [MeV] & ILC & CLIC  & FCC-ee \\ \hline
$\L [\fbinv]$              & 200  & 100 [200] & 200    \\ \hline
Statistical uncertainty &  10  & 20 [13] & 9  \\
Theoretical uncertainty (QCD) &  \multicolumn{3}{c}{40 -- 45}  \\
Parametric uncertainty $\alpha_s$  & 26 & 26 & 3.2 \\
Parametric uncertainty $y_t$ (HL-LHC) & \multicolumn{3}{c}{5}\\
Non-resonant contributions & \multicolumn{3}{c}{$<40$}\\
Experimental systematic uncertainty  &  \multicolumn{2}{c|}{15 -- 30} & 11 -- 20 \\
\hline
Total uncertainty       &  \multicolumn{3}{c}{40 -- 75} \\ \hline
\end{tabular}
\caption{Anticipated statistical and systematic uncertainties in the measurement of the top-quark threshold mass, $m_t^{{\rm PS}}$, from a threshold scan around 350 GeV obtained with a one-dimensional fit of the top-quark mass, keeping $\Gamma_t$, y$_t$, and $\alpha_s$ fixed. CLIC assumes a lower integrated luminosity than the other facilities. For comparison, the statistical precision achievable with 200 \fbinv for CLIC is also given. It should be noted that the results shown for ILC and FCC-ee assume a 8-point scan with a compressed energy range which improves sensitivity for $m_t^{{\rm PS}}$ at the expense of $y_t$ sensitivity. For the standard 10-point scan assumed for CLIC the statistical uncertainties would be 12 and 10 MeV for ILC and FCC-ee, respectively. The uncertainty due to the current world average for $\alpha_S$ is shown for ILC and CLIC, while for FCC-ee, the precision of $\alpha_s$ obtained with the run at the $Z$ pole (Tera-Z) is assumed. Concrete studies for CEPC are not yet available, but it can be assumed that uncertainties are similar as for FCC-ee.  See text for further details.}\label{tab:TOPHF-PSmtop}
\end{center}
\end{table}

Table~\ref{tab:TOPHF-PSmtop} shows the projected uncertainties for the threshold mass, $m_t^{{\rm PS}}$, for ILC, CLIC and FCC-ee obtained from the scan of the top-pair production threshold, assuming that all other parameters ($\Gamma_t$, y$_t$, $\alpha_s$) are fixed. The results for the different facilities are all based on the same study, using common assumptions for the reconstruction efficiency and background levels \cite{Seidel:2013sqa}. 
The machine-specific aspects are brought in by different luminosity spectra, and thus may not fully reflect the differences between experiments at the different accelerators and the experiments. The main systematic uncertainties are expected to be largely facility-independent, with small differences on theoretical and parametric uncertainties introduced by the energy range of the scan coupled with the shape of the threshold turn-on, which depends on the luminosity spectrum. 

For the parametric uncertainties, additional differences can arise when different precision for the underlying parameters are assumed. The current world average of $\alpha_s$ has an uncertainty of $9 \times 10^{-4}$, or 0.8\%. This corresponds to a parametric uncertainty on $\alpha_s$ of 26~MeV. The uncertainty on $\alpha_s$ is expected to be reduced by a factor two in the near term, and by a factor eight in the long term~\cite{dEnterria:2022hzv}. 
For FCC-ee, the high-statistics $Z$-pole running is expected to provide $\alpha_s$ with a precision of $1.2\,\times\, 10^{-4}$, which reduces the corresponding uncertainty on the top-quark mass to 3.2~MeV~\cite{Bernardi:2022hny}. 
For the other collider options, this uncertainty will reduce according to the expected ultimate precision on the strong coupling constant. Analogous to the strong coupling, the top Yukawa coupling also introduces a parametric uncertainty, which amounts to approximately 5~MeV for the projected precision of HL-LHC of 3.4\%. 
Non-resonant contributions such as single top production has been shown to be smaller than 40 MeV \cite{Fuster:2014hfw}

Experimental systematic uncertainties originate from different sources, which have been evaluated with varying degree of precision to date. Event selection and residual background uncertainties are expected to contribute on the level of 10 -- 20 MeV \cite{Seidel:2013sqa}. In the absence of a full threshold event generator these are based on generic assumptions for the accuracy of the signal efficiency and the precision of the knowledge of the residual background. For linear colliders, the more complex corrections for the luminosity spectrum result in uncertainties of less than 10 MeV \cite{Simon:2014hna}. Uncertainties on the beam energy directly enter into the mass determination. For FCC-ee, it is assumed that the beam energy can be determined to a precision of 5 MeV at the relevant beam energies \cite{Blondel:2019jmp}, which, translates into a 3 MeV uncertainty on the top quark mass. For linear colliders, conservative estimates assumed the beam energy systematic on the top-quark mass to be below 17 MeV \cite{Seidel:2013sqa}. Considering that the same integrated luminosity is assumed for both circular and linear colliders, and the fact that at the top-quark pair threshold the powerful method of resonant depolarisation will not be applicable for circular colliders, it is plausible that this uncertainty will be similar for both collider types. 

Overall, the total uncertainty is expected to range from 40 -- 75 MeV, with possibly slightly smaller uncertainties for FCC-ee than for linear colliders due to the absence of more complex corrections for the shape of the luminosity spectrum. In view of the dominant theory uncertainties the differences between different colliders are not significant in terms of the overall precision. With improvements in theory, particularly on QCD, and ultimate precision in the strong coupling constant, further reductions of the total uncertainty are possible.

\FloatBarrier
\subsection{Top-quark mass summary}
\label{sec:TOPHF-mtop-sum}

\begin{figure}[ht]
\centering
\includegraphics[width=0.6 \linewidth
]{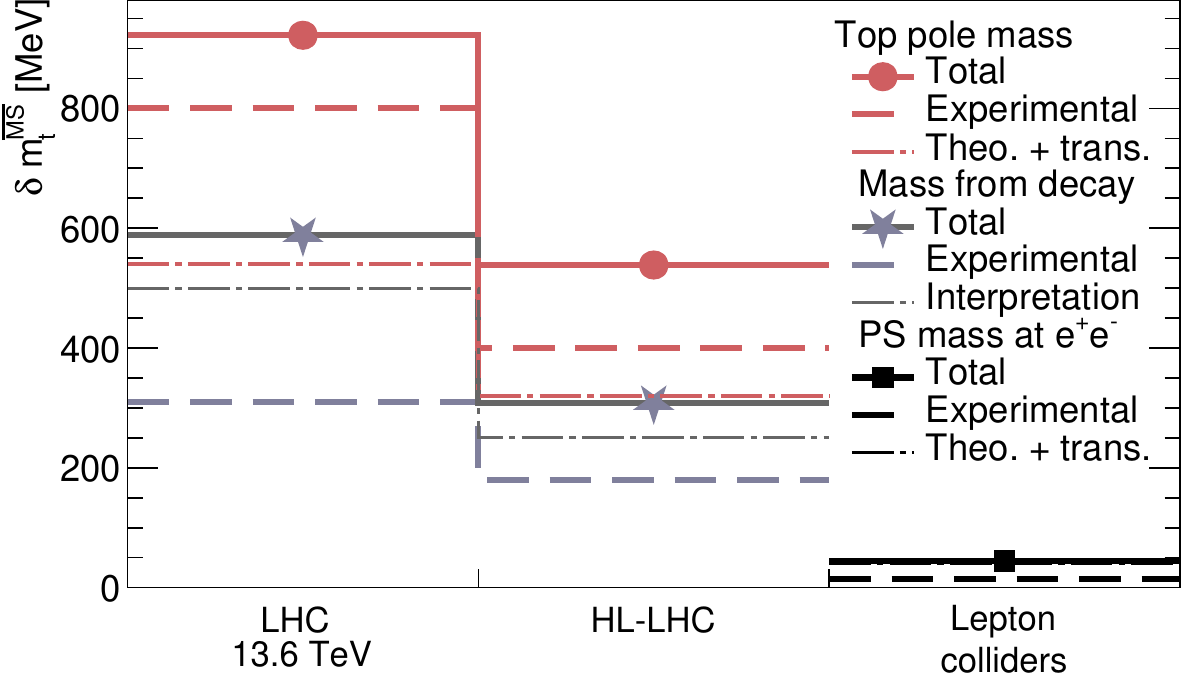}
\caption{Comparison of projected $\overline{\rm MS}$ top-quark mass determinations from decay and pole-mass measurements at the LHC and HL-LHC and from a PS mass measurement at a future lepton collider. The dashed-dotted lines show the approximate uncertainty in interpreting the mass from decay as $\overline{\rm MS}$ top-quark mass~\cite{Nason:2017cxd,Hoang:2020iah} (labeled \it{Interpretation}) or the combined uncertainty from theory and the conversion to the $\overline{\rm MS}$ scheme (labeled as \it{Theo.+trans.}).    
}
\label{fig:mtop_msbar}
\end{figure}

To be able to compare the projections for the different top-quark mass measurements, they have been converted to projections for the $\overline{\rm MS}$ top-quark mass, shown in Figure~\ref{fig:mtop_msbar}.  As described in Section~\ref{sec:TOPHF-mtop-theory}, the scheme conversion from the PS mass to the $\overline{\rm MS}$ mass is known to ${\cal O}(\alpha_s^4)$ precision and is at the level of 10-20 MeV while the pole mass suffers from an intrinsic renormalon ambiguity estimated to be at the level of 110-250 MeV. The uncertainty in the interpretation of the mass from decay (or MC mass) as pole mass is estimated to be around 500 MeV. Note that the LHC and HL-LHC projections do not take into account anticipated improvements due to new ideas for top-quark mass measurements described in Section~\ref{sec:TOPHF-mtop-new}.

\FloatBarrier
\subsection{New ideas for top-quark mass measurements}
\label{sec:TOPHF-mtop-new}

This section summarizes contributed white papers and some recent publications on new ideas for top-quark mass measurements. 

\subsubsection{Toward a model-{\em in}dependent 
measurement of the top-quark mass using \texorpdfstring{$B$}{B}-hadron decay length}
\label{sec:TOPHF-mtop-len}
Summary of white paper contribution~\cite{Agashe:2204.02928}

\begin{figure}[ht]
\centering
\includegraphics[width=0.48 \linewidth
,angle=0]{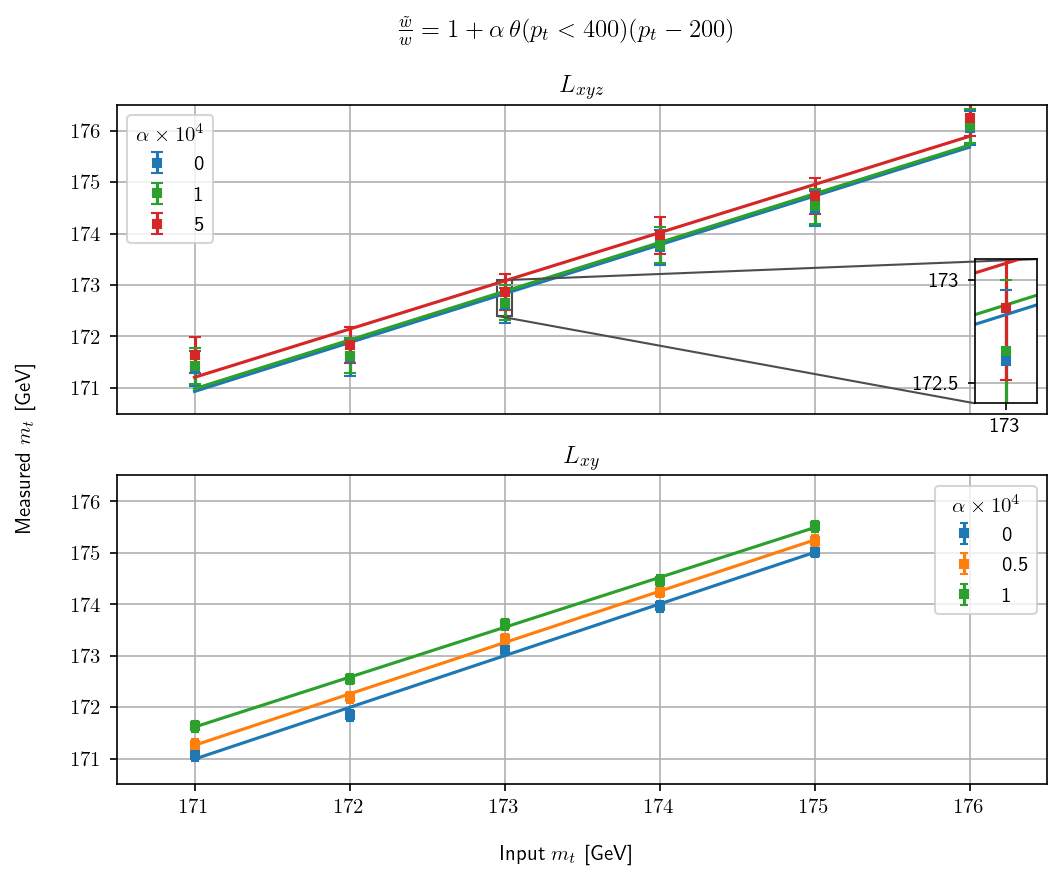}
\includegraphics[width=0.48 \linewidth
,angle=0]{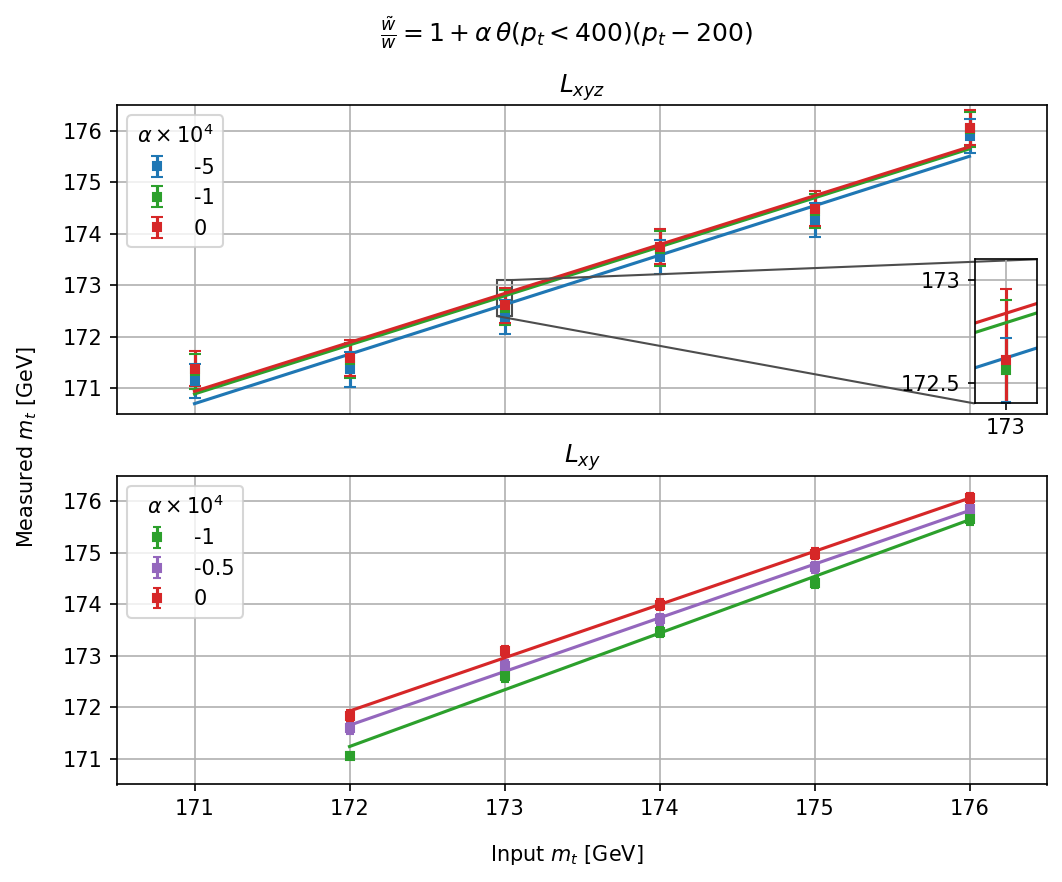}
\caption{Comparison of the extracted top-quark mass from pseudo-data with reweighted $p_{T,top}$ kinematics. Results for template fitting using the $L_{xy}$ observable with a hard-wired $b$-quark energy distribution (bottom panel), assuming SM production of the top quark, and the energy-peak method described in the text (top panel). Results on the left (right) correspond to hardening (softening) of the $p_{T,top}$ spectrum of the pseudo-data. Taken from Ref.~\cite{Agashe:2204.02928}.}
\label{fig:money-us-vs-Lxy}
\end{figure}

The ``energy-peak'' idea \cite{Agashe:2012bn} can furnish a measurement of the top-quark mass via the energy of the bottom quark from its decay, which, based on the ``parent-boost-invariance,'' is less sensitive to details of the production mechanism of the top quark (cf.~most other methods assume purely SM production of the top quarks, hence are subject to uncertainties therein, including a possible BSM contribution). The original proposal along this line was to simply use the $b$-jet energy as a very good approximation to the bottom-quark energy. This method has been successfully implemented by the CMS collaboration \cite{CMS-PAS-TOP-15-002}. However, the $b$-jet energy-peak method is afflicted by the jet-energy scale (JES) uncertainty. Fortunately, this drawback can be circumvented by using the decay length of a $B$-hadron contained in the $b$-jet as a proxy for the bottom-quark energy \cite{Hill:2005zy}. 
The new, interesting proposal presented in~\cite{Agashe:2204.02928}
is to then appropriately dovetail the above two ideas resulting in a 
''best of both worlds'' determination of the top-quark mass, i.e., based on a measurement of the $B$-hadron decay length, but improved by the energy-peak concept: this would be free of the JES uncertainty and largely independent of the top-quark production model. The main result is summarized in Fig.~\ref{fig:money-us-vs-Lxy}, where the error in the extraction of top-quark mass is shown using $B$-hadron decay length due to a re-weighting of the top-quark $p_T$ distribution: one can see that this energy-peak method is much less sensitive to this uncertainty than a similar analysis done by CDF \cite{CDF:2006irv} and CMS \cite{CMS-PAS-TOP-12-030}, but assuming SM production instead.

\subsubsection{Precision top-quark mass using soft drop jet mass}
\label{sec:TOPHF-mtop-soft}
Summary of a to-be-submitted journal paper relevant to future top-quark mass measurements~\cite{Hoang:2022xxx}

Boosted top quarks are an ideal setting for performing analytical resummation of top-quark mass sensitive observables. In the boosted limit, rigorous factorization formulae in the framework of SCET and HQET for event shape observables in the peak region, such as the jet mass, can be derived~\cite{Fleming:2007qr,Fleming:2007xt}, and can be analytically resummed to high accuracy~\cite{Bachu:2020nqn} with complete control over the top-quark mass scheme. Furthermore, the SCET framework also provides insights into describing the nonperturbative power corrections~\cite{Lee:2006fn} in terms of a few universal nonperturbative constants, and in a completely model independent fashion. 

However, the setup for $e^+e^-$ event shapes cannot be simply carried over to the case of the LHC. In addition to generalizing global event shapes to measurements on jets, one also has to account for the presence of the underlying event (UE) and pile up, which otherwise contaminate the measurement and spoil sensitivity to the top-quark mass. 
This approach can be extended to the LHC by considering inclusive jets where rigorous factorization formulae~\cite{Kang:2016mcy} can be established. To overcome the effects of underlying event and pile-up soft drop grooming can be employed on jets prior to the jet mass measurement.

Soft drop grooming~\cite{Larkoski:2014wba} is a procedure of systematically removing soft radiation and contamination from the underlying event and pile up from jets, while retaining ability to perform analytical resummation of IRC safe observables measured on the groomed jet. Among those observables is the jet mass $M_J$ which
is defined by starting with the constituents of the jet of radius $R$ and summing only over those constituents that remain in the groomed jet, $J_{sd}$: $M_J^2=(\sum_{J_sd} p_i^\mu)^2$.
This so-called soft drop jet mass can be analytically resummed to high accuracy~\cite{Frye:2016aiz,Kang:2018jwa,Benkendorfer:2021unv,Kardos:2020gty}. 
 Furthermore, there exists a field theory based formalism to describe the hadronization corrections in this observable~\cite{Hoang:2019ceu} in terms of a few 
${\cal O}(\Lambda_{\rm QCD})$ constants, $\{\Omega_{1\kappa}^{\circ\!\!\circ}, \Upsilon_{1,0}^\kappa, \Upsilon_{1,1}^\kappa\}$, that only depend on the jet initiating parton $\kappa$ (quark or gluon), and perturbatively calculable coefficients $C_{n}(m_J^2, z_{\rm cut}, \beta, p_{T\rm jet}, \eta_J)$~\cite{Pathak:2020iue} that capture the entire functional dependence of these hadronization corrections on the jet mass, kinematic variables and soft drop parameters.

 \begin{figure}[ht]
    \centering
    \includegraphics[width=0.6\textwidth]{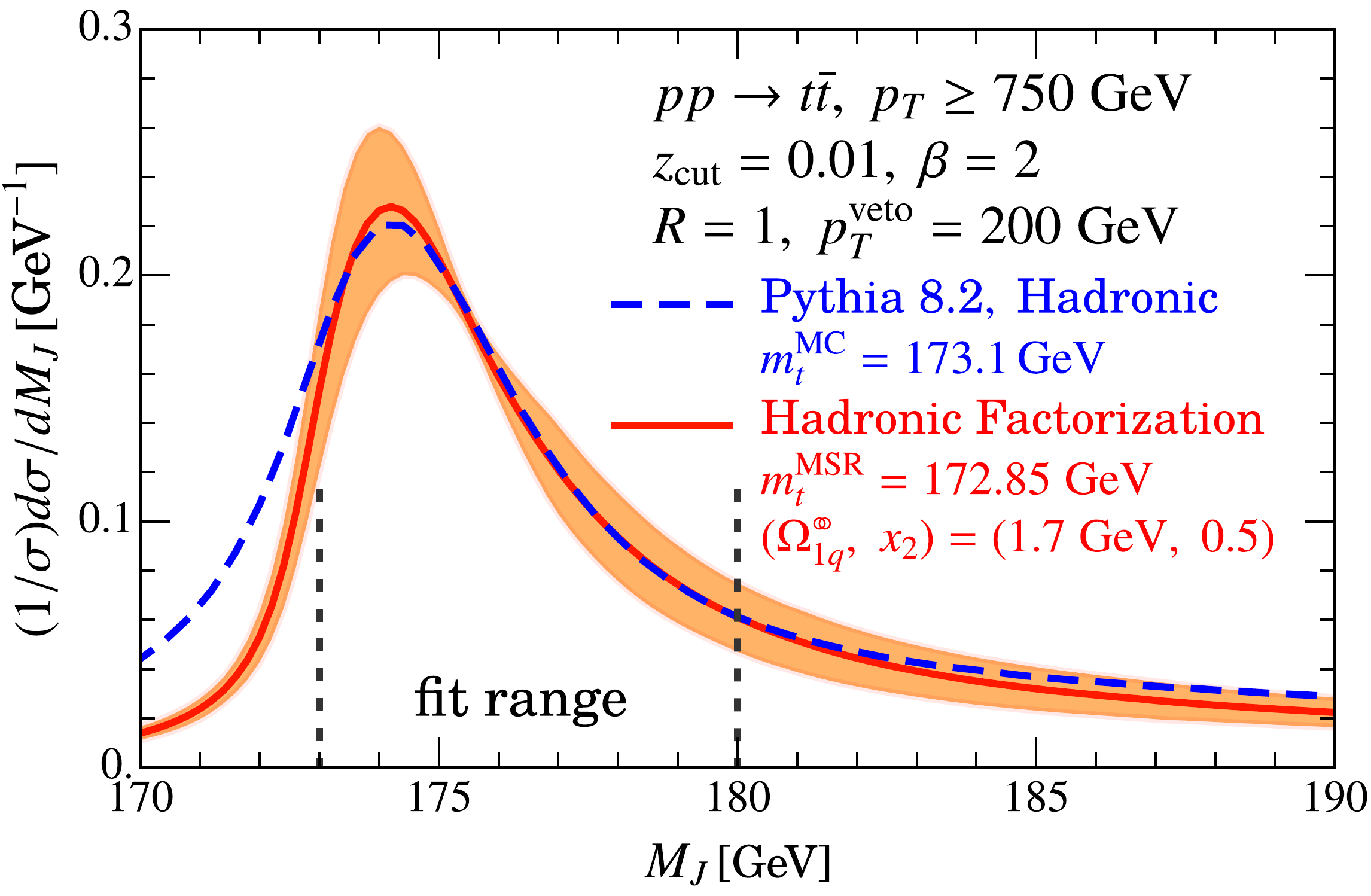}
    \caption{Groomed top-quark jet mass ($M_J$) spectrum at NLL~\cite{Hoang:2017kmk} compared with \textsc{Pythia}8.}
        \label{fig:pptop}
\end{figure}
In Ref.~\cite{Hoang:2017kmk}, soft drop jet mass measurement on a sample of inclusive boosted top jets was proposed as a top-quark mass sensitive observable for precision top-quark mass determination, see Fig.~\ref{fig:pptop}. The extension of light quark and gluon groomed jet mass calculations to the case of top quarks involves a careful study of how the top-quark decay products can interfere with the soft drop grooming algorithm, and thus modify both the perturbative and the nonperturbative~\cite{Hoang:2017kmk} structure of the theory prediction. Fortunately, in the limit of light soft drop grooming the decay products screen the peak region from grooming and simplify the theory treatment, while maintaining robustness against soft contamination. This also results in effectively only a single parameter $\Omega_{1t}^{\circ\!\!\circ}$ being relevant for the leading hadronization corrections. Hence, the method does not rely on MC estimates of hadronization effects. Likewise higher order corrections can be systematically incorporated~\cite{Hoang:2022xxx}.
Consequently, a concrete comparison of unfolded experimental data and hadron-level theory predictions calculated in a definite top-quark mass scheme and becomes foreseeable. 

As a first step, a calibration of the MC top-quark mass parameter in \textsc{Pythia}8 has been carried out at hadron level~\cite{ATLAS:2021urs} in collaboration with ATLAS which has demonstrated the feasibility of this program at the HL-LHC. The aim is to sufficiently improve perturbative control and sensitivity of the observable to achieve an ${\cal O}$(1 GeV) precise top-quark mass measurement overcoming the theoretical issues associated with direct measurements of the MC top-quark mass parameter.
Future improvements will involve extensions of the formalism for nonperturbative corrections to account for effects of the underlying event~\cite{Ferdinand:2022xxx} as well as constraining them by exploiting groomed light quark and gluon jets. 
\subsubsection{Precision top-quark mass using energy correlators}
\label{sec:TOPHF-mtop-corr}
Summary of a journal submission~\cite{Holguin:2022epo} relevant to future top-quark mass measurements

In~\cite{Holguin:2022epo}, the measurement of the top-quark mass was explored using statistical correlators of energy flow operators~\cite{Sveshnikov:1995vi,Tkachov:1995kk,Korchemsky:1999kt,Bauer:2008dt,Hofman:2008ar,Belitsky:2013xxa,Belitsky:2013bja,Kravchuk:2018htv}. These correlators directly measure moments of the QCD energy-momentum tensor, naturally suppressing effects from soft physics due to the energy weights employed. This feature elegantly overcomes complications from soft resummation, hadronization, and underlying event contamination --- which can be major challenges for measurements made in the LHC environment. The three-point correlator (EEEC), on a QCD jet, can be computed through the $n$th-weighted cross-section
\begin{align} \label{eq:measurementLHC}
&\frac{\dd \Sigma_{\mathrm{EEEC}}}{\dd p_{T,\text{jet}} \dd \zeta_{12} \dd \zeta_{23} \dd \zeta_{31}} = \\ 
&\sum_{i,j,k\,\in\, {\rm jet}} \int \dd \sigma_{ijk}  \frac{(p_{T,i})^{n}(p_{T,j})^{n}(p_{T,k})^{n}}{(p_{T,\text{jet}})^{3n}} 
\delta\left(\zeta_{12} - \hat{\zeta}_{ij}\right)\delta\left(\zeta_{23} - \hat{\zeta}_{ik}\right)\delta\left(\zeta_{31}- \hat{\zeta}_{jk}\right)\,, \nonumber
\end{align}
for $n \geq 1$. Here $\sigma_{ijk}$ is the inclusive cross-section for the production of individual hadrons $i,j,k$ in a final-state jet with transverse momentum $p_{T,\mathrm{jet}}$ (we consider here inclusive top-quark jets decaying hadronically). Here $\zeta_{ij} = \Delta R_{ij}$ are the relative angles between the final state hadrons. This measurement is under theoretical control both when computed on all hadrons or whilst restricting to charged particles~\cite{Chen:2020vvp,Li:2021zcf,Jaarsma:2022kdd}. 

\begin{figure}
  \centering
    \includegraphics[width=0.6\textwidth]{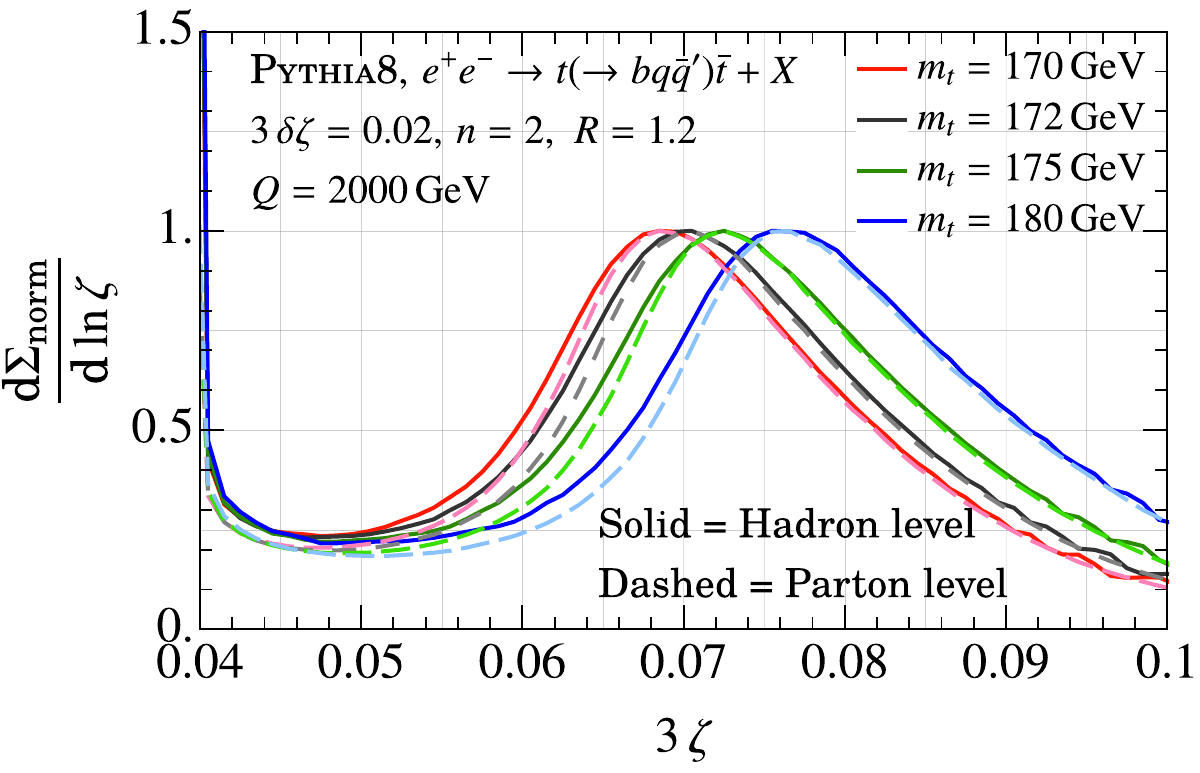} 
  \caption{The $n=2$ three-point correlators on boosted top quarks in $e^+e^-$ showing a clear peak at $\zeta \sim 3m_t^2/Q^2$. All graphs are normalized to peak height.}\label{fig:sensitivity}
\end{figure}

These correlation functions have historically seen interest in the Conformal Field Theory community, wherein they typically exhibit a featureless power law scaling characteristic of an asymptotically free theory. However, the EW decay of the heavy top quark breaks this scaling and imprints itself as a distinct peak in the three-point correlation function at angle $\zeta_{\rm peak} = \sum_{i,j} \zeta_{ij} \sim m_t^2/p_{T,\mathrm{jet}}^2$. This angle is determined by the boost of the top quark --- making the measurement theoretically cleanest when performed on a sample of highly boosted top quarks. In~\cite{Holguin:2022epo}, it was demonstrated that by taking various limits of the parameter-space of the EEEC the top-quark mass peak can be enhanced, providing high sensitivity to $m_{t}$. Evidence (found through a MC study, see Fig.~\ref{fig:sensitivity}) was also presented which demonstrated that the enhanced peak region can be adequately described in fixed order perturbation theory in a definite top-quark mass scheme. 

In the LHC environment, the determination of the $p_{T,\rm jet}$ spectrum is impacted by detector sensitivity, underlying event, and hadronization. This in turn affects the determination of the $m_{t}$ from the peak location\footnote{An error $\delta_{p}$ in the $p_{T,\rm jet}$ leads to an error on $m_{t}$ of $\delta_{p}m_{t}/p_{T,\rm jet}$. For intuition, presently quoted uncertainties by ATLAS \cite{ATLAS:2020cli} on the jet-energy scale amount to about $1\%$ for jets with $p_T$ between $200$GeV and $2000$GeV.}. Without controlling for the $p_{T,\rm jet}$ spectrum, errors on the top-quark mass extraction will likely be comparable with those from other measurements determined by hard physics, for instance the total top-quark production cross section. However, in~\cite{Holguin:2022epo} it was demonstrated that modelling of the $p_{T,\rm jet}$ spectrum could be used to greatly improve the uncertainty. Furthermore, there is also good reason to expect that the dependence on the $p_{T,\rm jet}$ spectrum could be model independently reduced or even overcome entirely. Regions of the EEEC measurement parameter space are also sensitive to the $W$-mass at a characteristic angle $\sim m_W^2/p_{T,\mathrm{jet}}^2$ \cite{Holguin:toappear}. This allows for a measurement of the top-quark mass as a function of the, more precisely known $W$-mass rather than as a function of the $p_{T,\rm jet}$ spectrum.


\subsubsection{Top-quark mass extraction from \texorpdfstring{$t\bar{t}j +X$}{tt+X} events at the LHC: theory predictions}
\label{sec:TOPHF-mtop-ttj}
Summary of white paper contribution~\cite{Alioli:2022ttk}

Many analyses have focused on the extraction of the top quark mass, using different direct and indirect methods. Among the latter, we consider in particular the possibility of determining the top quark mass by comparing experimental data for specific differential cross-sections with the corresponding theory predictions. One of the advantages of this method is the possibility to extract the top-quark mass in well defined mass renormalization schemes, used to perform the theoretical calculations of the cross-sections (see Section~\ref{sec:TOPHF-mtop-theory}. A distribution particularly sensitive to the top-quark mass value, and thus well suited for its determination, is the so-called $\mathcal{R}$ distribution in $t\bar{t}j + X$ hadroproduction events~\cite{Alioli:2013mxa, Fuster:2017rev}. $\mathcal{R}$ is built from the $\rho_s$ distribution, which, in turn, is inversely proportional to the invariant mass of the $t\bar{t}{j}$ system $s_{t\bar{t}j}$, i.e.
\begin{equation}\label{eq:TOPHF-mtr}
  {\cal R}(m_t^R,\rho_s)= 
  \frac{1}{{\ensuremath{\sigma_{t\bar t + \textnormal{\scriptsize 1-jet}}}}} 
  \frac{d{\ensuremath{\sigma_{t\bar t + \textnormal{\scriptsize 1-jet}}}}}{d\rho_s}(m_t^R, \rho_s), \,\,\, \mathrm{with}\,\,\,
  \rho_s \,=\, \frac{2 m_0} {\sqrt{{\ensuremath{s_{t\bar t j}}}}} \, .
\end{equation}
where $m_t^R$ denotes the top-quark mass value in the $R$ renormalization scheme, whereas $m_0$ is a constant parameter of the order of the top-quark mass itself (here $m_0$ is fixed to the value 170 GeV throughout). It has been shown that the shape of the $\mathcal{R}$ distribution is extremely sensitive to $m_t^R$ and that this sensitivity increases when using samples of $t\bar{t}j +X $ events instead of samples of $t\bar{t} + X$ events. Events for the latter process, in fact, could also be used for building a $\rho_s$ and $\mathcal{R}$ distribution, with definition similar to Eq.~\ref{eq:TOPHF-mtr} but replacing $s_{t\bar{t}j}$ with $s_{t\bar{t}}$ and $\sigma_{t\bar{t} + \textnormal{\scriptsize 1-jet}}$ with $\sigma_{t\bar{t}}$.  
In the following the discussion concentrates on $\rho_s$ and $\mathcal{R}$ distributions in $t\bar{t}j + X$ production in $pp$ collisions at the LHC. These distributions were already used in some experimental analyses for top-quark mass extraction~\cite{ATLAS:2015pfy, CMS:2016khu, ATLAS:2019guf} with data from collisions at $\sqrt{S} = 7$ and 8~TeV and are still used in further ongoing analyses with data at $\sqrt{S} = 13$~TeV. 
Predictions for these distributions and the associated theoretical uncertainties will be presented here, based on the studies of Refs.~\cite{Alioli:2022lqo, Alioli:2022ttk}. These studies include QCD radiative corrections at NLO and focus on the case of stable top quarks. In fact, the experimental collaborations have developed sophisticated methods to reconstruct the top quarks from their decay products and they indeed apply these techniques in their $t\bar{t}j + X$ analyses devoted to the $m_t^R$ extraction~\cite{ATLAS:2019guf,CMS:2019esx}. The results obtained so far have shown that off-shell and spin-correlation effects do not produce modifications of the $\mathcal{R}$ distribution substantial enough to induce relevant shifts in the value of the extracted top quark mass, with respect to the case where these effects are neglected. On the other hand, parton shower emissions turn out to be more relevant and they have to be taken fully into account in the reconstruction of the top quarks. All the available predictions have been collected on a website, \texttt{https://ttj-phenomenology.web.cern.ch/},
from where they can be downloaded as tables of numerical values, ready for use in the experimental analyses or for further phenomenological studies. 

\begin{figure}[htb]
  \begin{center}
\includegraphics[width=0.43\textwidth]{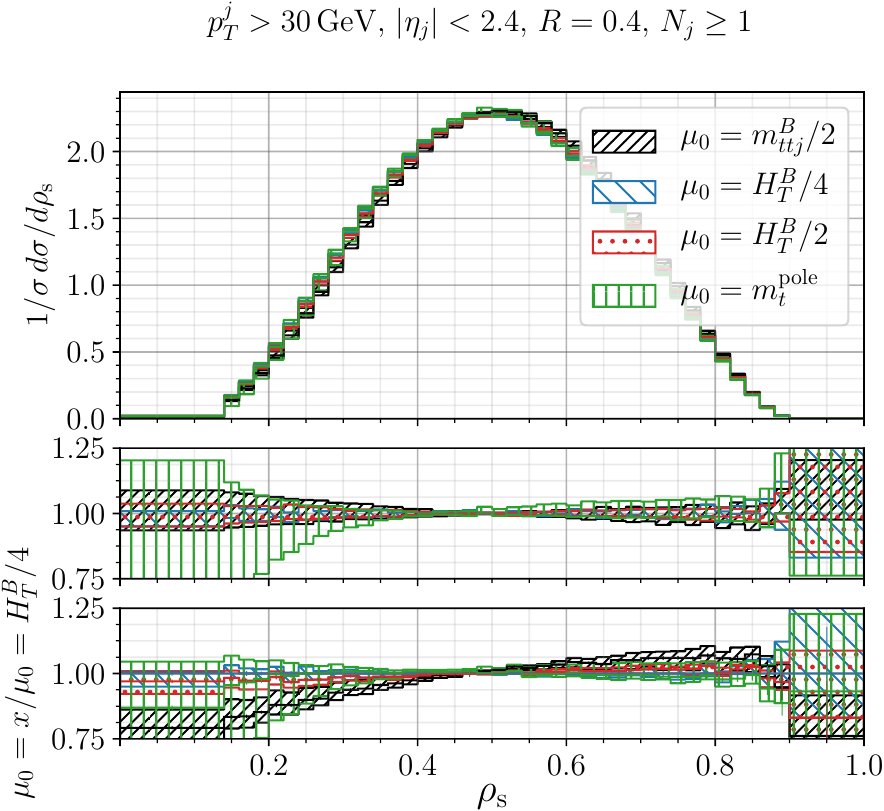}
\includegraphics[width=0.46\textwidth]{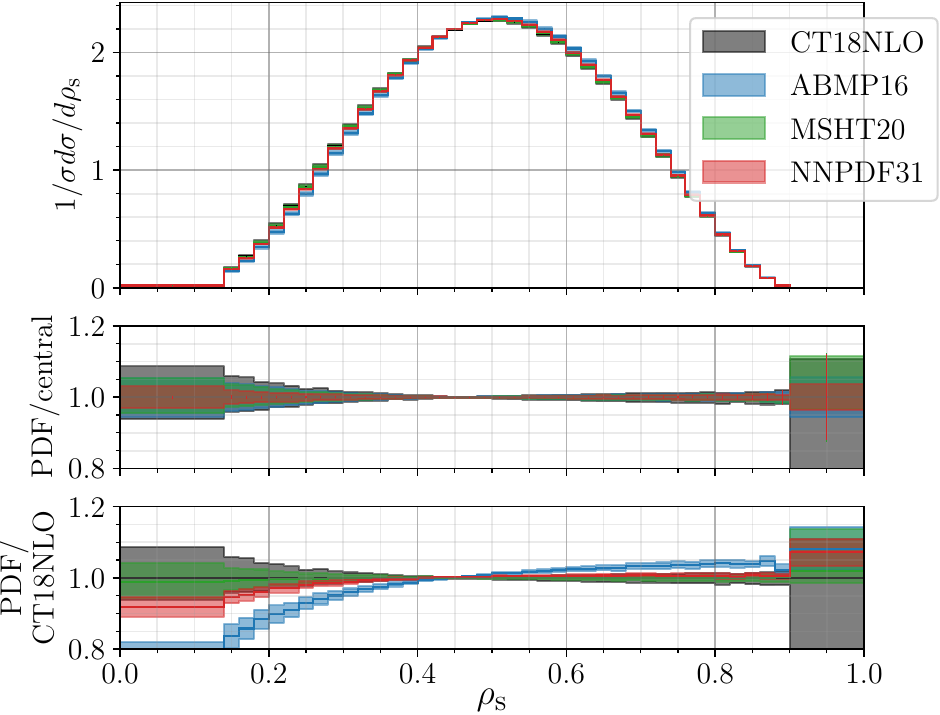}\\
\caption{
NLO QCD normalized $\rho_\mathrm{s}$, i.e. $\mathcal{R}$, distribution of the process $pp \rightarrow t\bar{t}j +X$ at $\sqrt{s}=13$~TeV, including:
scale uncertainties ({\it left panel}), with the scales $\mu_0=m_t^{\mathrm{pole}}$ (green), $m_{t\bar{t}j}^B/2$ (black), $H_T^B/2$ (red) and $H_T^B/4$ (blue). The scale variation uncertainty bands are obtained by taking the envelope of the seven-point scale variation graphs, while the prediction obtained with $K_R = K_F = 1$ is shown as solid line;
approximate PDF uncertainties ({\it right panel}),
obtained in a computation with LO matrix-elements and NLO PDFs and $\alpha_S$ evolution, using as input the dynamical scale $\mu_0=H_T^B/4$ and the CT18NLO (black), ABMP16 (blue), MSHT20 (green) and NNPDF3.1 (red) NLO PDF sets. Each PDF uncertainty is calculated as recommended by the authors of the corresponding PDF fit. The CT18NLO PDF uncertainty is rescaled from the 90\% Confidence Level to the 68\% Confidence Level as provided by the other PDF fits.
}
\label{fig:TOPHF_rho_norm}
\end{center}
\end{figure}

Present experimental analyses are especially focused on relatively large $\rho_s$ values. The progressive accumulation of high-statistics experimental data will make possible to extend the $\rho_s$ interval, progressively covering more extreme (i.e. lower and larger) $\rho_s$ values, hence exploiting the mass sensitivity on a broader range of $\rho_s$. However, using the forthcoming high-statistics data in an extended $\rho_s$ interval requires high-accuracy predictions, deep understanding of the perturbative behaviour of the calculation and of its dependence on further inputs like $\alpha_s(M_Z)$, the PDFs and the jet reconstruction procedure. Systematic studies have been performed in Ref.~\cite{Alioli:2022lqo} and recapitulated for this study in~\cite{Alioli:2022ttk}.
In the following, we summarize the main observations and recommendations resulting from these studies: 
\begin{itemize}
\item 
in the computation of the $\mathcal{R}$ distribution use dynamical scales: in particular the choice $\mu_0 = H_T/4$ turns out to be particularly interesting because of the perturbative convergence and minimization of the size of scale uncertainties. While the static scale $\mu_0 = m_t^R$ can still be regarded as a good choice for large $\rho_s$ values, the aforementioned dynamical scale choice is proven to perform definitely better in case of $\rho_s < 0.4$, as shown in Fig.~\ref{fig:TOPHF_rho_norm}, left panel. 

\item use state-of-the-art PDF fits, with particular attention to the large-$x$ region. This region, where present PDFs are still quite uncertain, is in fact definitely spanned when computing predictions at small $\rho_s$ values. NLO PDF uncertainties on $\mathcal{R}$ using a number of currently available PDF sets are shown in Fig.~\ref{fig:TOPHF_rho_norm}, right panel.

\item use short-distance top-quark mass renormalization schemes, free of renormalon ambiguities, as a viable alternative to the on-shell scheme. At large $\rho_s$, where threshold effects become relevant, the ${{\rm \overline{MS}}}$ scheme does not prove to be competitive, whereas the MSR scheme is expected to be a viable choice over the whole $\rho_s$ range, both in line of principle and according to results of this study. The predictions for $t\bar{t}j + X$  cross-sections in the MSR scheme represent the first example of calculations in this scheme for this process. More studies and analyses on the systematics and potential benefits (and/or shortcomings) inherent the extraction of the top-quark mass in this scheme from $t\bar{t}j + X$ events are indeed welcome.

\item develop methodologies to go beyond NLO accuracy in predictions 
for $t\bar{t}j + X$ (differential) cross-sections with stable top quarks: including NNLO radiative corrections and the effects of resummation of different kinds of large logarithms might be important especially in those regions where scale uncertainties are particularly large. 

\item further refine the experimental top-quark reconstruction procedures. It is expected that matching calculations for $t\bar t j + X$ production with full off-shell effects to parton shower approaches will be particularly useful in this respect. As discussed in Section~\ref{sec:TOPHF-XS-LHC-ttX}, full off-shell NLO QCD calculations for the $t\bar t j + X$ process are available~\cite{Bevilacqua:2015qha,Bevilacqua:2016jfk}, and it has been shown that in the fiducial phase-space regions proper modeling of the decay of top quarks plays a key role~\cite{Bevilacqua:2017ipv}.

\end{itemize}
In order to facilitate analyses following these directions, it is planned to go on keeping up-to-date the aforementioned website of $t\bar{t}j + X$ predictions, in such a way to reflect the latest theoretical and experimental developments from both the authors of this contribution and other groups.

\subsubsection{Dependence of the top-quark mass measured in top-quark pair production on the parton distribution functions at the LHC
and future colliders}
\label{sec:TOPHF-mtop-pdf}
Summary of white paper contribution~\cite{Gombas:2203.08064}

The PDF uncertainty is the largest systematic uncertainty in the extraction of the top-quark mass. It affects the experimental measurements (kinematic dependence of the acceptance) and the theoretical cross-section predictions, where it is the largest contribution to the uncertainty. 
White paper~\cite{Gombas:2203.08064} studies the PDF uncertainty and its effect on the top quark pole mass at the HL-LHC and the future FCC-hh collider. The measurements of the top-quark pole mass can be improved by simultaneously updating the PDF best fit while fitting the top quark mass~\cite{Kadir:2020yml}. The impact of the PDF uncertainty is studied in $\ttbar$ events and in top-pair plus jet events ($ttj$, see Section~\ref{sec:TOPHF-mtop-ttj}).

\begin{figure}[ht]
  \begin{center}
\includegraphics[width=0.48\textwidth]{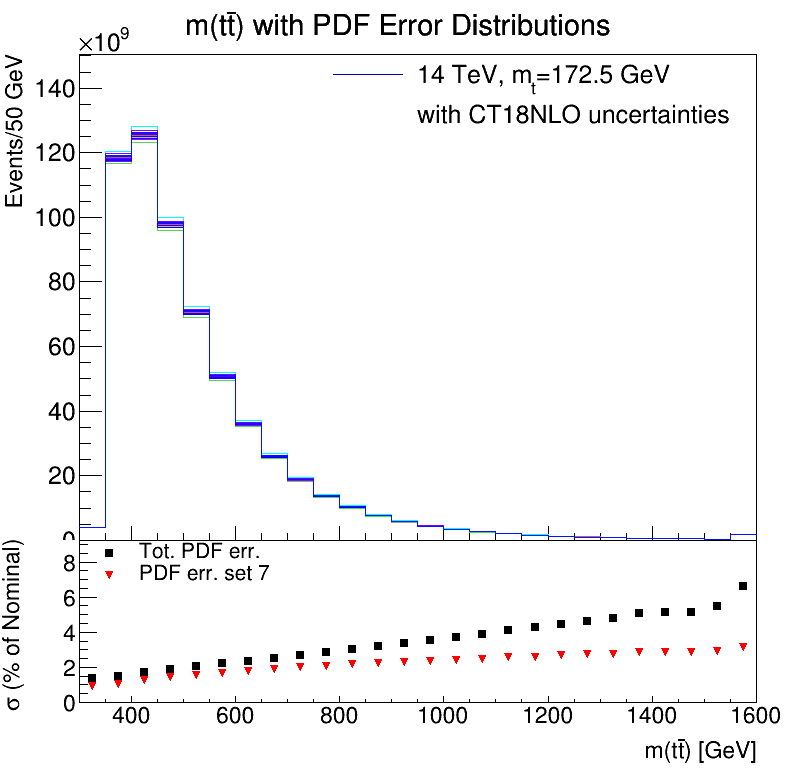}
\includegraphics[width=0.48\textwidth]{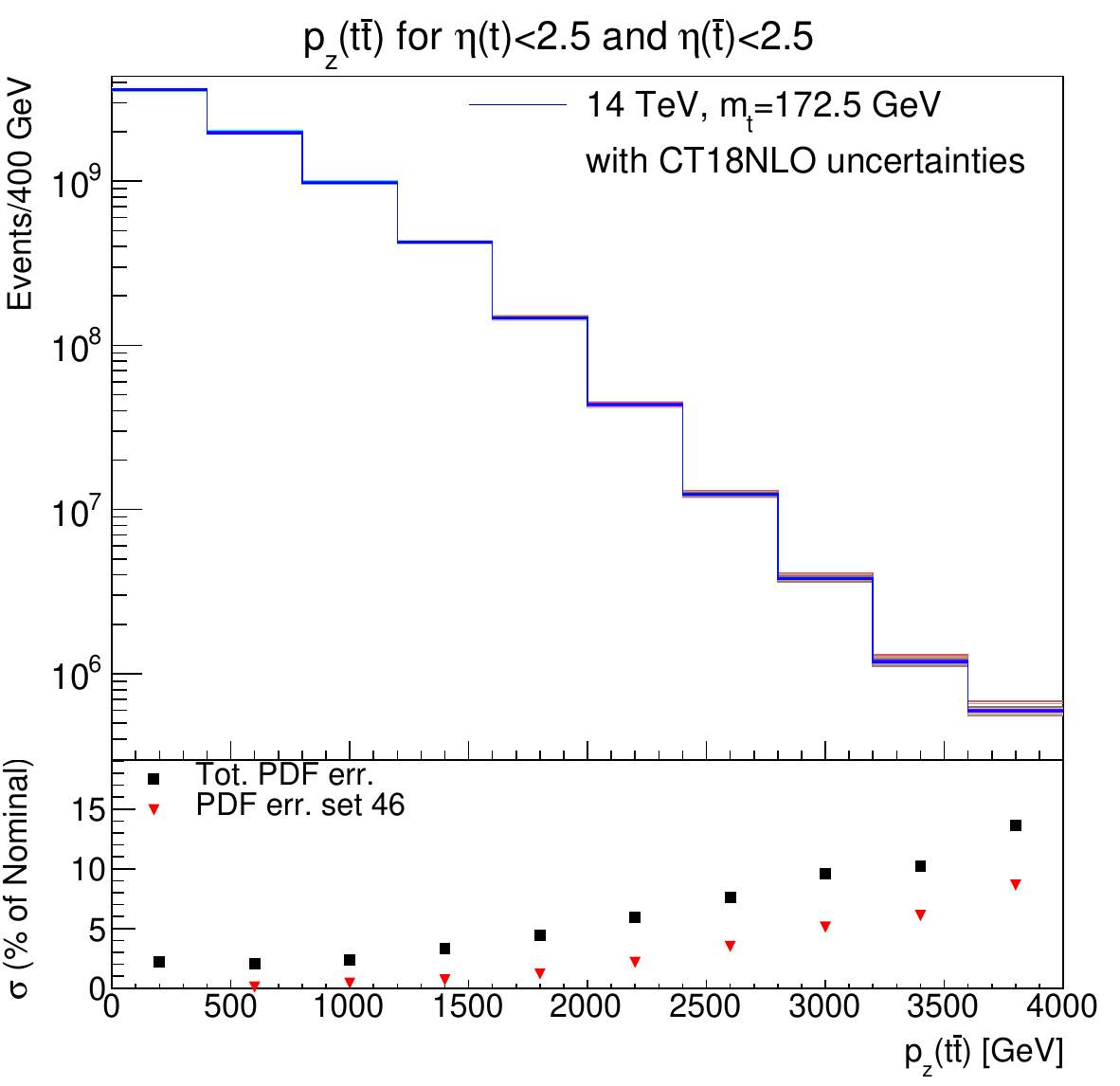}
\caption{ Distribution of the (left) invariant mass of the $\ttbar$ system and (right) the longitudinal momentum $p_z$ of the $\ttbar$ system at the 14 TeV HL-LHC. The $p_Z$ distribution includes only events for which the top and antitop quarks both are in the central part of the detector where they can be reconstructed (rapidity~$ < 2.5$). The ratio panel shows the total PDF uncertainty for the CT18NLO PDF set, and the contribution from largest of the CT18NLO PDF Eigenvectors. From~\cite{Gombas:2203.08064}.
}
\label{fig:TOPHF_mtoppdf1}
\end{center}
\end{figure}

Figure~\ref{fig:TOPHF_mtoppdf1} shows the invariant mass distribution of the $\ttbar$ system, the distribution that is commonly used to extract the top quark pole mass in $\ttbar$ events because the threshold region provides sensitivity to the top-quark mass. Figure~\ref{fig:TOPHF_mtoppdf1} also shows the $p_Z$ distribution of the $\ttbar$ system, requiring that both top quarks are in the central part of the detector (rapidity~$< 2.5$). The PDF uncertainties are also shown, they are 2\% in the threshold region of the mass distribution and up to 15\% for the $p_Z$ distribution, though different PDF Eigenvectors mainly affect the two~\cite{Hou:2019efy}.

\begin{figure}[ht]
  \begin{center}
\includegraphics[width=0.48\textwidth]{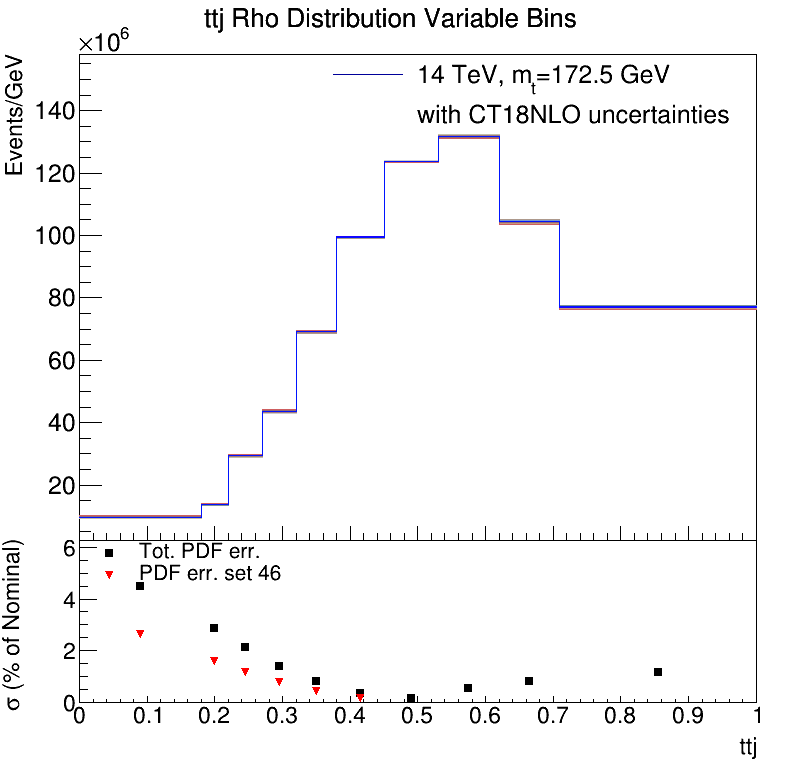}
\includegraphics[width=0.48\textwidth]{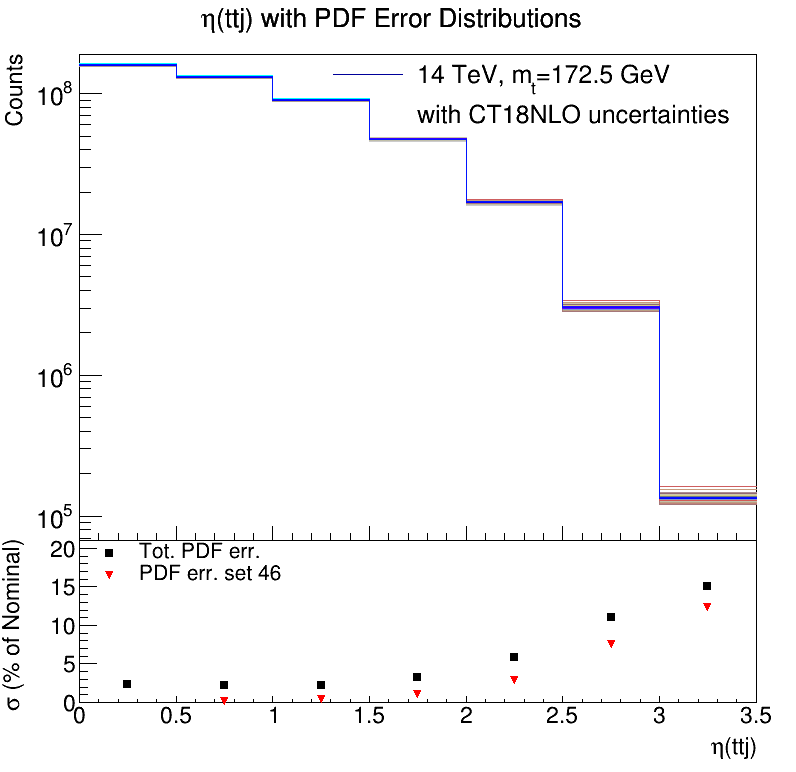}
\caption{ Distribution of the (left) the $\rho$ variable and (right) the rapidity of the $ttj$ system at the 14 TeV HL-LHC. The ratio panel shows the total PDF uncertainty for the CT18NLO PDF set, and the contribution from largest of the CT18NLO PDF Eigenvectors. From~\cite{Gombas:2203.08064}.
}
\label{fig:TOPHF_mtoppdfttj1}
\end{center}
\end{figure}

Figure~\ref{fig:TOPHF_mtoppdfttj1} shows the distribution of the $\rho$ variable, defined in Section~\ref{sec:TOPHF-mtop-ttj}, for $ttj$ events. The region of high $\rho$ has excellent mass sensitivity, but also a PDF uncertainty of a little less than 2\%. This can be reduced by fitting PDFs, for example to the rapidity distribution of the $ttj$ system, also shown in Figure~\ref{fig:TOPHF_mtoppdfttj1}.

\begin{figure}[ht]
  \begin{center}
\includegraphics[width=0.48\textwidth]{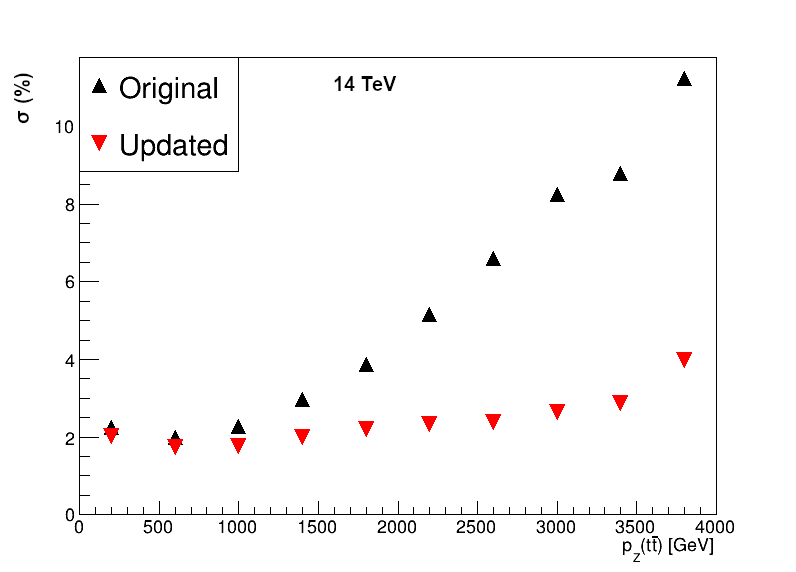}
\includegraphics[width=0.48\textwidth]{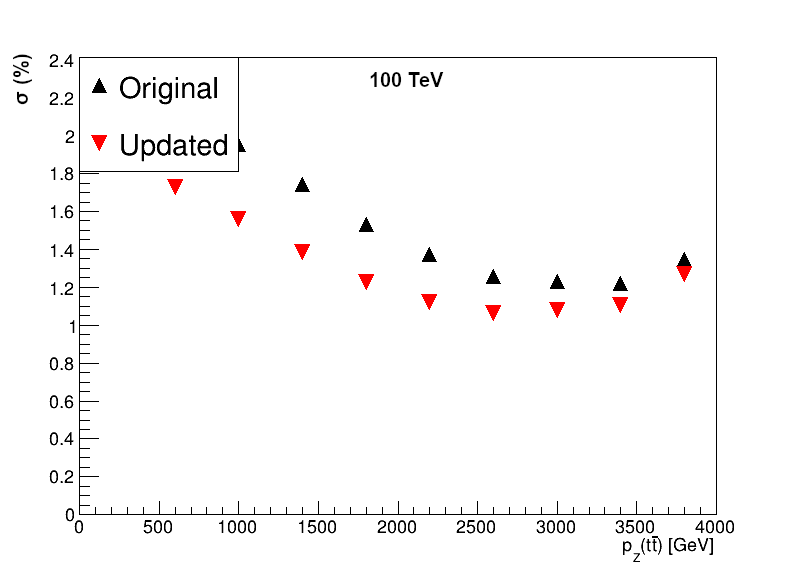}
\caption{Distribution of the improvement of the PDF uncertainty on the top-quark mass in \ttbar events from the fit to the differential $p_z$ distribution shown in Figure~\ref{fig:TOPHF_mtoppdf1}, for (left) the 14~TeV HL-LHC and (right) the 100~TeV FCC-hh. From~\cite{Gombas:2203.08064}.
}
\label{fig:TOPHF_mtoppdf2}
\end{center}
\end{figure}

The $p_z$ distribution is used to reduce the PDF uncertainties in $\ttbar$ events using ePump~\cite{Schmidt:2018hvu}, assuming an uncorrelated experimental uncertainty of 1\% in each bin of the data distributions, representing what should achievable by the end of the HL-LHC. Figure~\ref{fig:TOPHF_mtoppdf2} shows how the PDF uncertainties are reduced for this choice, at the 14~TeV HL-LHC and the 100~TeV FCC-hh. 

With this simple fitting procedure, the uncertainty from PDFs on the top quark can be reduced by up to 20\% in $\ttbar$ and $ttj$ events.

\FloatBarrier
\subsubsection{Optimising top-quark pair-production
threshold scan at future \texorpdfstring{$e^+ e^-$}{e+e-} colliders}
\label{sec:TOPHF-mtop-genetic}
Summary of white paper contribution~\cite{Nowak:2103.00522}

One of the important tasks for future $e^+e^-$ colliders is to measure the top-quark mass and width in a scan of the top-pair production threshold. This scan requires large statistics samples at multiple CM energies and thus takes a lot of time and other resources. The shape of the cross-section in the turn-on region as a function of CM energy is well known theoretically~\cite{Hoang:1999zc}. Therefore, the time and resources required can be optimized by selecting the CM energies and luminosity to collect at each CM energy. 

\begin{figure}[!h!t]
  \begin{center}
\includegraphics[width=0.49\textwidth]{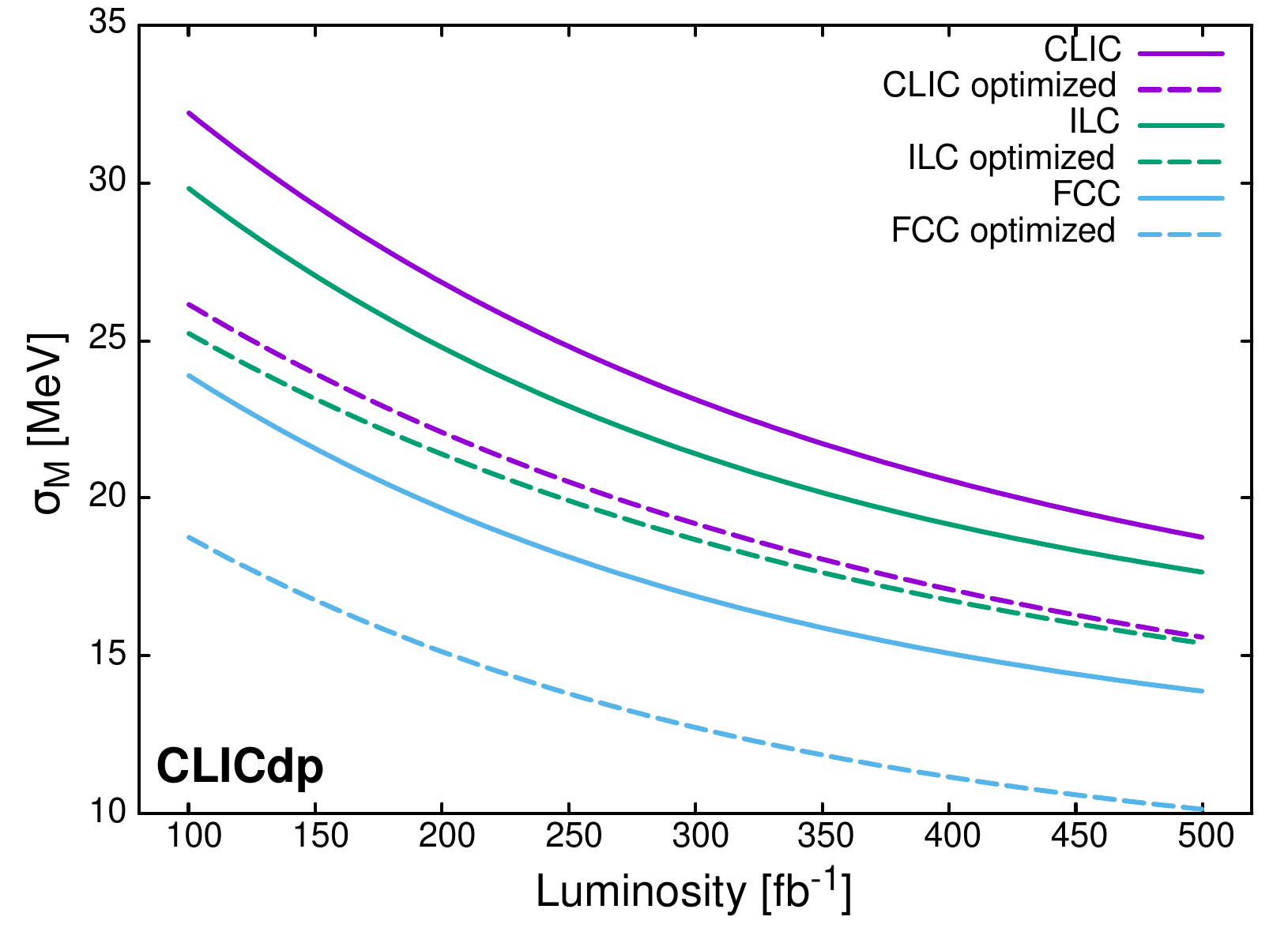}
\includegraphics[width=0.49\textwidth]{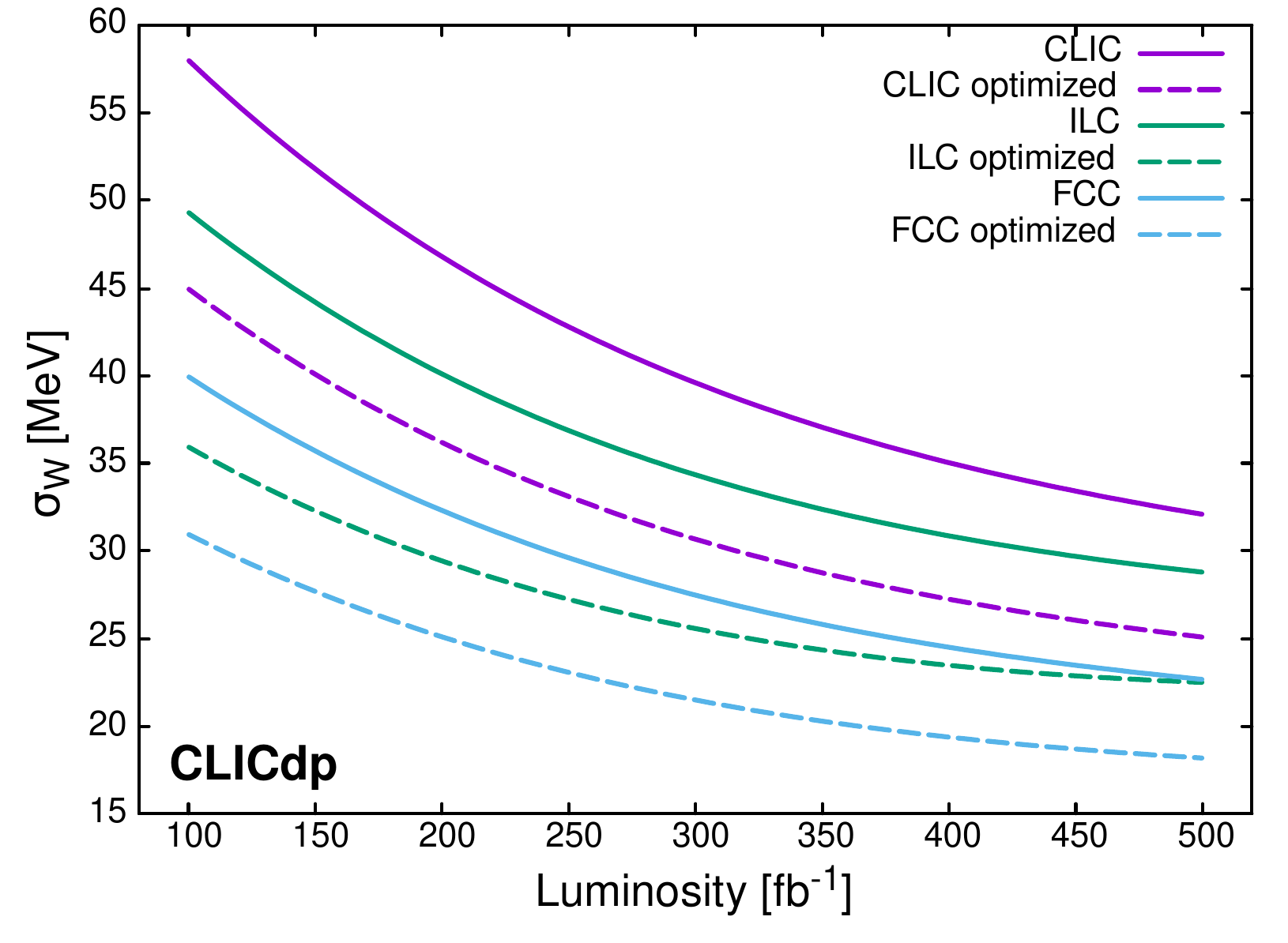}
\caption{Scan points for the top quark mass and width measurements optimized by a genetic algorithm, overlayed on the $\ttbar$ cross-section as a function of CM energy at a $e^+e^-$ collider. From~\cite{Nowak:2103.00522}.
}
\label{fig:TOPHF_mtop_gene1}
\end{center}
\end{figure}

However, while the shape of the pair-production cross section at the threshold is generally well known, it depends also on other model parameters, such as the top Yukawa coupling, and the measurement is a subject to many systematic uncertainties. The top-quark mass determination from the threshold scan at CLIC is therefore optimized using a genetic algorithm~\cite{Nowak:2103.00522}. The most general approach is used with all relevant model parameters and selected systematic uncertainties included in the fit procedure. Expected constraints from prior measurements are also taken into account. The top-quark mass can be extracted with precision of the order of 30 to 40 MeV, including considered systematic uncertainties, already for 100 fb$^{-1}$ of data collected at the threshold. Figure~\ref{fig:TOPHF_mtop_gene1} shows the result of the optimization: one point below the threshold, two in the turn-on region, and one above.

\begin{figure}[!h!t]
  \begin{center}
\includegraphics[width=0.49\textwidth]{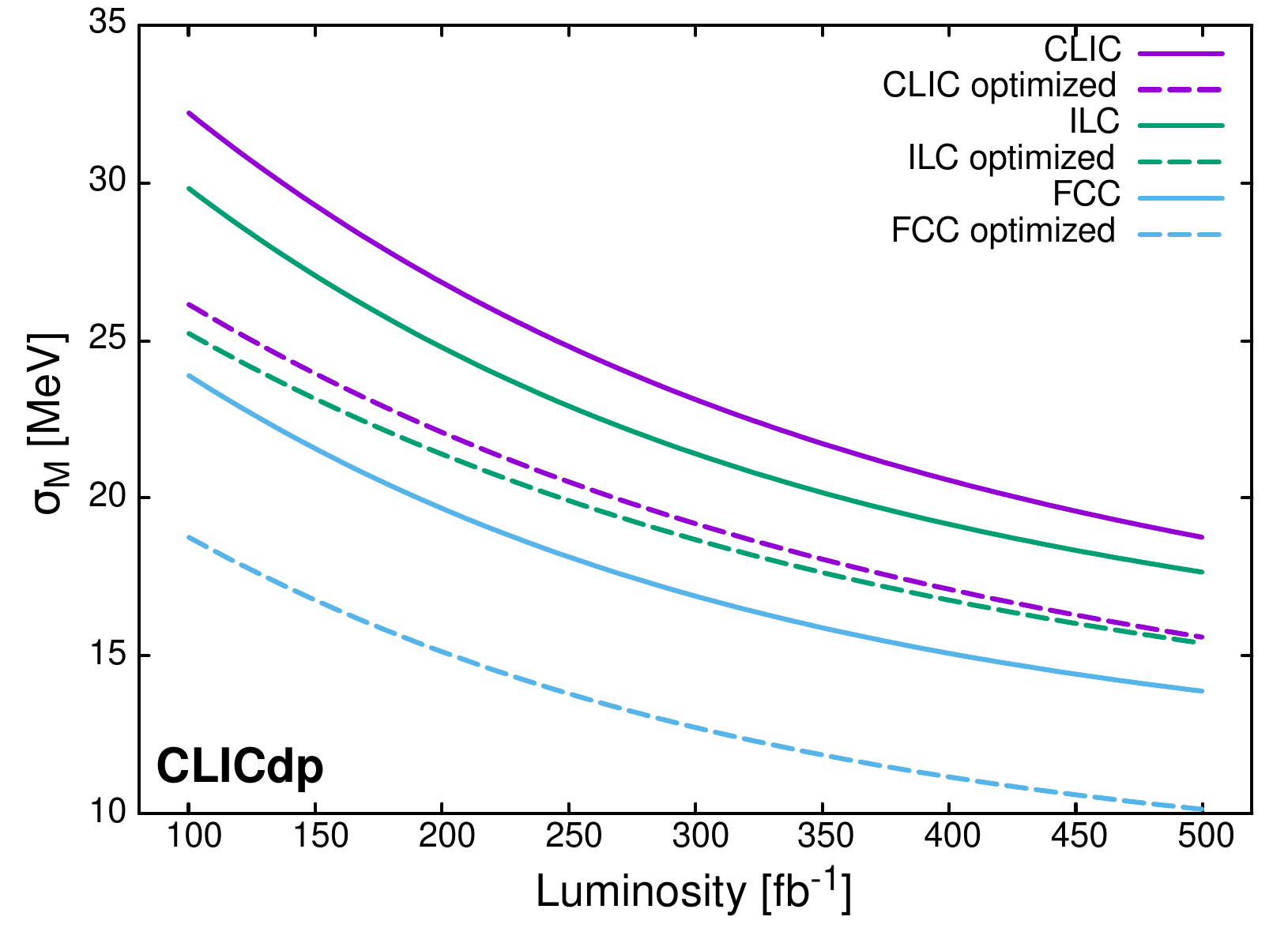}
\includegraphics[width=0.49\textwidth]{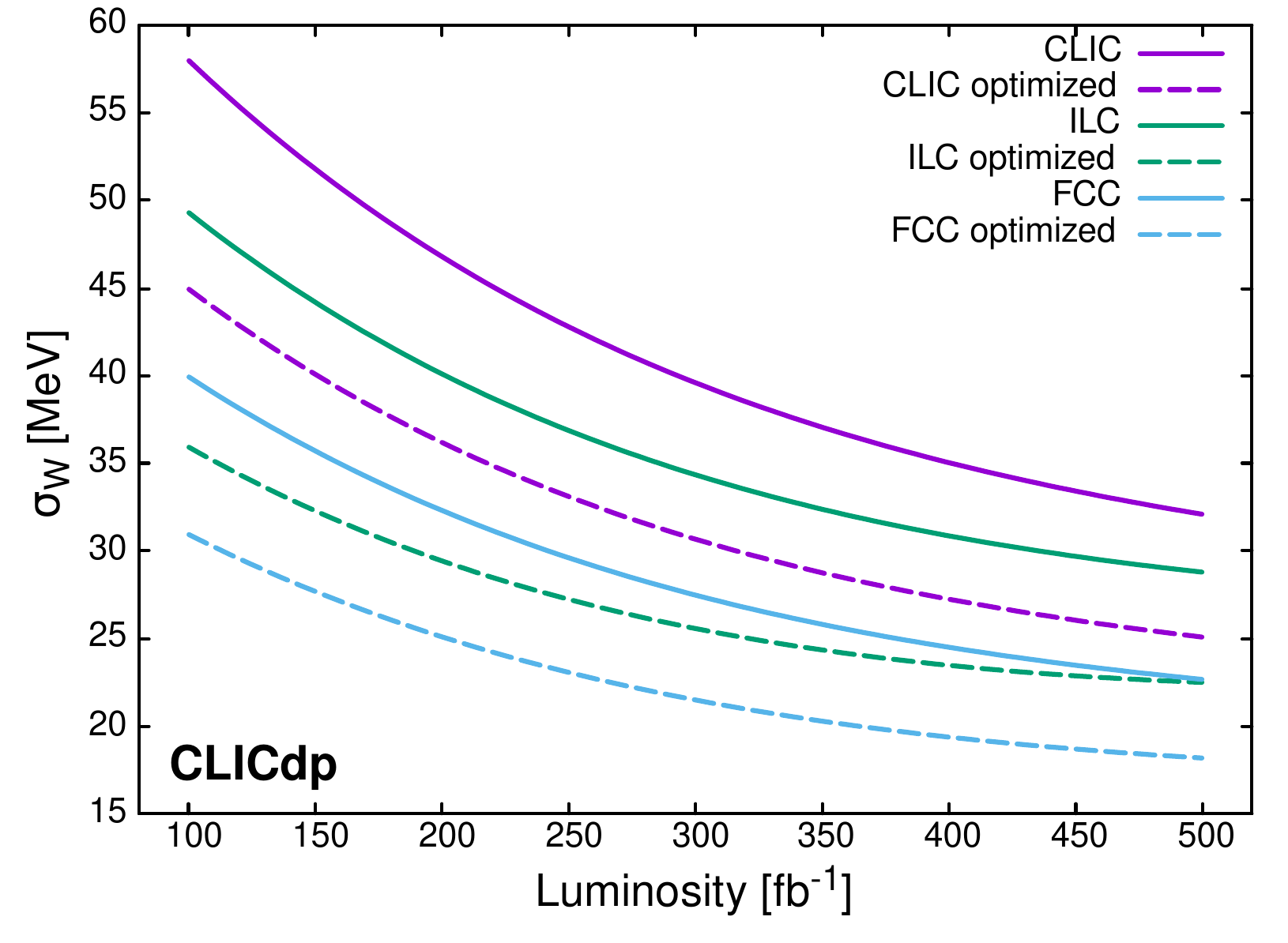}
\caption{Precision that can be obtained for the top quark (left) mass and (right) width in the top-quark mass scan at a $e^+e^-$ collider as a function of integrated luminosity collected. From~\cite{Nowak:2103.00522}.
}
\label{fig:TOPHF_mtop_gene2}
\end{center}
\end{figure}

The resulting expected uncertainty on the top-quark mass and width as a function of total integrated luminosity is shown in Figure~\ref{fig:TOPHF_mtop_gene2}. Note that the uncertainty includes some but not all of the systematic uncertainties, see Section~\ref{sec:TOPHF-mtop-ee} for details.

The scan parameters were also optimized to also include the top-quark Yukawa coupling in the scan, this requires ten points, the additional CM energies mainly in the plateau region~\cite{Nowak:2103.00522}. 
The genetic algorithm improved the statistical uncertainty of the mass measurement by about 20\%. 


\FloatBarrier
\FloatBarrier
\section{Top-quark width}
\label{sec:TOPHF-mtop-width}

In this section we focus on the top-quark width, $\Gamma_t$. The width has been measured from single top-quark production at the Tevatron~\cite{D0:2012hgn} and the LHC~\cite{CMS:2014mxl,CMS:2012xhh,CMS:2020vac} as well as in differential distributions~\cite{Herwig:2019obz,Baskakov:2018huw}. These measurements are the most precise, though they assume that there is no new physics affecting the decay of the top quark. The width is also measured directly from the mass peak in top quark decays at the Tevatron and the LHC~\cite{CDF:2013xca,ATLAS:2017vgz,ATLAS:2019onj}. These measurements are limited by the experimental resolution, the uncertainty on the width obtained by the ATLAS measurement is 500~MeV.

\subsection{Top-quark width measurements at hadron colliders}

Table~\ref{tab:TOPHF-width-had} compares the current measurements of the top quark width and the projections to the HL-LHC and the FCC-hh. The measurements are based on single top-quark production in the $t$-channel. The projections are based on extracting the top-quark width from differential distributions of single top $tW$~production~\cite{Herwig:2019obz,Baskakov:2018huw}. An analysis of early Run~2 data gives a width of 300~MeV, larger than from single top at 8~TeV. This is expected to improve significantly once the full Run~2 results are available and combined between methods and experiments. The projections~\cite{Baskakov:2018huw} for HL-LHC and FCC-hh assume a precision of the $tW$ measurement of 10\% and 5\%, respectively, including theoretical and experimental uncertainties. This requires computing $tW$ differentially at N$^3$LO including resummation, for example, see Section~\ref{sec:TOPHF-XS} and reducing PDF uncertainties. 
The experimental uncertainties affecting the width extraction are dominated by systematic uncertainties due to the jet energy scale and the modeling of $\ttbar$ and single top events. Assuming these will be reduced by a factor two, and that the statistical uncertainties are negligible, leads to the HL-LHC projection shown in Table~\ref{tab:TOPHF-width-had}. 

\begin{table}[t]
\begin{center}
\begin{tabular}{l|c|c|c|c|c}
$\delta \Gamma_t$ [MeV] &  Tevatron & LHC Run 1 & LHC Run 2 & HL-LHC & FCC-hh \\ \hline
$\sqrt{s}$ [TeV] &      1.96 & 7/8 & 13 & 14 & 100\\
$\L [\fbinv]$    &      5.4 & 20 & 140 & 3,000 & 30,000 \\ \hline
Statistical uncert. & N/A & 20 & -- & $\approx 0$ & $\approx 0$   \\
Systematic uncert.  & N/A & 140 & -- & 65 & 25\\ \hline
Total uncert.       & 470 & 140 & 300 & 65 & 25 \\ 
\end{tabular}
\caption{Current (Tevatron~\cite{D0:2012hgn}, LHC Run 1~\cite{CMS:2014mxl}, and LHC Run 2~\cite{Herwig:2019obz}) and anticipated (based on Ref.~\cite{Baskakov:2018huw}) statistical and systematic uncertainties in the measurement of the total top-quark width at hadron colliders. A significant reduction of the Run~2 uncertainty is expected once the measurements by ATLAS and CMS are completed and have been combined. }\label{tab:TOPHF-width-had}
\end{center}
\end{table}

\subsection{Top-quark width measurements at \texorpdfstring{$e^+ e^-$}{e+e-} colliders}

At a lepton collider, the width can be measured in a multi-parameter fit when scanning the top-production threshold for the mass measurement, see Section~\ref{sec:TOPHF-mtop-ee}. The systematic uncertainties for this scan have not yet been evaluated in detail. The projected statistical uncertainties for ILC, CLIC and FCC-ee shown in Table~\ref{tab:TOPHF-width-lep} are based on threshold scan studies where for ILC, 11 energy points and two beam polarization combinations with 10 $\fbinv$ each are used~\cite{Horiguchi:2013wra}; for CLIC, 10 energy points with 10 $\fbinv$ each~\cite{CLICdp:2018esa} are used; and for FCC-ee, 200 $\fbinv$ split evenly across 8 energy points \cite{Bernardi:2022hny} are used. The systematic uncertainties have only been evaluated for FCC-ee, it can be expected that the NNNLO uncertainty is similar for ILC and CLIC, while the $E_{CM}$ and $\alpha_s$ uncertainties should be larger at ILC and CLIC, similar to Table~\ref{tab:TOPHF-PSmtop}.

\begin{table}[t]
\begin{center}
\begin{tabular}{l|c|c|c}
$\delta \Gamma_t$ [MeV] &  ILC & CLIC & FCC-ee\\ \hline
$\sqrt{s}$ [TeV] & 0.340-0.350 & $(2 m_t^{{\rm PS}})^{+0.006}_{-0.003}$ & 0.340-0.345\\
$\L [\fbinv]$  & 220 & 100 & 200\\ \hline
Statistical uncert. &  21 & 20 & 45\\
Systematic uncert.  & ? & ? & 3 ($E_{CM}$),5 ($\alpha_s$),40 (NNNLO)\\ \hline
Total uncert.       &  ? &  ? & 60\\
\end{tabular}
\caption{Anticipated statistical and systematic uncertainties in the measurement of the total top-quark width at  $e^+ e^-$ colliders (ILC \cite{Horiguchi:2013wra}, CLIC~\cite{CLICdp:2018esa}, and FCC-ee~\cite{Bernardi:2022hny}). The systematic uncertainties have only been evaluated for FCC-ee, it can be expected that the NNNLO uncertainty is similar for ILC and CLIC, while the $E_{CM}$ and $\alpha_s$ uncertainties should be larger at ILC and CLIC, similar to Table~\ref{tab:TOPHF-PSmtop}.
}\label{tab:TOPHF-width-lep}
\end{center}
\end{table}

\FloatBarrier
\section{Top-quark production processes}
\label{sec:TOPHF-XS}

Top-quark production processes are important, in particular for measurements of top-quark properties (mass~\ref{sec:TOPHF-mtop}, width~\ref{sec:TOPHF-mtop-width}, couplings, spin correlations), direct searches for signals of new physics (e.g., resonances) and interpretations in effective field theory (see Section~\ref{sec:TOPHF-EW-couplings}). 
Prospects for measurements of top-pair production at $e^+e^-$ colliders which opens up for centre-of-mass energies that exceed twice the top-quark mass are discussed in Section~\ref{sec:TOPHF-EW-eett}.

\subsection{Top-quark pair and single top-quark production at the LHC: a brief review of theory calculations}
\label{sec:TOPHF-XS-LHC-theo}


Top-quark production at LHC energies proceeds predominantly via the strong production of top-antitop pairs. At leading order (LO), the partonic channels are quark-antiquark annihilation, $q{\bar q} \to t{\bar t}$, and gluon fusion, $gg \to t{\bar t}$. Additional channels, such as those with $gq$ and $g{\bar q}$ in the initial state, start appearing at next-to-leading order (NLO) in QCD and beyond. The gluon fusion channel is numerically dominant, accounting for around 90\% of the total cross section at the LHC.  

The NLO QCD corrections for inclusive $t{\bar t}$ production have been known for over thirty years~\cite{Nason:1987xz,Nason:1989zy,Beenakker:1988bq,Beenakker:1990maa} and NLO electroweak (EW) corrections became available soon after~\cite{Beenakker:1993yr} (see also~\cite{Kuhn:2005it,Bernreuther:2006vg,Kuhn:2006vh,Hollik:2007sw}). The matching of fully differential NLO QCD calculations to parton showers in MC@NLO~\cite{Frixione:2002ik,Frixione:2003ei} and POWHEG~\cite{Nason:2004rx,Frixione:2007vw} became available
about 10 years later. The next-to-next-to-leading-order (NNLO) QCD corrections have been calculated close to a decade ago~\cite{Barnreuther:2012wtj,Czakon:2012zr,Czakon:2012pz,Czakon:2013goa}. Top-quark differential distributions with NNLO QCD~\cite{Czakon:2016ckf,Czakon:2016dgf,Catani:2019hip} and NLO EW corrections appeared five years ago~\cite{Czakon:2017wor}. Differential distributions at NNLO in the $\bar{\text MS}$ mass renormalization scheme have been provided in~\cite{Catani:2020tko}. Apart from these advances in the calculation of pair production of stable top quarks, in the past few years, several off-shell calculations including top decays became available at NLO QCD~\cite{Denner:2010jp,Bevilacqua:2010qb,Denner:2012yc,Frederix:2013gra,Cascioli:2013wga,Denner:2017kzu}, matched to a parton shower~\cite{Campbell:2014kua}, and at NLO EW~\cite{Denner:2016jyo}.  NNLO QCD calculations including top-quark decays in the narrow-width approximation have been performed in~\cite{Gao:2017goi} (approximate) and in \cite{Czakon:2020qbd}. A framework for matching NNLO calculations of top-quark pair production with parton showers has recently been presented in ~\cite{Mazzitelli:2020jio,Mazzitelli:2021mmm}.  

Resummations of soft-gluon corrections for $t{\bar t}$ cross sections and differential distributions reached next-to-leading-logarithm (NLL) accuracy over twenty five years ago with the calculation of the one-loop soft anomalous dimension matrices \cite{Kidonakis:1996aq,Kidonakis:1997gm} for this process, while two-loop calculations \cite{Kidonakis:2009ev,Ferroglia:2009ep,Ferroglia:2009ii,Kidonakis:2010dk} of these soft anomalous dimensions later allowed the performance of resummation to next-to-next-to-leading-logarithm (NNLL) accuracy. Results for inclusive pair-production cross sections at NNLL can be found in Refs.~\cite{Kidonakis:2009ev,Ferroglia:2009ep,Ferroglia:2009ii,Kidonakis:2010dk,Czakon:2009zw,Ahrens:2011mw,Kidonakis:2011zn}, and a combination of NNLL accuracy in both threshold and Coulomb corrections has been achieved in Ref.~\cite{Beneke:2011mq}. 

Recently, in Ref.~\cite{Czakon:2019txp} NNLO QCD and NLO EW calculations have been combined with double resummation at NNLL' accuracy of threshold logarithms and small-mass logarithms and results have been presented for the $t\bar t$ invariant-mass, top-quark transverse-momentum and rapidity distributions. 
    
While the next-to-next-to-next-to-leading-order (N$^3$LO) QCD corrections have not been fully calculated, approximate N$^3$LO (aN$^3$LO) results that include third-order soft-gluon corrections derived from NNLL resummation are now available for total cross sections and (single and double) top-quark differential distributions in transverse momentum and rapidity \cite{Kidonakis:2014isa,Kidonakis:2014pja,Kidonakis:2015ona,Kidonakis:2019yji}. The uncertainties due to scale and PDF in the total cross-section predictions are at the level of 2.5\% and 1.6\% ,respectively, for the aN$^3$LO calculations, see Section~\ref{sec:TOPHF-XS-LHC-tt-theo}. 
In order to reach the $\approx 1$\% level required for precision top-quark mass and EFT couplings and many other measurements, higher-order QCD and EW corrections need to be computed, and the PDF uncertainties need to be significantly reduced.

The production of single top quarks provides opportunities for the direct study of the electroweak properties of the top quark. 
Single-top production may proceed via the $t$-channel, the $s$-channel, or the associated production of a top quark with a $W$-boson ($tW$ production). The $t$-channel processes, $qb \to q't$ and ${\bar q} b \to {\bar q}' t$, involve the exchange of a spacelike $W$-boson. The $t$-channel cross section is numerically the largest at LHC energies and can be used to extract the value of the CKM matrix element $V_{tb}$~\cite{Fang:2018lgq}. 
Measurements of $t$-channel production cross sections (inclusively and differentially) are sensitive to PDFs~\cite{Nocera:2019wyk} due to the similarity to deep inelastic scattering, while they are also sensitive to the bottom-quark mass through the bottom-quark PDF~\cite{Campbell:2021qgd,Nocera:2019wyk}. These measurements will be a helpful addition to future global QCD analyses of proton PDFs. Single top-quark production in the $t$-channel is also used to measure the top-quark polarization and test for anomalous $tWb$ couplings~\cite{Schwienhorst:2010je,ATLAS:2017ygi,CMS:2016uzc,Neumann:2019kvk,Cao:2021wcc,ATLAS:2022vym,CMS:2015cyp}, see also Section~\ref{sec:TOPHF-EW-couplings}.

The $s$-channel processes, $q{\bar q}' \to {\bar b} t$, involve a timelike $W$ boson, and the cross section is the smallest at LHC energies. The $tW$ process proceeds via $bg \rightarrow tW^-$. The $tW$ cross section is the second largest at LHC energies and allows a direct measurement of the top-quark width, see Section~\ref{sec:TOPHF-mtop-width}.

NLO QCD corrections for the $t$- and $s$-channels \cite{Harris:2002md} and for $tW$ production \cite{Zhu:2002uj} have been known for twenty years. NNLO QCD corrections have been calculated more recently for the $t$-channel \cite{Brucherseifer:2014ama,Berger:2016oht,Berger:2017zof,Campbell:2020fhf}, and also for the $s$-channel \cite{Liu:2018gxa}. Very recently, non-factorizable corrections of $\cal{O}(\alpha_s^2)$ have been calculated in~\cite{Bronnum-Hansen:2021pqc,Bronnum-Hansen:2022tmr}. Two-loop master integrals needed for a NNLO prediction for $tW$ production have been calculated in~\cite{Long:2021vse,Chen:2022yni}. 
    
Soft-gluon resummation was achieved for all single-top channels at NLL accuracy via one-loop calculations of the soft anomalous dimension matrices over fifteen years ago in \cite{Kidonakis:2006bu}, and was later improved to NNLL accuracy via two-loop calculations for the $s$-channel in Ref.~\cite{Kidonakis:2010tc}, the $tW$ channel in Ref.~\cite{Kidonakis:2010ux}, and the $t$-channel in Ref.~\cite{Kidonakis:2011wy}. More recently, the three-loop soft anomalous dimension for $tW$ production was calculated in Ref.~\cite{Kidonakis:2019nqa} (with additional partial three-loop results for $t$- and $s$-channel single-top production). Furthermore, results with aN$^3$LO soft-gluon corrections for the $tW$ total cross section and the top-quark and transverse-momentum and rapidity distributions were presented in~\cite{Kidonakis:2016sjf}, with further updated aN$^3$LO results for these quantities and also for $W$-boson differential distributions given in Ref.~\cite{Kidonakis:2021vob}.

The production of a single top quark in association with a $Z$ boson or a Higgs boson or a photon has also been studied. These processes involve three-particle final states at leading order, including a light quark. The NLO QCD corrections for $tqZ$ production were calculated in Ref.~\cite{Campbell:2013yla}, and for $tqH$ production in Ref.~\cite{Campbell:2013yla,Demartin:2015uha}. Soft-gluon resummation has also been performed for $tqH$~\cite{Forslund:2020lnu,Forslund:2021evo} and $tqZ$~\cite{Forslund:2020lnu} production as well as for $tq\gamma$~\cite{Kidonakis:2022ocq} production.

\subsection{Higher-order corrections for \texorpdfstring{$t\bar t$}{tt} production in high energy \texorpdfstring{$pp$}{pp} collisions}
\label{sec:TOPHF-XS-LHC-tt-theo}
Summary of white paper contribution~\cite{Kidonakis:2203.03698} 

\begin{figure}[!h!tbp]
\centering
\includegraphics[width=10cm]{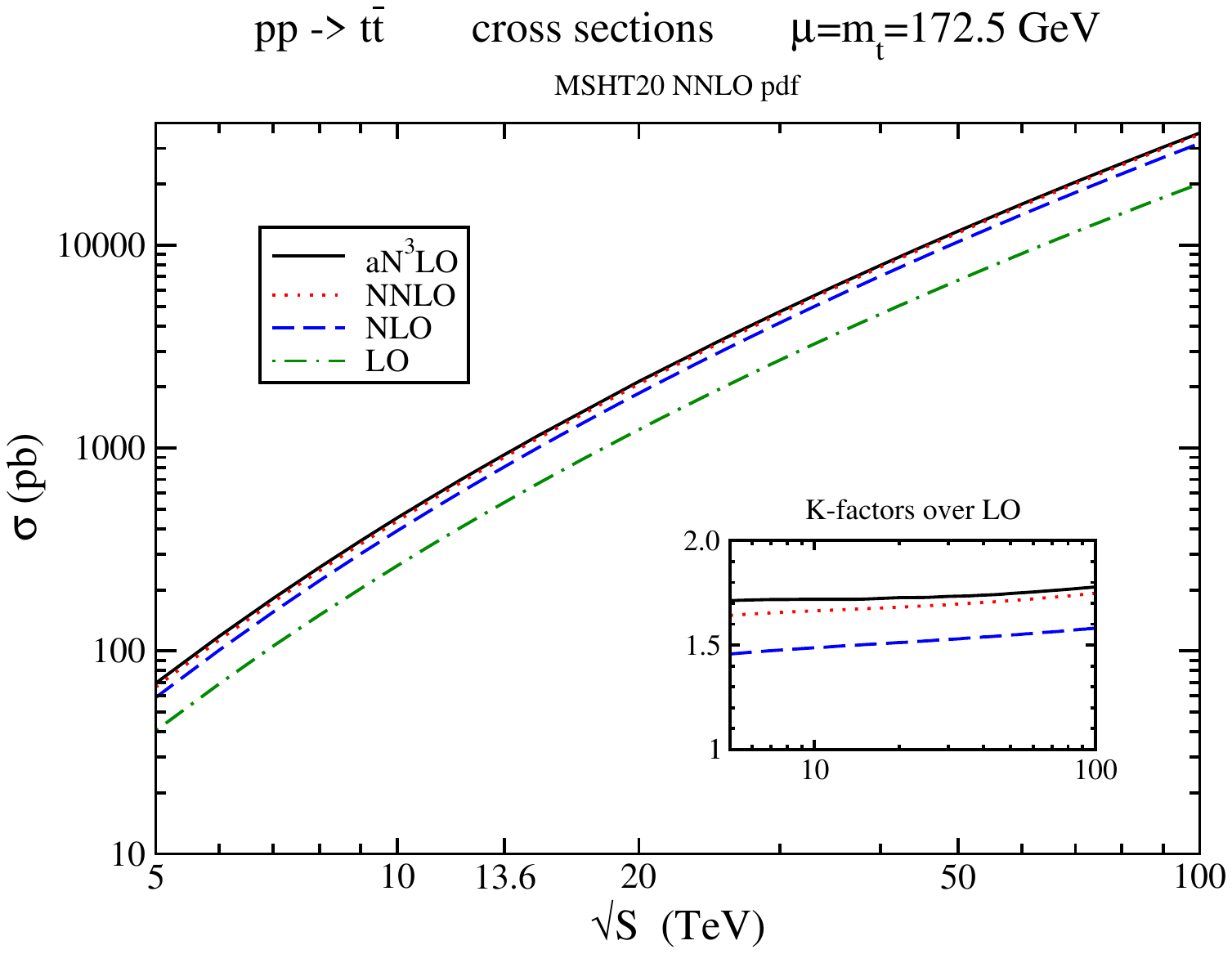}
\caption{The total cross sections at LO, NLO, NNLO, and aN$^3$LO for $t{\bar t}$ production at $pp$ collider energies. }
\label{fig:ttaN3LO}
\end{figure}

The theoretical formalism of soft-gluon resummation \cite{Kidonakis:1996aq,Kidonakis:1997gm,Kidonakis:2009ev,Kidonakis:2010dk} provides powerful techniques for calculations of QCD perturbative corrections at higher orders, and it has produced very accurate predictions for $t{\bar t}$ production through approximate N$^3$LO (aN$^3$LO)~\cite{Kidonakis:2014isa,Kidonakis:2014pja,Kidonakis:2015ona,Kidonakis:2019yji}. The soft-gluon corrections are numerically dominant and provide excellent approximations to the complete NLO and NNLO corrections not only at Tevatron and LHC energies (see e.g. the reviews in Refs. \cite{Kidonakis:2013zqa,Kidonakis:2018ybz}) but also at much higher collider energies, and the additional aN$^3$LO corrections are significant and provide improved theoretical predictions. For a top-quark mass $m_t=172.5$ GeV, the aN$^3$LO cross section (where aN$^3$LO=NNLO+soft-gluon N$^3$LO corrections) at 13 TeV is $839^{+23}_{-18}{}^{+17}_{-11}$ pb, at 13.6 TeV it is $928^{+25}_{-20}{}^{+18}_{-12}$ pb, and at 14 TeV it is $990^{+27}_{-22}{}^{+19}_{-13}$ pb, where the first uncertainty is from scale variation $m_t/2 \le \mu \le 2m_t$ while the second is from MSHT20 NNLO pdf \cite{Bailey:2020ooq}. Figure~\ref{fig:ttaN3LO} shows the LO, NLO, NNLO, and aN$^3$LO cross sections for $pp$ collider energies ranging from 5 TeV to 100 TeV \cite{Kidonakis:2203.03698}. The inset plot displays the $K$-factors, i.e. the ratios of the NLO, NNLO, and aN$^3$LO cross sections to the LO ones.

\begin{table}[htb]
\begin{center}
\begin{tabular}{|c|c|c|c|c|c|c|c|c|} \hline
\multicolumn{9}{|c|}{$K$-factors for $t{\bar t}$ production in $pp$ collisions} \\ \hline
$K$-factor & \hspace{-2mm} 7 TeV \hspace{-2mm} & \hspace{-2mm} 8 TeV \hspace{-2mm} & \hspace{-2mm} 13 TeV \hspace{-2mm} & \hspace{-3mm} 13.6 TeV \hspace{-3mm} & \hspace{-2mm} 14 TeV \hspace{-2mm} & \hspace{-2mm} 27 TeV \hspace{-2mm} & \hspace{-2mm} 50 TeV \hspace{-2mm} & \hspace{-2mm} 100 TeV \hspace{-2mm} \\ \hline
NLO/LO  & 1.47 & 1.48 & 1.50 & 1.50 & 1.50 & 1.52 & 1.55 & 1.58 \\ \hline
NNLO/LO & 1.65 & 1.66 & 1.67 & 1.67 & 1.67 & 1.69 & 1.71 & 1.75 \\ \hline
aN$^3$LO/LO & 1.72 & 1.72 & 1.72 & 1.72 & 1.72 & 1.73 & 1.75 & 1.78 \\ \hline
aNLO/NLO & 1.01  & 1.00  & 0.99  & 0.99  & 0.99 & 0.97  & 0.95 & 0.92 \\ \hline
aNNLO/NNLO & 1.01 & 1.01 & 1.00 & 1.00 & 1.00 & 1.00 & 0.99 & 0.98 \\ \hline
\end{tabular}
\caption{The $K$-factors in $t{\bar t}$ production (with $\mu=m_t$) at different perturbative orders in $pp$ collisions with various values of $\sqrt{S}$, with $m_t=172.5$ GeV and MSHT20 NNLO pdf.
\label{tab:NKxstop}
}
\end{center}
\end{table}

Table~\ref{tab:NKxstop} shows various $K$-factors for $t{\bar t}$ production for several $pp$-collider energies \cite{Kidonakis:2203.03698}. The ratio aN$^3$LO /LO  is larger than the ratio NNLO/LO, which indicates significant contributions from third-order soft-gluon corrections. The dominance of the soft-gluon contributions for all energies, and thus the excellence of the soft-gluon approximation at NLO and NNLO, is easily seen by the aNLO/NLO ratio (where aNLO=LO+soft-gluon NLO corrections) and the aNNLO/NNLO ratio (where aNNLO=NLO+soft-gluon NNLO corrections), which remain very close to one. Although the dominance of these corrections at LHC energies has been known for a long time (and reviewed in Refs. \cite{Kidonakis:2013zqa,Kidonakis:2018ybz}), their continuing importance at very high energies is noteworthy and was not necessarily expected. Similar conclusions were drawn for the importance of the soft-gluon corrections for $tW$ production through 100~TeV energy in Ref.~\cite{Kidonakis:2021vob}.

\FloatBarrier
\subsection{Experimental aspects of \texorpdfstring{$pp \to t\bar t$}{pp to tt}}
\label{sec:TOPHF-XS-LHC-ttexp}

Top-quark pair production in hadron machines is interesting to study for several reasons. First, it is the leading source of top quarks, and second, top pair production is one of the main background processes to most searches for new physics. Hence its precise modeling is essential for a reliable extraction of the new physics signal. Third, the top quark itself has a chance to originate from a decay of new particles, e.g. heavy replica of the Higgs boson. Such a process would result in a resonant and interference signal on top of the SM continuum production. And last but not least, it is an important test of  perturbative QCD, which should work well at these high energies, as well as of electroweak (EW) theory. 

\begin{figure}[!h!tb]
\centering
\includegraphics[width=0.6\textwidth]{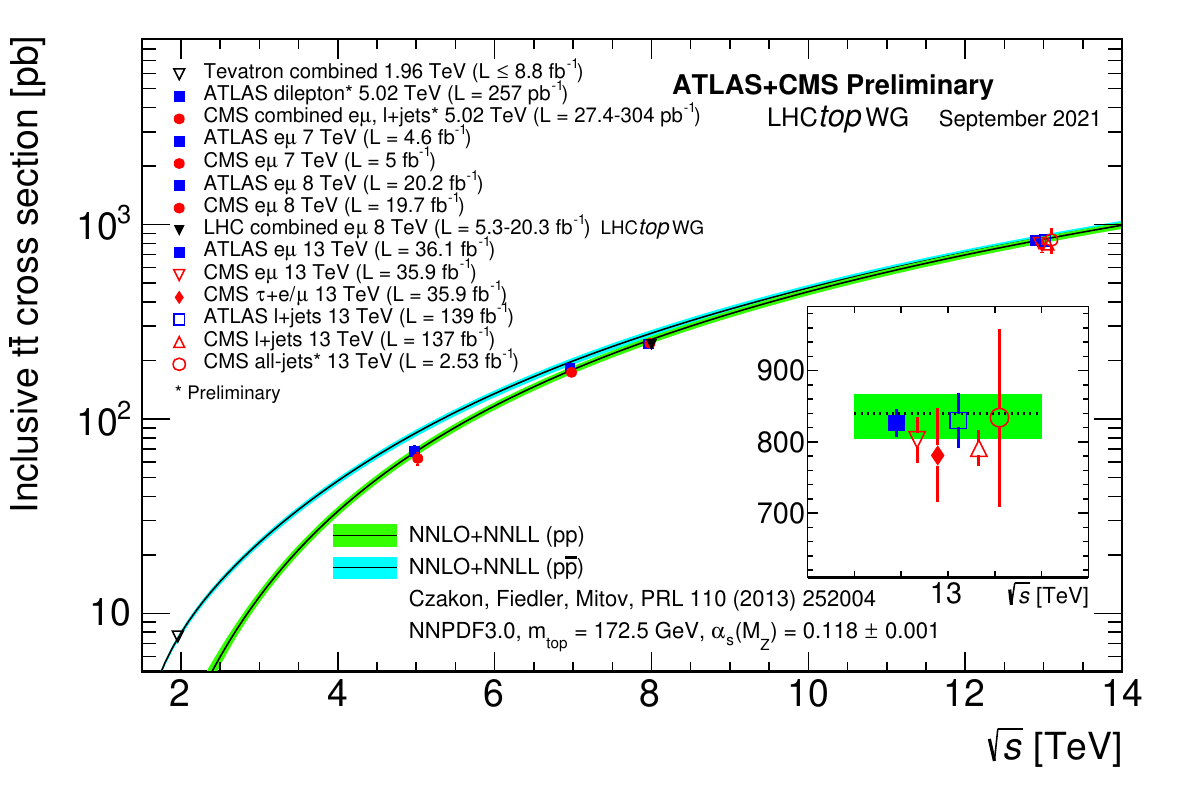}
\caption{Cross-section measurement for $\ttbar$ production at the LHC at different CM energies. From~\cite{LHCtopWG}.}
\label{fig:ttXsecs}
\end{figure}

The $\ttbar$ cross-section measurements at the Tevatron and the LHC are summarized in Figure~\ref{fig:ttXsecs}. The cross-sections have been measured to better than 5\% uncertainty. The ATLAS+CMS combination at 8~TeV has an uncertainty of only 2.5\%. The largest contributions to this uncertainty are from the luminosity determination and theory modelling of the $\ttbar$ system. Reducing these uncertainties by another factor two should be possible by the end of the HL-LHC~\cite{Azzi:2019yne}, achieving an uncertainty around 1\%.

In the study of top-quark pair production in hadron machines theory and experiment are closely connected, motivating and cross checking each other. For example, the large charge asymmetry in $t\bar{t}$ production observed at the Tevatron~\cite{D0:2007ihv},~\cite{CDF:2008nne} was not explained by the NLO level calculations~\cite{Kuhn:1998kw} leading to a suite of papers suggesting beyond the SM explanations (see e.g. ~\cite{Kamenik:2011wt}). Yet, the inclusion of the EW corrections~\cite{Bernreuther:2012sx} and the extension of the calculations to NNLO~\cite{Czakon:2014xsa} demonstrated a good agreement with a combined result from CDF and D0~\cite{CDF:2017cvy}. 

\begin{figure}[!h!tb]
\centering
\includegraphics[width=0.45\textwidth]{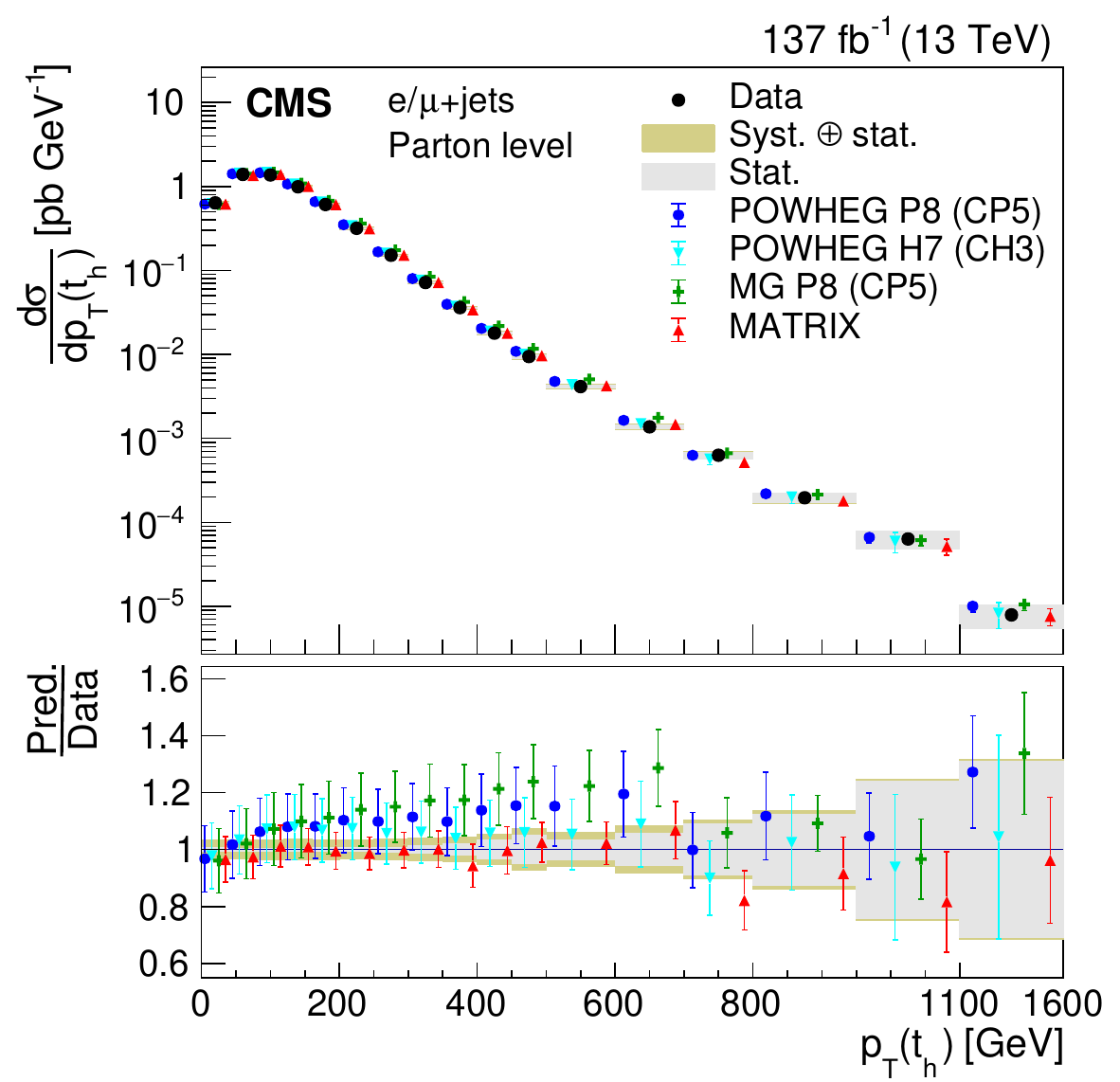}
\includegraphics[width=0.54\textwidth]{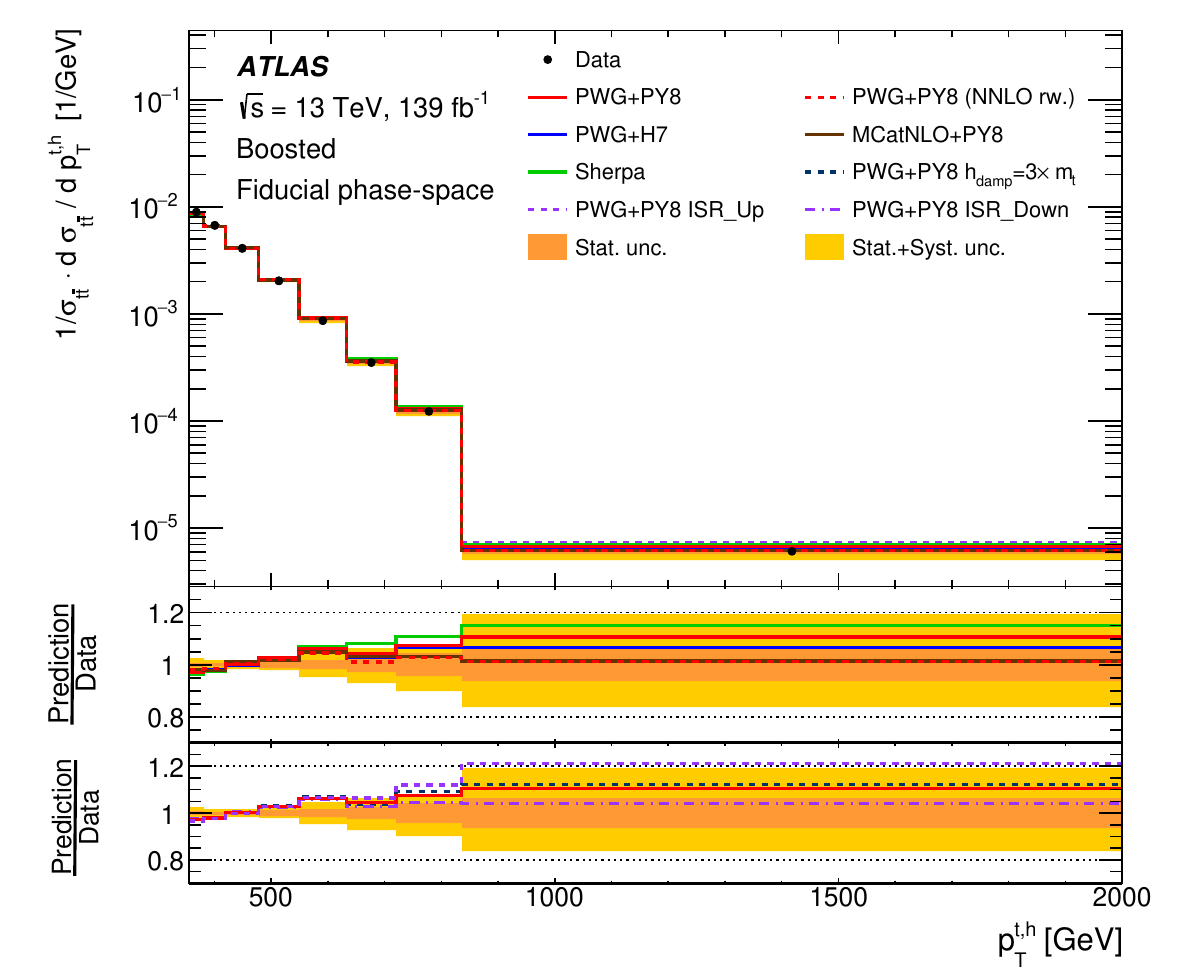}
\caption{The normalized
differential cross section of \ttbar\ production as a function of top quark $p_{\mathrm{T}}$
compared to several models at the parton level measured by CMS(left) and ATLAS(right).}
\label{fig:ttXsect}
\end{figure}

Similarly, original data from both ATLAS and CMS demonstrated a disagreement in the distribution in the top-quark $p_T$ with the predictions based on NLO simulations. Latest results at the parton level obtained by CMS compared to NNLO level MATRIX simulations~\cite{Grazzini:2017mhc}, \cite{Czakon:2015owf} demonstrated a much better agreement~\cite{CMS:2021vhb} as shown in Figure~\ref{fig:ttXsect}. Similar measurement performed by ATLAS~\cite{ATLAS:2022xfj}  was compared to different simulations with the best agreement demonstrated by POWHEG~\cite{Alioli:2010xd} +PYTHIA8~\cite{Sjostrand:2014zea} reweighted to NNLO prediction~\cite{Czakon:2017wor} with EW corrections.  

\begin{figure}
\centering
\includegraphics[width=0.38\textwidth]{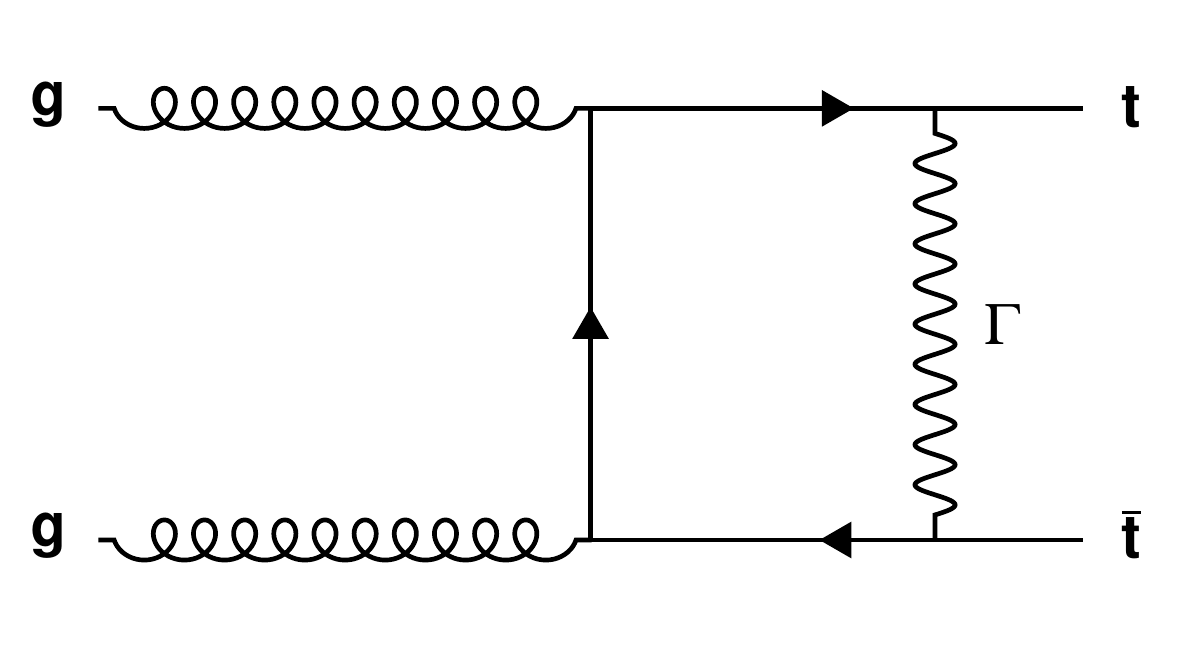} \\
\includegraphics[width=0.48\textwidth]{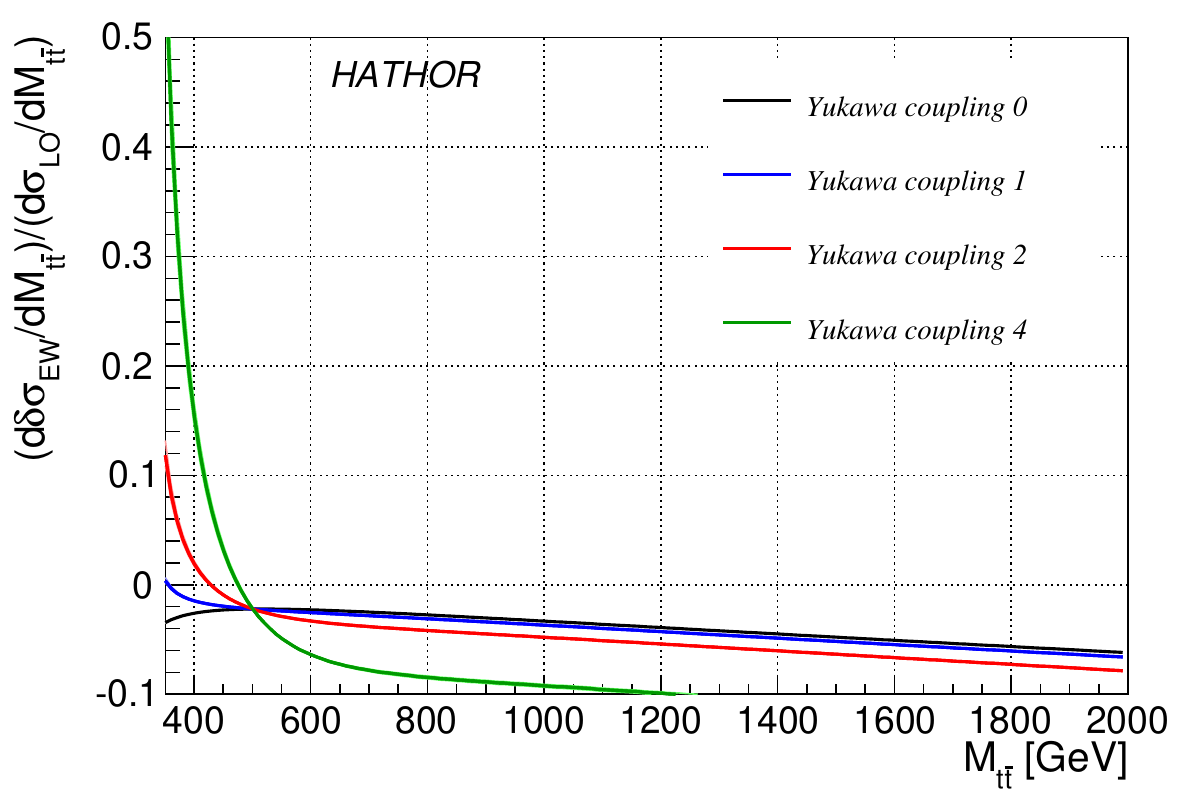}
\includegraphics[width=0.48\textwidth]{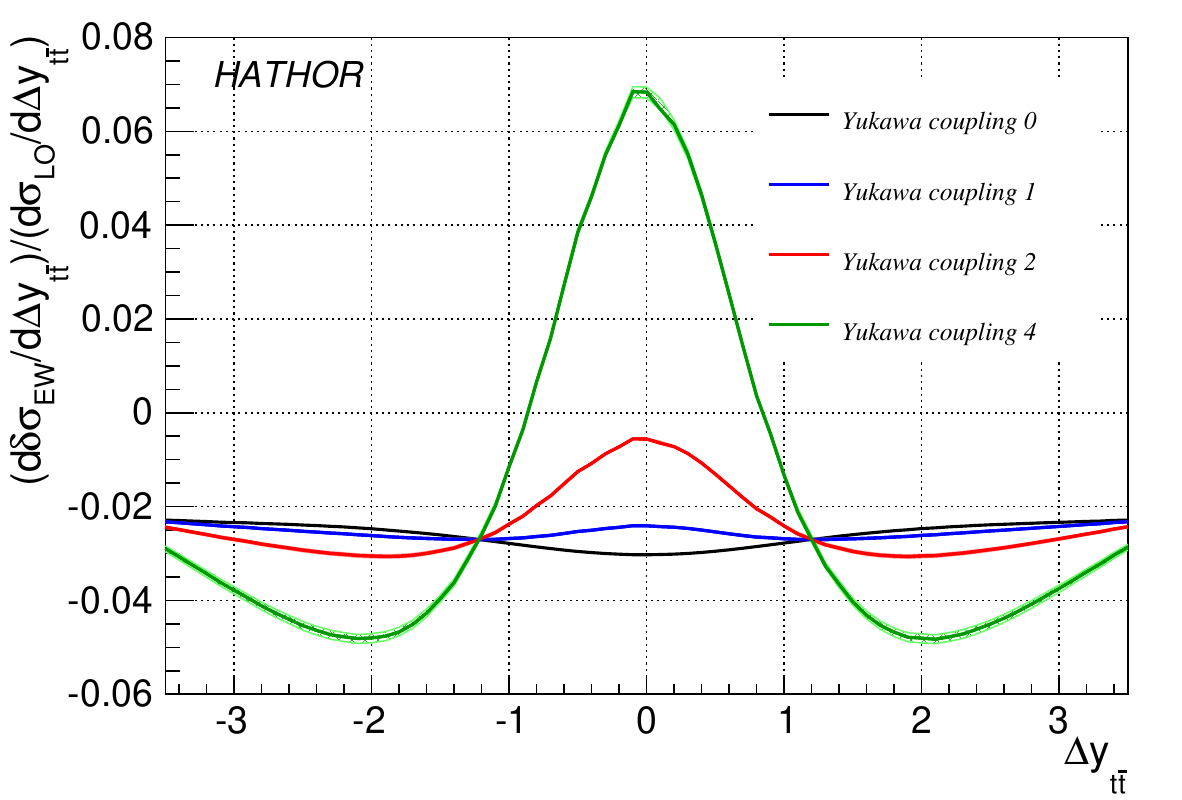}
\caption{ 
 (Top) An example of Feynman diagrams of \ttbar\ production with virtual exchange of EW  bosons $\Gamma$, including Higgs boson. (Bottom) The dependence of EW corrections to \ttbar\ production for different values of $y_t$ on $M_{\ttbar}$ (left) and $\Delta y_{\ttbar}$ (right), generated by \textsc{hathor}~\cite{Aliev:2010zk}.}
\label{fig:Yt_theory}
\end{figure}

The realization that the EW corrections are important in $t\bar{t}$ production resulted in a new method of the evaluation of the top quark Yukawa coupling, $y_t$ using the kinematics of the top and and anti-top quark near the production threshold~\cite{Kuhn:2013zoa}. An example of a Feynman diagram that includes a virtual exchange of EW bosons  including Higgs boson is shown  in the left of Figure~\ref{fig:Yt_theory}.   The interference of such diagram with the tree-level process is proportional $y_t^2$. Hence, increased values of $y_t$ would  result in significant distortions of the distribution in the invariant mass of top and anti-top quarks, as well as the difference in rapidity between the two as shown in Figure~\ref{fig:Yt_theory} (center and right). Comparing data to predictions generated with different values of  $y_t$ allowed CMS to constrain it to be less than $1.67$~\cite{CMS:2019art} in lepton+jets channel and  $1.54$~\cite{CMS:2020djy} in the dilepton channel at the 95\% CL. These results, though not as sensitive as $Y_t$ measurement from a combined analysis of Higgs boson production in gluon fusion and  $t\bar{t}H$ channels, have an advantage of not being dependent on the assumption about the values of the other Higgs boson couplings. The production of four top quarks discussed in Section.~\ref{sec:TOPHF-4top-production} has the same advantage, but so far is not as sensitive. At the same time while the four top quark production process is statistically limited, measurement of $y_t$ from the $t\bar{t}$ kinematics is partially  limited by systematic uncertainties, among which the leading source of theoretical uncertainties are the modeling of parton shower and color reconnection. Future improvements in the $y_t$ measurements require a better understanding of these uncertainties, or development of reliable techniques to constrain them in situ.

\FloatBarrier
\subsection{Experimental aspects of single top-quark production and \texorpdfstring{$V_{tb}$}{Vtb}}
\label{sec:TOPHF-XS-LHC-stexp}

In the SM, single top-quark production is a charged-current electroweak process that involves the $tWb$ vertex in the production of the top quark and in its decay, with only negligible contributions from $tWd$ and $tWs$ couplings, and even smaller contributions from Flavor-Changing Neutral Currents (FCNC, see Section~\ref{sec:TOPHF-BSM})~\cite{Giammanco:2017xyn}. Precise measurements of single top-quark cross sections are motivated by their sensitivity to new physics that modifies either the production or the decay vertex or both~\cite{AguilarSaavedra:2008zc}. The single top-quark production cross section under the assumption of SM-type left-handed couplings is proportional to the square of the Cabibbo-Kobayashi-Maskawa (CKM)~\cite{Cabibbo:1963yz,Kobayashi:1973fv} matrix element $V_{tb}$~\cite{Alwall:2006bx,Lacker:2012ek}.

\begin{figure}[!h!tb]
\centering
\includegraphics[width=0.8\textwidth]{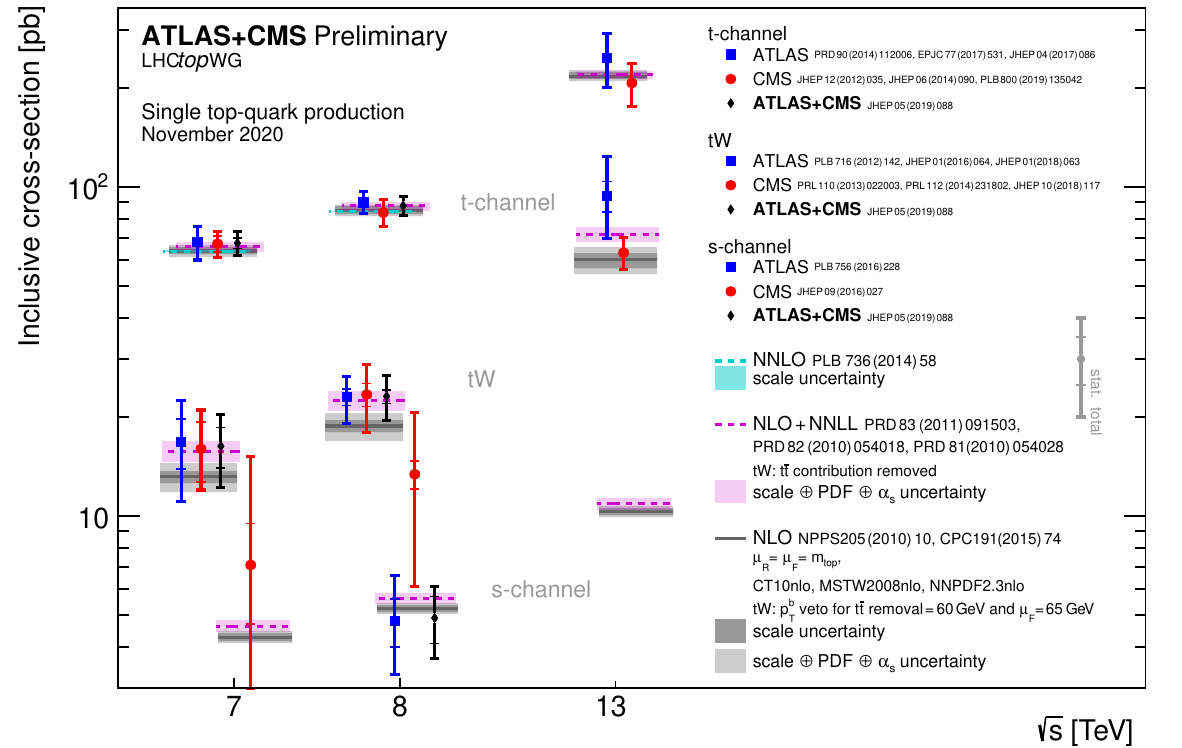}
\caption{Cross-section measurement for single-top quark production at the LHC at different CM energies. From~\cite{LHCtopWG}. }
\label{fig:singletxs}
\end{figure}

The single top cross-section measurements at the LHC are summarized in Figure~\ref{fig:singletxs}. 
The $t$-channel mode has the largest production cross section and smallest uncertainty, 6.6\% in the 8~TeV LHC combination~\cite{atlas:2019hhu}. The largest uncertainties are due to the theory modeling of the single top signal and the $\ttbar$ background. The measurements at 13~TeV in Run~2 have not yet been finalized, and the combination has not been completed. It is expected that the uncertainties will be somewhat smaller due to the improved theory modeling. For the HL-LHC, these uncertainties are projected to be reduced by another factor two, to an uncertainty of about 2\%. However, this requires significant improvements in the modeling of the top-quark final state and parton shower~\cite{QCDreport}.

The sensitivity of single-top quark production to the CKM matrix elements $V_{tu}$, $V_{ts}$ and $V_{tb}$ can be enhanced through measurements of ratios and differential distributions~\cite{Fang:2018lgq}. This study obtains an uncertainty on $V_{tb}$ of about 1\% with 8~TeV data. The  expected sensitivity for $V_{tb}$ at the HL-LHC is about 0.15\%~\cite{Fang:2018lgq}.

At a lepton collider, single top-quark production in the SM occurs through lepton-photon scattering~\cite{Escamilla:2017pvd}, a process that can be observed at collision energies above 0.5~TeV. The CKM matrix element $V_{tb}$ by contrast is included in global EW fits; it is proportional to the left-handed coupling of the top quark to the $W$ boson and $b$ quark. In this way, it can be constrained to significantly better than 1\% in precision $\ttbar$ measurements at lepton colliders, see Ref.~\cite{EWreport}.


\FloatBarrier
\subsection{\texorpdfstring{$t\bar{t}X, X=j,\gamma,Z,W^\pm,H,b\bar{b}$}{ttX, X = photon, Z, W, H, bb}: review of theory predictions for LHC and HL-LHC}\label{sec:TOPHF-XS-LHC-ttX}

The study of $t\bar{t}+X$ processes with $X=\gamma,Z,W^\pm,H,t\bar{t}$ is of
unique interest in exploring the connection between gauge, scalar and flavor
dynamics in the SM, especially in looking for anomalies that
could point to physics beyond the SM (BSM).  Indeed, the associated
production of top-quark pairs with bosons allows also for a direct measurement
of the top-quark interactions with EW gauge bosons as well as the Higgs boson.
The interpretation of potential anomalies in top-quark couplings both in
specific models or in terms of more general effective interactions represents a
prominent avenue towards discovering indirect evidence of new physics at
present and future colliders. At the same time, besides their physics
potential, these processes give rise to complex hadronic and multi-lepton
signatures that enter many other measurements as backgrounds and thus need to
be known precisely in order to disentangle signal and background contributions
accurately.

The LHC offers the unique opportunity to study the production of top-quark
pairs in association with additional gauge bosons and heavy particles. At the
high-luminosity of the HL-LHC many such processes will be measured with an
accuracy better than 10\% and will provide quite stringent constraints on
anomalous top-quark couplings. Only high-energy lepton colliders (ILC at 1000
GeV, CLIC, Muon Collider) and future hadron colliders (HE-LHC, FCC-hh) will be
able to improve on HL-LHC measurements.
\begin{figure}[ht!]
    \centering
    \includegraphics[width=0.7\textwidth]{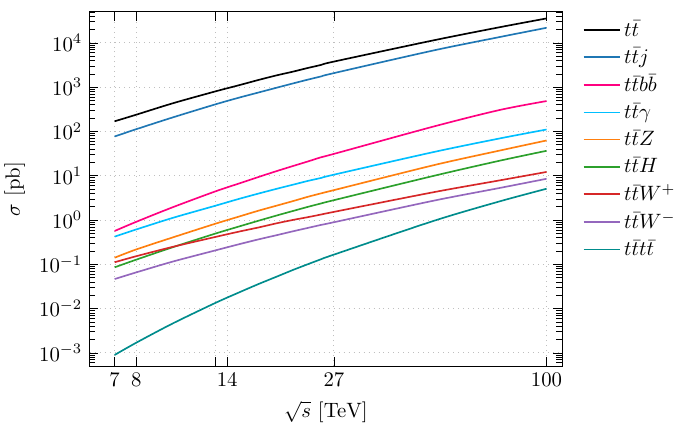}
    \caption{Total cross sections for various $pp\to t\bar{t}X$ processes as a function of the 
    center-of-mass energy $\sqrt{s}$. The $pp\rightarrow t\bar{t}$ is also shown for reference. 
    Light objects are subject to the following cuts $p_T > 25$ GeV,
    $|\eta| < 2.5$ and jets are clustered using the anti-$k_T$ algorithm with $R=0.4$. }
    \label{fig:ttV_xsecs}
\end{figure}
To illustrate the case of hadron colliders, Fig.~\ref{fig:ttV_xsecs} shows the
total cross section for various associated production channels ($pp\rightarrow
t\bar{t}X$) as a function of the center-of-mass energy $\sqrt{s}$. It is
evident that the associated $t\bar{t}X$ production with additional gauge bosons
and heavy particles comprises very rare production modes with distinctive final-state signatures. These present interesting signals that provide valuable inputs for example for EFT fits (Section~\ref{sec:TOPHF-EW-couplings}) and contribute large backgrounds to top-Higgs coupling measurements~\cite{Higgsreport}. A dedicated experimental and theoretical effort is needed to reach uncertainties of order 1\% in the inclusive measurements and less than 5\% in differential measurements, and similar precision in the theoretical calculations.

As the measurement of these top-quark couplings become more and more precise in
the future, the theory predictions have to match the accuracy of the
experiments in order to harness the full power of the collected data to
distinguish SM from BSM physics.  Delivering theoretical predictions that could
compare with the precision of the LHC and the HL-LHC experiments involves
considering not only higher-order QCD and EW effects but also understanding how
to accurately model the complexity of high-multiplicity hadron-collider events.
With this respect, recent years have seen a huge progress in developing novel
theoretical techniques to boost the precision of SM calculations for
fully-decayed $t\bar{t}X$ processes to a new level. This has involved
calculating higher-order QCD and EW corrections to the fully-decayed signatures
arising from $t\bar{t}X$ processes, therefore taking into account full
off-shell and spin-correlation effects, as well as matching fixed-order
calculations to Monte Carlo parton-shower event generators used in the
comparison of theoretical predictions with data. Thanks to this theoretical
progress, it is now possible not only to use better SM predictions but also to
understand where future progress is most needed. 

In the following, each process is elaborated on in more detail, including the importance of the given production process, the current state-of-the art
for theoretical predictions, and which improvements might be necessary in order
to harness the full potential of the upcoming HL-LHC run.

\subsubsection{\texorpdfstring{$pp\to t\bar{t}j$}{pp to ttj}}\label{sec:TOPHF-ttj-theory}
The production of a top-quark pair in association with an additional hard jet
has the largest cross section among all associated production processes.
Because of its large rate the $pp\to t\bar{t}j$ process constitute a sizable
background for various SM measurements as well as for BSM searches.
For example, the $pp\to t\bar{t}j$ process is the dominant background to Higgs
boson production in VBF processes in the $H\to W^+W^-$ decay
channel~\cite{Rainwater:1999sd,Kauer:2000hi}. It is also a main background for
various BSM signatures involving $W^+W^-$ production along with additional
jets~\cite{Mangano:2008ha,Englert:2008tn,Englert:2008wp,Campanario:2010mi} or
new heavy resonances~\cite{Gresham:2011dg,ATLAS:2012mjy,CMS:2012gqb}. Besides,
being a sizable background process, the process can be utilized to constrain
anomalous dipole moments of the top-quark from modified spin-correlation
effects~\cite{Buchmuller:1985jz,
Aguilar-Saavedra:2008nuh,Aguilar-Saavedra:2014iga,CMS:2014bea,CMS:2016piu} and
to study enhanced charge
asymmetries~\cite{Berge:2013xsa,Antunano:2007da,Berge:2016oji,Berge:2012rc,
Alte:2014toa,Basan:2020btr}. Furthermore, the top-quark mass can be inferred
from the $t\bar{t}j$ production
process~\cite{Alioli:2013mxa,Fuster:2017rev,Alioli:2022lqo} as has been already
demonstrated by the LHC
experiments~\cite{ATLAS:2015pfy,CMS:2016khu,ATLAS:2019guf}. See Section~\ref{sec:TOPHF-mtop-ttj} for a discussion of theory predictions for the extraction of the top-quark mass from $t\bar{t}j +X$ events at the LHC.

While NLO QCD corrections for on-shell $t\bar{t}j$ were first calculated in
Refs.~\cite{Dittmaier:2007wz,Dittmaier:2008uj}, NLO EW contributions have
become available only recently in Ref.~\cite{Gutschow:2018tuk}. The effects of
parton showers on theoretical predictions have been studied by matching
fixed-order NLO calculations to parton
showers~\cite{Kardos:2011qa,Alioli:2011as,Czakon:2015cla}, that also allow to
include approximate top-quark decays. Moreover, higher-order QCD and EW
corrections have been included via multi-jet
merging~\cite{Hoche:2016elu,Gutschow:2018tuk}. In
Refs.~\cite{Melnikov:2010iu,Melnikov:2011qx} NLO QCD corrections to radiative
top-quark decays have been investigated for the first time within the framework
of the Narrow-Width-Approximation (NWA).  Finally, NLO QCD predictions
including full off-shell effects have been
computed~\cite{Bevilacqua:2015qha,Bevilacqua:2016jfk,Bevilacqua:2017ipv} to
asses the importance of single and non-resonant contributions as well as
higher-order corrections to top-quark decays. Ultimately theoretical
predictions are dominated by uncertainties originating from missing
higher-order corrections, which have been estimated to be of the order of
$10-20\%$ at the differential level.

Due to its considerable size the $pp\to t\bar{t}j$ process and its contribution
to a wide range of physics studies it will become even more important for the
HL-LHC run to obtain more accurate theoretical predictions by including even
higher-order corrections. Currently, first steps have been taken towards NNLO
QCD corrections for on-shell $t\bar{t}j$ production~\cite{Badger:2022mrb} to
improve further our understanding of this production channel. See also Section~\ref{sec:TOPHF-mtop-pole}, which discusses the top pole mass extraction from $ttj$.

\subsubsection{\texorpdfstring{$pp\to t\bar{t}\gamma$}{pp to tt+photon}}\label{sec:TOPHF-ttgamma}
The $pp\to t\bar{t}\gamma$ process is of high interest at the LHC as it allows
to probe directly the electric charge of the top quark~\cite{Baur:2001si} as
well as the structure of the $t\bar{t}\gamma$ interaction vertex, which can constrain possible new physics effects via
effective field theory approaches~\cite{BessidskaiaBylund:2016jvp}. The latter
gives key insights into possible BSM scenarios that predict anomalous electric
dipole moments for top-quarks~\cite{Bouzas:2012av,Fael:2013ira,
Aguilar-Saavedra:2014vta,Schulze:2016qas,Etesami:2016rwu,Baur:2004uw}.
On the
theoretical side, the accurate description of fiducial signatures of $pp\to
t\bar{t}\gamma$ is one of the most challenging tasks, because once top-quark
decays are included fiducial signatures receives large contributions from
radiative top-quark decays via the decay chain: $pp\to t\bar{t}$ followed by
e.g. $t\to Wb\gamma$. The latter poses challenges for the experiments to
reconstruct the top-quark momenta~\cite{Dobur:2021jqf}, as these contributions
can be as large as $50\%$ of the total
signal~\cite{Melnikov:2011ta,Bevilacqua:2019quz} depending on the definition of
the fiducial phase space volume.

For the on-shell $pp\to t\bar{t}\gamma$ production NLO QCD~\cite{Duan:2009kcr,
Duan:2011zxb} and NLO EW~\cite{Duan:2016qlc} corrections are well studied.
Recently, even the full NLO corrections, i.e. the full tower of possible
$\mathcal{O}(\alpha^n\alpha_s^k)$ corrections, have been computed in
Ref.~\cite{Pagani:2021iwa}. The dominant theoretical uncertainties are given by
missing higher-order QCD corrections and amount to roughly $\pm12\%$ at the
integrated and $\pm 20\%$ at the differential level. Compared to that EW
corrections are fully contained within the residual QCD scale uncertainty
unless the high-energy tail of dimensionful observables, such as $p_T(\gamma)$,
are considered. Furthermore, the process $pp\to t\bar{t}\gamma\gamma$ has been
studied as well in
Refs.~\cite{Kardos:2014pba,vanDeurzen:2015cga,Maltoni:2015ena,Pagani:2021iwa}.
Top-quark decays have been first considered via matching to parton
showers~\cite{Kardos:2014zba} and later on in a more systematic way in the NWA
that includes full NLO QCD corrections to the decay
\cite{Melnikov:2011ta,Bevilacqua:2019quz}. The impact of top-quark decays on
the charge asymmetry has been investigated in detail in
Ref.~\cite{Bergner:2018lgm}. Finally, for the di-lepton decay channel
theoretical predictions at NLO QCD accuracy including full off-shell effects
have become available recently in
Refs.~\cite{Bevilacqua:2018woc,Bevilacqua:2018dny}.  The latter improve on the
previous results by including non-factorizable contributions and further
reducing the theoretical uncertainty due to scale variations.

While current measurements in the di-lepton
channel~\cite{ATLAS:2020yrp,CMS:2022lmh} are already challenging the accuracy
of theoretical predictions, it is expected that the HL-LHC will be able to
further reduce the total  uncertainty to $3\%~(7\%)$ for the di-lepton
(single-lepton) decay channel~\cite{ATL-PHYS-PUB-2018-049}. These projections
are based on the assumption that theoretical uncertainties can be further
reduced by a factor of two. Ultimately, the main uncertainties of these
measurements are theory dominated. In the single-lepton signature the
$t\bar{t}$ predictions are the main uncertainty, while in the $e\mu$ di-lepton
channel the $t\bar{t}\gamma$ signal and $t\bar{t}$ background contributions
have the largest impact and are of similar size.  Therefore, NNLO QCD
corrections for $pp\to t\bar{t}\gamma$ become necessary in order to exploit the
full potential of future data sets, including in EFT fits, discussed in Section~\ref{sec:TOPHF-EW-couplings}.

\subsubsection{\texorpdfstring{$pp\to t\bar{t}Z$}{pp to ttZ}}
Similar to $t\bar{t}\gamma$, the $pp\to t\bar{t}Z$ production process allows to
directly probe the top-quark interactions with electroweak gauge bosons. For
example, in the presence of additional heavy gauge bosons such as a $Z^\prime$
or vector-like leptons the $t\bar{t}Z$ vertex might develop dipole
contributions~\cite{Greiner:2014qna,Cox:2015afa,
Kumar:2015tna,Kim:2016plm,Fox:2018ldq,Alvarez:2020ffi,Bissmann:2020lge}. In the
case of heavy new physics the $t\bar{t}Z$ process can be utilized to provide
complementary information with respect to for example $t\bar{t}\gamma$ in order
to provide model independent constraints in effective field theory
approaches~\cite{Baur:2004uw,Baur:2005wi,Berger:2009hi,Rontsch:2014cca,Rontsch:2015una,Buckley:2015lku,
BessidskaiaBylund:2016jvp,Schulze:2016qas,Hartland:2019bjb,Maltoni:2019aot,Durieux:2019rbz,Brivio:2019ius,
Afik:2021xmi}. On the other hand, the production of $t\bar{t}Z$ is an important
background process in the context of many SM measurements, as it gives rise to
multi-lepton signatures of high complexity. Therefore, its accurate theoretical
description has direct impact on SM measurements such as $pp\to t\bar{t}H$
\cite{ATLAS:2017ztq,CMS:2020mpn,ATLAS-CONF-2019-045} and $pp\to tZ$
\cite{ATLAS:2020bhu,CMS:2021rvz}.

The first calculation of NLO QCD corrections for the inclusive $pp\to
t\bar{t}Z$ production has been reported first in Ref.~\cite{Lazopoulos:2008de}
and later revisited in Refs.~\cite{Maltoni:2015ena,Kardos:2011na}.
Additionally, NLO EW corrections have been computed in
Ref.~\cite{Frixione:2015zaa}. The achieved theoretical accuracy at NLO QCD is
at the level of $\pm 12\%$ for the inclusive cross section and of the order of
$\pm 20\%$ for differential cross section distributions. A further improvement
of the theoretical description of inclusive cross sections at the differential
level has been achieved via soft-gluon resummation at the
next-to-next-to-logarithmic level (NNLL) in Refs.
\cite{Broggio:2017kzi,Kulesza:2018tqz,Kulesza:2020nfh}. In a complementary
approach by matching the fixed-order NLO QCD predictions with parton showers
\cite{Garzelli:2011is,Garzelli:2012bn,Maltoni:2015ena} an improved description
of fiducial signatures has been achieved. Recently, more refined theoretical
predictions became available by taking into account also the $\gamma/Z$
interference contributions in the matching to parton
showers~\cite{Ghezzi:2021rpc}. Finally, also the impact of full off-shell
effects has been studied for the $t\bar{t}Z$ process, where top quarks are
decaying leptonically and the $Z$ boson decays either via $Z\to \nu\nu$
\cite{Bevilacqua:2019cvp,Hermann:2021xvs} or $Z\to \ell\ell$
\cite{Bevilacqua:2022nrm}. The inclusion of NLO QCD corrections to top-quark
decays further improves on the theoretical description and uncertainties of
about $\pm 8\%$ can be obtained for integrated fiducial cross sections and of the order of $\pm 10\%$ at the differential level.

In Ref.~\cite{CMS-PAS-FTR-18-036} the possible constraints on anomalous
top-quark couplings at the HL-LHC have been investigated. Here the main source
of uncertainties are the $t\bar{t}\gamma$ and $WZ$ cross sections, where it has
been already assumed that the theoretical uncertainties reduce by a factor of
$2$. Furthermore, current $pp\to t\bar{t}Z$ measurements~\cite{ATLAS:2020cxf}
are already at the same level of accuracy as theoretical predictions at
NLO+NNLL accuracy, therefore it will be mandatory to include even higher-order
QCD corrections in the near future. The global EFT fit also assumes that theory uncertainties are reduced by a factor two, and that experimental uncertainties are reduced by a factor five, see also Section~\ref{sec:TOPHF-EW-couplings}.

\subsubsection{\texorpdfstring{$pp\to t\bar{t}H$}{pp to ttH}}
The associated production of a Higgs boson with a pair of top quarks ($t\bar{t}
H$) offers the unique opportunity to unambiguously obtain a measurement of the
top-quark Yukawa coupling, the largest and probably most likely to depend on
new physics responsible for the nature of Yukawa interactions in the SM. The prospects for measuring the top quark Yukawa coupling are detailed in Ref.~\cite{Higgsreport}. However, the theoretical challenges in modeling $t\bar{t}H$ are similar to the other processes considered here, thus they are discussed here.

As can be seen in Fig.~\ref{fig:ttV_xsecs}, the $t\bar{t}H$ production cross
section is one of the smallest $t\bar{t} X$ cross sections, as it amounts to
only $1\%$ of the total Higgs boson cross section~\cite{LHCHiggsCrossSectionWorkingGroup:2016ypw}.  Nonetheless, the
$pp\to t\bar{t}H$ process has been observed at the LHC
\cite{CMS:2018uxb,ATLAS:2018mme}.

The process has been studied extensively in the theory community. For stable
top quarks and Higgs bosons theoretical predictions at NLO QCD accuracy can be
found in Refs.~\cite{Reina:2001sf,
Reina:2001bc,Beenakker:2001rj,Dawson:2002tg,Beenakker:2002nc,Dawson:2003zu},
while NLO QCD corrections to $pp\to t\bar{t}Hj$ have been investigated in
Ref.~\cite{vanDeurzen:2013xla}. Afterwards, NLO EW corrections for $t\bar{t}H$
have been studied as well in
Refs.~\cite{Zhang:2014gcy,Frixione:2014qaa,Frixione:2015zaa}.  Further
improvements of the theoretical predictions have been achieved by including
threshold resummation \cite{Kulesza:2015vda,Broggio:2015lya,Broggio:2016lfj,
Kulesza:2017ukk,Broggio:2019ewu,Kulesza:2020nfh} up to next-to-next-to-leading
logarithmic (NNLL) accuracy. While at fixed-order theoretical uncertainties of
about $\pm 10\%$ can be achieved, soft-gluon resummation allows to reduce these
further down to $\pm 6\%$.
Top-quark decays have been included in various approaches. For instance, the
$pp\to t\bar{t}H$ process has been matched to parton showers
\cite{Frederix:2011zi,Garzelli:2011vp,Maltoni:2015ena,Hartanto:2015uka} that
allow for top-quark and Higgs boson decays at LO accuracy.  Furthermore, decays
have been included at fixed-order within the NWA in Ref.~\cite{Zhang:2014gcy}.
Another approach at fixed-order is based on full matrix elements including
Breit-Wigner propagators for resonant particles as well as non-factorizable
contributions and predictions at NLO QCD accuracy have been first reported in
Ref.~\cite{Denner:2015yca} and later complemented with NLO EW corrections in
Ref.~\cite{Denner:2016wet}.  Afterwards, NLO QCD predictions for the full
off-shell $pp \to t\bar{t}H$ production have been extended first by including
the Higgs boson decay via the NWA at NLO QCD accuracy~\cite{Stremmer:2021bnk}
and later in Ref.~\cite{Hermann:2022vit} by studying the impact of CP violating
Higgs couplings~\cite{Artoisenet:2013puc}. Finally, the first efforts to extend
the inclusive $t\bar{t}H$ production to NNLO QCD accuracy have been taken in
Refs.~\cite{Catani:2021cbl,Chen:2022nxt,Brancaccio:2021gcz}.

According to current HL-LHC projections the $t\bar{t}H$ signal will be observed
in the $H\to\gamma\gamma$ decay channel, however, its measurement will be
dominated by theory uncertainties of the order
$20\%$~\cite{ATL-PHYS-PUB-2014-012}.  For the $H\to b\bar{b}$ decay channel, an uncertainty of the order of $13\%$ on the production cross section and similar precision for a measurement of the top-quark Yukawa coupling is expected~\cite{CMS-PAS-FTR-21-002}. In order to realize both scenarios, the
theoretical description of $t\bar{t}H$ signatures has to advance and NNLO QCD
corrections for $t\bar{t}H$ production and Higgs boson decays become
inevitable.

With the overwhelmingly large data set at the end of the HL-LHC run even
previously unthinkable analyses become feasible. For instance, with the direct
measurement of $pp\to t\bar{t}H$ one can constrain the trilinear Higgs
coupling~\cite{CMS-PAS-FTR-18-020}, while, in contrast, the non-observation of processes like $pp\to HH$ and $pp\to t\bar{t}HH$ can only be translated into an upper bound on the trilinear coupling~\cite{CMS-PAS-FTR-21-010}.

\subsubsection{\texorpdfstring{$pp\to t\bar{t}W$}{pp to ttW}}\label{sec:ttw_theory}
The production of a $W$ boson in association with a top-quark pair gives rise
to some of the most spectacular collider signatures at the LHC. Due to multiple
resonant decays of top-quarks and $W$~bosons, the $t\bar{t}W$ process is
accessible in a plethora of different experimental signatures. The study of the $pp\to t\bar{t}W$ process at the LHC is of utmost importance as it represents a main background to most analyses in multi-lepton decay channels, for example top-quark pair production in association with a Higgs boson~\cite{Maltoni:2015ena,CMS:2018uxb,
ATLAS:2018mme,CMS:2020mpn}. It has a direct impact on the measurement of the
top-quark Yukawa coupling~\cite{Higgsreport} and searches for new physics~\cite{Bose:2022obr,ATLAS:2016dlg,CMS:2016mku,CMS:2017tec,ATLAS:2017tmw}. Furthermore, the $t\bar{t}W$ process is a main background in searches for four top-quarks~\cite{CMS:2019rvj,ATLAS:2020hpj}, see also Section~\ref{sec:TOPHF-4top-production}.  It is also one of the rare production modes that give rise to same-sign lepton
signatures in the Standard Model. The corresponding production cross section
can be enhanced in various BSM models such as supersymmetry, extra dimensions,
heavy top-quark partners, extended Higgs sectors as well as heavy Majorana
neutrinos~\cite{Barnett:1993ea,Guchait:1994zk,Baer:1995va,Maalampi:2002vx,
Cheng:2002ab,Dreiner:2006sv,Contino:2008hi,vonBuddenbrock:2016rmr, vonBuddenbrock:2017gvy,vonBuddenbrock:2018xar,Buddenbrock:2019tua, Almeida:1997em,Bai:2008sk}.
In addition, the $t\bar{t}W$ process plays an important role in global SMEFT
fits~\cite{EWreport,Brivio:2019ius,deBlas:2022ofj} and studies of the top-quark charge asymmetry~\cite{Maltoni:2014zpa}.

Due to its importance the process has received a lot of attention in recent
years to further improve on its theoretical accuracy. The first fixed-order predictions taking into account NLO QCD corrections in production and decay were computed~\cite{Badger:2010mg,Campbell:2012dh}. The leading NLO EW corrections have been first investigated~\cite{Frixione:2015zaa, Frederix:2018nkq} and then also the impact of formally subleading mixed QCD and EW corrections have been studied~\cite{Dror:2015nkp,Frederix:2017wme}.  The resummation of
soft-gluon effects has been addressed~\cite{Li:2014ula,Broggio:2016zgg,Kulesza:2018tqz,Broggio:2019ewu}, which, however, in this case only marginally improves upon fixed-order predictions. Recently, NLO QCD and EW corrections, including full off-shell effects in the multi-lepton decay channel, have also become available~\cite{Bevilacqua:2020pzy,Denner:2020hgg,Bevilacqua:2020srb,
Denner:2021hqi}. A complementary approach to an improved description of
fiducial signatures is to match the on-shell $t\bar{t}W$ process
with parton showers~\cite{Garzelli:2012bn,Maltoni:2014zpa,
Maltoni:2015ena,Frederix:2020jzp,FebresCordero:2021kcc}, where further
higher-order QCD corrections can be included via multi-jet
merging~\cite{vonBuddenbrock:2020ter, Frederix:2021agh}. For the first time,
a detailed comparison between fixed-order full off-shell calculations and parton-shower-matched predictions has been presented in~\cite{Bevilacqua:2021tzp}, and a summary is provided in Section~\ref{sec:TOPHF-ttX-ttW}.

For the HL-LHC run, it will become mandatory to include at least NNLO QCD
corrections for the production part in order to reach the expected factor two improvement, see Section~\ref{sec:TOPHF-EW-couplings}. Sizable corrections are expected as this
is the first time the $gg$ initiated production channel opens up. Furthermore, for signatures involving hadronic $W$~boson decays, also higher-order QCD corrections should be taken into account in the description of these decays.
The inclusion of these higher-order corrections is of utmost importance,
especially in light current tensions between theoretical predictions and
measurements~\cite{ATLAS:2019nvo,CMS-PAS-HIG-19-008,CMS-PAS-TOP-21-011}.

\subsubsection{\texorpdfstring{$pp\to t\bar{t}b\bar{b}$}{pp to ttbb}}
The $pp\to t\bar{t}b\bar{b}$ production process is of great importance at the
LHC and even more for the HL-LHC as it constitutes the dominant QCD background
to $t\bar{t}H$ production in the $H\to b\bar{b}$ decay channel~\cite{CMS:2018hnq,ATLAS:2017fak} and thus has a direct impact on the top-quark Yukawa measurement~\cite{Higgsreport}. Furthermore, $t\bar{t}b\bar{b}$ is also one of the main backgrounds in searches for $t\bar{t}t\bar{t}$
production~\cite{CMS:2017nnq,ATLAS:2018kxv}, see Section~\ref{sec:TOPHF-4top-production}.  Besides its role as a background process, the $t\bar{t}b\bar{b}$ process is interesting by itself as it probes
QCD dynamics in top-quark pair production in a truly multi-scale environment. 

Calculations of NLO QCD corrections for the on-shell production are well known~\cite{Bredenstein:2008zb,Bredenstein:2009aj,
Bevilacqua:2009zn,Bredenstein:2010rs,Worek:2011rd,Bevilacqua:2014qfa}.
Including these higher-order corrections allows to reduce theoretical
uncertainties to the level of $30\%$. Fiducial signatures are additionally
impacted by the intricate QCD dynamics of the $g\to b\bar{b}$ splittings in
parton shower evolutions.  This effect has been studied in great detail by
matching $t\bar{t}b\bar{b}$ to parton showers using either massless $b$
quarks~\cite{Kardos:2013vxa,Garzelli:2014aba} or massive
ones~\cite{Cascioli:2013era,Bevilacqua:2017cru,Jezo:2018yaf}. In this context,
also the associated production with an additional light jet, $pp\to
t\bar{t}b\bar{b}j$, has been investigated~\cite{Buccioni:2019plc} at
fixed-order to further improve the understanding of QCD radiation in the
$t\bar{t}b\bar{b}$ process. Due to the immense complexity of these
calculations, only recently predictions including full off-shell effects have
become available for the di-lepton decay
channel~\cite{Bevilacqua:2021cit,Denner:2020orv} which further reduce the
theoretical uncertainties down to the $20\%$ level. To assess the importance of single- and non-resonant contributions, radiative top-quark decays, i.e. $t\to
Wb b\bar{b}$, as well as the size of NLO QCD corrections to these decays a
dedicated comparison with the predictions obtained from the NWA has been reported~\cite{Bevilacqua:2022twl}. The latter study concludes that the
NWA performs excellently for this process with differences below the
percent-level even for differential cross section distributions. Also the
impact of the $g\to b\bar{b}$ splitting in the top-quark decays turns out to be
negligible.

At the HL-LHC, the $t\bar{t}H(H\to b\bar{b})$ measurements will be dominated by theoretical uncertainties on the background processes~\cite{CMS-PAS-FTR-21-002}. Even though the computation of
higher-order QCD corrections at the NNLO level is out of reach for the
foreseeable future, it would be very interesting, especially in the context of precision QCD studies of $pp\to t\bar{t}b\bar{b}$ itself.

\subsection{Modeling uncertainties of \texorpdfstring{$t\bar{t}W^{\pm}$}{ttW} multilepton signatures}
\label{sec:TOPHF-ttX-ttW}
Summary of white paper contribution~\cite{Bevilacqua:2021tzp}

In light of recent discrepancies between the modelling of
$t\bar{t}W^\pm$ signatures and measurements reported by the LHC experimental collaborations, Ref.~\cite{Bevilacqua:2021tzp} investigates in
detail theoretical uncertainties for multi-lepton signatures. 
Results from the state-of-the-art full off-shell calculation
and its Narrow Width Approximation (NWA) are compared to results obtained from the
on-shell $t\bar{t}W^\pm$ calculations, with approximate
spin-correlations in top-quark and $W^\pm$ decays, matched to parton
showers. The off-shell calculation is based on matrix elements for the
fully decayed final state $pp \to \ell^+ \nu_\ell \, \ell^-
\bar{\nu}_\ell \, \ell^\pm \nu_\ell \, b\bar{b}$ at ${\cal
O}(\alpha_s^3 \alpha^6)$ (denoted by $t\bar{t}W^\pm$ QCD) and at
${\cal O}(\alpha_s \alpha^8)$ (denoted by $t\bar{t}W^\pm$ EW).  All
double-, single and non-resonant Feynman diagrams for top quarks and
$W^\pm$ gauge bosons are included, together with interference and
finite-width effects. In the NWA approach, on the other hand, top
quarks and $W^\pm$ gauge bosons are restricted to on-shell
states. Consequently, the full matrix elements are approximated by the
double resonant $t\bar{t}W^\pm$ contributions and the cross section is
factorised into a production and a decay stage.  The NWA predictions
can be further classified into two categories:  The full
NWA, if next-to-leading order (NLO) QCD corrections are included in
the $t\bar{t}W^\pm$ production as well as in the subsequent top-quark
decays, and the NWA with LO decays, if NLO QCD
corrections are incorporated only in the production stage.  These three types of theoretical predictions for
$t\bar{t}W^\pm$ at ${\cal O}(\alpha_s^3 \alpha^6)$ have first been
reported in Ref. \cite{Bevilacqua:2020pzy}. For the parton-shower
matched results (denoted by NLO+PS) for the $pp \to t\bar{t}W^\pm$
process, NLO QCD corrections are only included in the production
stage, while top-quark decays are provided at LO accuracy retaining
spin correlations.  Furthermore, top-quark and $W$ boson virtualities
are modelled according to Breit-Wigner distributions. However, single-
and non-resonant top-quark and $W$ gauge boson contributions as well
as all interferences and NLO QCD spin correlations are still missing.
The \textsc{Powheg-Box} implementation presented in Ref.
\cite{FebresCordero:2021kcc} is used as well as results have been generated in the
\textsc{MC@NLO} framework as provided by MG5${}_{-}$aMC@NLO
\cite{Alwall:2014hca}.

All these predictions for $pp \to t\bar{t}W^\pm$ are
quite different in nature and therefore inherently affected by
distinctive theoretical uncertainties.  With all of them being available for study in 
Ref. \cite{Bevilacqua:2021tzp}, is a unique opportunity to understand their similarities
and differences in more detail. Additionally the size
of full off-shell effects for both the $t\bar{t}W^\pm$ QCD and
$t\bar{t}W^\pm$ EW contribution can be quantified. Ref.~\cite{Bevilacqua:2021tzp} concentrates on the multi-lepton
signature. The reason is twofold. On the one hand, it is the cleanest
signature, on the other hand, it yields the strongest discrepancies
when compared to ATLAS and CMS measurements \cite{ATLAS-CONF-2019-045,CMS-PAS-TOP-21-011}.  As
both full off-shell and NLO+PS predictions have their own virtues, it
would be beneficial not only to compare them, but also to combine them
into a single sample to increase the modelling precision for
$t\bar{t}W^\pm$. For this purpose, the authors of \cite{Bevilacqua:2021tzp} propose a simple method of
approximating the full off-shell effects in parton-shower matched
calculations for on-shell $t\bar{t}W^\pm$ production. Specifically, the
following additive combination is employed:
\begin{equation}
 \frac{d\sigma^\mathrm{th}}{dX} = \frac{d\sigma^\mathrm{NLO+PS}}{dX} + 
 \frac{d\Delta\sigma_\textrm{off-shell}}{dX}\;, \quad \textrm{with} \quad
 \frac{d\Delta\sigma_\textrm{off-shell}}{dX} =  
 \frac{d\sigma^\mathrm{NLO}_\textrm{off-shell}}{dX} - \frac{d\sigma^\mathrm{NLO}_\mathrm{NWA}}{dX}\;,
\label{eqn:improve}
\end{equation}
where fixed-order full off-shell effects are simply added to a
parton-shower based computation. The differential correction,
$d\Delta\sigma_\textrm{off-shell}/dX$, includes the single and
non-resonant contributions as well as interference effects.  To
estimate the theoretical uncertainty of the improved predictions
the scale variations and matching uncertainties are computed independently
and combined as follows:
\begin{equation}
 \delta^\mathrm{th} = \sqrt{ \left(\delta^\mathrm{NLO+PS}_\textrm{scale}\right)^2
 + \left(\delta^\mathrm{NLO+PS}_\textrm{matching}\right)^2
 + \left(\delta^{\Delta\sigma}_\textrm{scale}\right)^2
 }\;,
\label{eqn:error}
\end{equation}
where $\delta^{\Delta\sigma}_\textrm{scale}$ is the estimated
uncertainty of  $d\Delta\sigma_\textrm{off-shell}/dX$.  
%
\begin{figure}[!h!tbp]
  \begin{center}
    \includegraphics[width=0.32\textwidth]{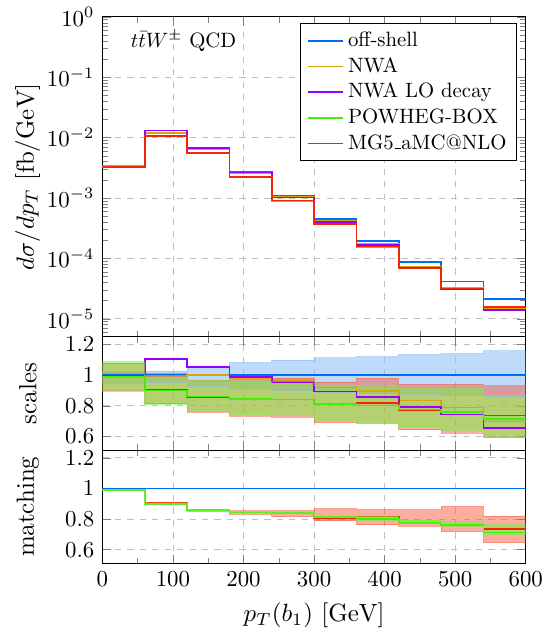}
    \includegraphics[width=0.32\textwidth]{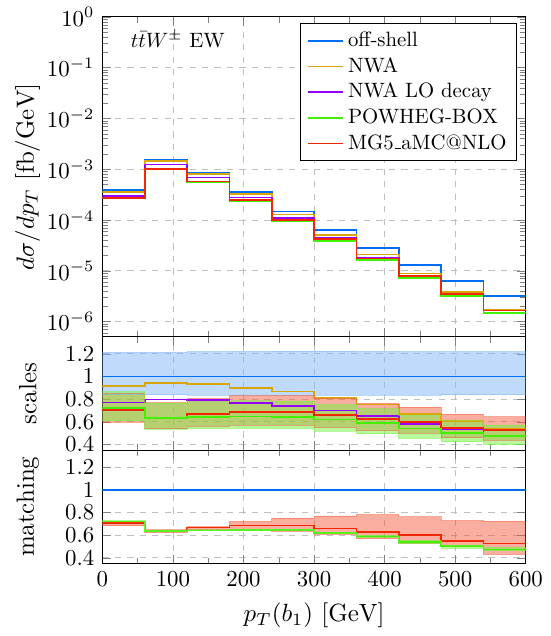}
 \includegraphics[width=0.32\textwidth]{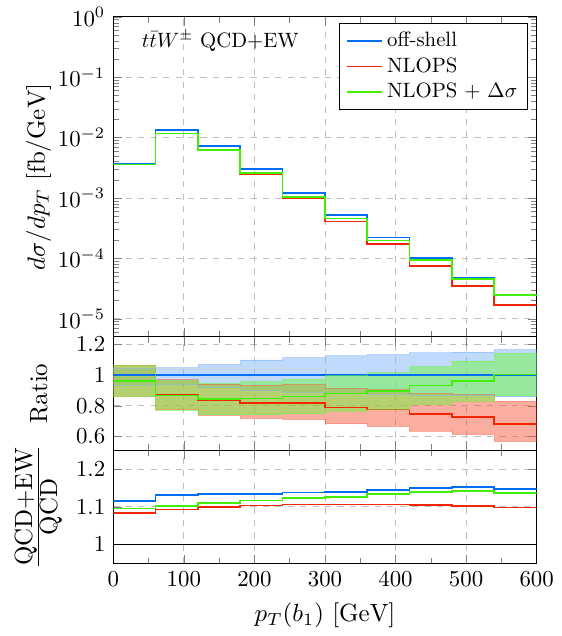}
\end{center}
\caption{\it Differential cross section distributions for 
$pp \to t\bar{t}W^\pm$ in the multi-lepton final state at  
${\cal O}(\alpha_s^3 \alpha^6)$ and at ${\cal O}(\alpha_s
\alpha^8)$  for the LHC  at $\sqrt{s}=13$ TeV as a
function of the transverse momentum of the hardest $b$ jet. 
The upper panels show the absolute NLO QCD predictions for 
$t\bar{t}W^\pm$ QCD and $t\bar{t}W^\pm$ EW production. Also
presented is  the combined result, $t\bar{t}W^\pm$ QCD+EW. For
the first two plots the uncertainty bands correspond to
independent variations of the renormalisation and factorisation
scales (middle panel) and of the matching parameters (bottom
panel). For the rightmost plot the uncertainty bands correspond to
scale variations (off-shell) and total uncertainties (NLO+PS)
(middle panel). The differential impact of the $t\bar{t}W^\pm$
EW contribution is shown as well (bottom panel).}
\label{fig:ttw_modeling}
\end{figure}

As an example, in Figure \ref{fig:ttw_modeling} the transverse momentum
distribution of the hardest $b$ jet is shown for $t\bar{t}W^\pm$ QCD and
$t\bar{t}W^\pm$ EW. The upper panels depict the central differential
distribution for the various predictions employed in our study. The
middle panel illustrates the scale uncertainty bands stemming from
independent variations of factorisation and renormalisation
scales. All curves are normalised to the central prediction of the
off-shell calculation. The bottom panel shows the matching
uncertainties for the parton-shower based predictions. As can be seen, there is
a shape differences between the various predictions over the whole
plotted range. At $p_T \approx 600$ GeV, NLO+PS results as well as NWA
predictions differ from the full off-shell calculation by up to $35\%$
and $50\%$, respectively for $t\bar{t}W^\pm$ QCD and $t\bar{t}W^\pm$
EW.

Additionally, in Figure \ref{fig:ttw_modeling}, combined results are presented for
$t\bar{t}W^\pm$ QCD+EW. In this case, one can examine how well the improved
predictions (denoted by ${\rm NLOPS} + \Delta \sigma$) capture full
off-shell effects. For $t\bar{t}W^\pm$ QCD+EW the upper panel depicts
the central predictions, the middle one the ratio to the full
off-shell prediction, while the bottom one shows for each prediction
the impact of $t\bar{t}W^\pm$ EW over the $t\bar{t}W^\pm$ QCD
result. Additionally, the middle panel shows uncertainty bands
for the full off-shell and the NLO+PS calculation. Finally, total
uncertainties for the ${\rm NLOPS} +\Delta \sigma$ results are given
as well. One observes that in the bulk of the distribution the
corrections due to $d\Delta\sigma_\textrm{off-shell}/dX$ are small as
it should be because these phase-space regions are dominated by the
double resonant $t\bar{t}W^\pm$ production.  On the other hand, the
tails of the distribution receive sizable corrections up to even
$50\%$ with respect to the NLOPS result. One can also observe, that the
theoretical ${\rm NLOPS}+\Delta \sigma$ prediction captures the large
off-shell corrections in the tails. Furthermore, one can notice that
the $t\bar{t}W^\pm$ EW contribution is rather large and quite
flat. Indeed, $10\%-15\%$ corrections are obtained on top of the
dominant $t\bar{t}W^\pm$ QCD contribution.

One could already learn a lot by a detailed comparison of unfolded
$t\bar{t}W^\pm$ data with the various theoretical predictions
presented here and in Ref.  \cite{Bevilacqua:2021tzp}.  In fact,
fixed-order theoretical predictions can and have already been directly
compared to unfolded LHC data.  For instance, for $t\bar{t}$
production NNLO QCD theoretical predictions in the full NWA have been
employed in comparisons with ATLAS and CMS data for various
observables build from top-quark decay products
\cite{Czakon:2020qbd,CMS:2022uae}. Full off-shell NLO QCD predictions for
$t\bar{t}\gamma$ from Ref.  \cite{Bevilacqua:2018woc} have already
been employed by the ATLAS collaboration in comparisons with the
measurements of inclusive and differential cross-sections of combined
$t\bar{t}\gamma$ and $tW\gamma$ production in the $e\mu$ channel
\cite{ATLAS:2020yrp}. They improved the predictions for leptonic
observables, as well as for the description of the prompt photon. In
the next step, similar comparisons can be made for the $t\bar{t}W^\pm$
process in the multi-lepton channel. In a similar fashion, full
off-shell effects approximately incorporated in the NLO computation of
on-shell $t\bar{t}W^\pm$ production matched to parton showers can be
used to improve the modeling of LHC data. Finally, full off-shell
calculations for $t\bar{t}W^\pm$ should be matched to parton shower
programs using methods that allow for the consistent treatment of
resonances. Such an approach has already been worked out for the
simpler case, namely for $t\bar{t}$ and $tW$ production
\cite{Jezo:2015aia,Jezo:2016ujg} and used in comparisons to ATLAS data
\cite{ATLAS:2018ivx}.

\FloatBarrier
\FloatBarrier
\subsection{Four-top production}\label{sec:TOPHF-4top-production}

The production of four top-quarks is one of the rarest SM processes that is expected to be discovered by the LHC experiments. 
The $pp\to t\bar{t}t\bar{t}$ process allows to probe various BSM extensions of the SM, as its cross section can receive significant enhancements~\cite{Lillie:2007hd,Kumar:2009vs,Cacciapaglia:2011kz,Perelstein:2011ez,
Aguilar-Saavedra:2011mam,Beck:2015cga,Alvarez:2016nrz,Alvarez:2019uxp,
Alvarez:2021hxu,Carpenter:2021vga}. Prominent examples for such scenarios are supersymmetric gluino-pair production~\cite{Nilles:1983ge,Farrar:1978xj,Toharia:2005gm,
Afik:2021xmi,LHCNewPhysicsWorkingGroup:2011mji}, top-quark pair production in
association with a Higgs boson in two-Higgs doublet
models~\cite{Dicus:1994bm,Craig:2015jba,Craig:2016ygr}, top-philic Dark Matter
scenarios~\cite{Boveia:2016mrp,Albert:2017onk} and composite-Higgs
models~\cite{Giudice:2007fh,Pomarol:2008bh}.
The $t\bar{t}t\bar{t}$ process is also particularly sensitive to the top-quark
Yukawa coupling and CP properties of the Higgs
boson~\cite{Cao:2016wib,Cao:2019ygh}.  

The four top-quark production
process plays a special role in constraining BSM physics via effective field theory approaches, since it is very sensitive to four-fermion operators~\cite{Banelli:2020iau,Zhang:2017mls,Aguilar-Saavedra:2018ksv}. Examples of Feynman diagrams for this process are shown in  Figure~\ref{fig:Atlas_4t} (left). While the largest contributions are from QCD processes, there are small contributions mediated by the Higgs boson. 

\subsubsection{\texorpdfstring{Theoretical aspects of $pp\to t\bar{t}t\bar{t}$}{pp to tttt}}
\label{sec:TOPHF-tttt-theory}

The dominant NLO QCD corrections to 4-top production have been first computed in
Ref.~\cite{Bevilacqua:2012em} and afterwards. With the advent of automated NLO EW corrections, the subleading EW corrections have been investigated~\cite{Frederix:2017wme}. NLO QCD corrections are large, of the
order of $20-60\%$, depending on the scale choices, while the EW channels contribute a $+10\%$ corrections at the inclusive level.  The sizable enhancement of the EW contributions originate from the so-called Sommerfeld
enhancement~\cite{Kuhn:2013zoa,Beneke:2015lwa} of heavy quark production near threshold. The leading QCD process has been matched to parton showers for the
first time~\cite{Maltoni:2015ena}, while recently refined predictions including full NLO QCD corrections as well as the dominant EW tree-level
contributions have been matched to parton showers~\cite{Jezo:2021smh}
using the POWHEG BOX framework. Currently, the theoretical uncertainties of the inclusive cross section are estimated to be of the order of $\pm 20\%$ due to
missing higher-order corrections and up to $\pm 5\%$ due to PDF uncertainties.

In order to reach the experimental precision of order 10\% (see Section~\ref{sec:TOPHF-tttt-exp}), the theoretical predictions necessarily have to improve in the future. It is hard to imagine that NNLO QCD corrections will become available any time soon. However, improvements due to the inclusion of threshold resummation might be
achievable.  Further advances of the theoretical predictions could be obtained if a framework for parton shower matching for mixed QCD-EW contributions can be established.

\subsubsection{\texorpdfstring{Experimental aspects of $pp\to t\bar{t}t\bar{t}$}{pp to tttt}}
\label{sec:TOPHF-tttt-exp}

The 4-top analyses at the HL-LHC benefit from the increase in collision energy. The cross section increases by a factor of approximately 1.3 when increasing the collision energy from 13 to 14 TeV. It increases by two orders of magnitude when going to the FCC-hh 100~TeV collider, see Figure~\ref{fig:ttV_xsecs}.

\begin{figure}[!h!tb]
\begin{minipage}{0.35\textwidth}
\begin{center}
\includegraphics[width=0.65\textwidth]{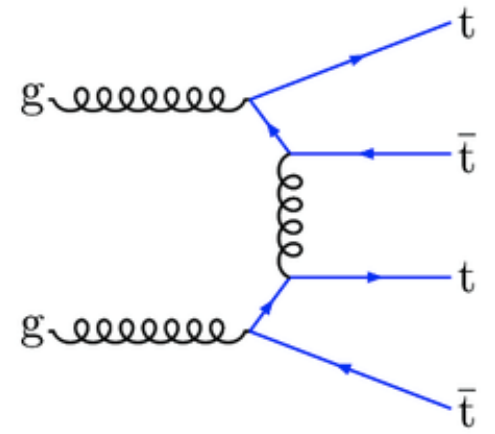} 
\includegraphics[width=0.65\textwidth]{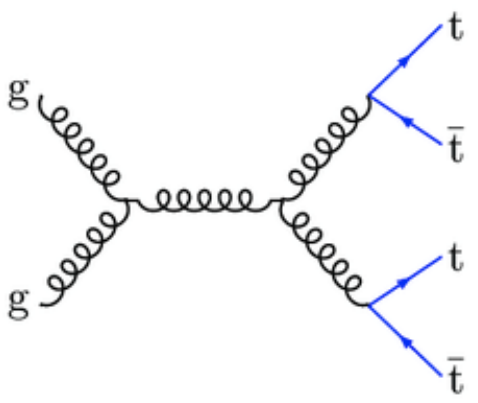}
\includegraphics[width=0.65\textwidth]{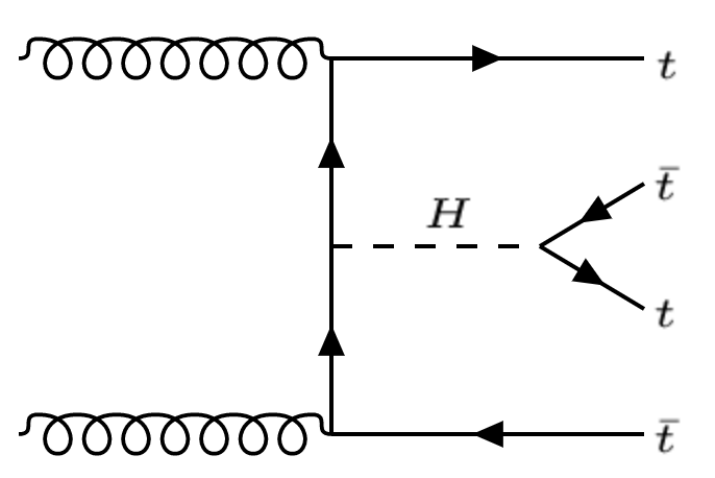}
\end{center}
\end{minipage}
\begin{minipage}{0.55\textwidth}
\includegraphics[width=0.99\textwidth]{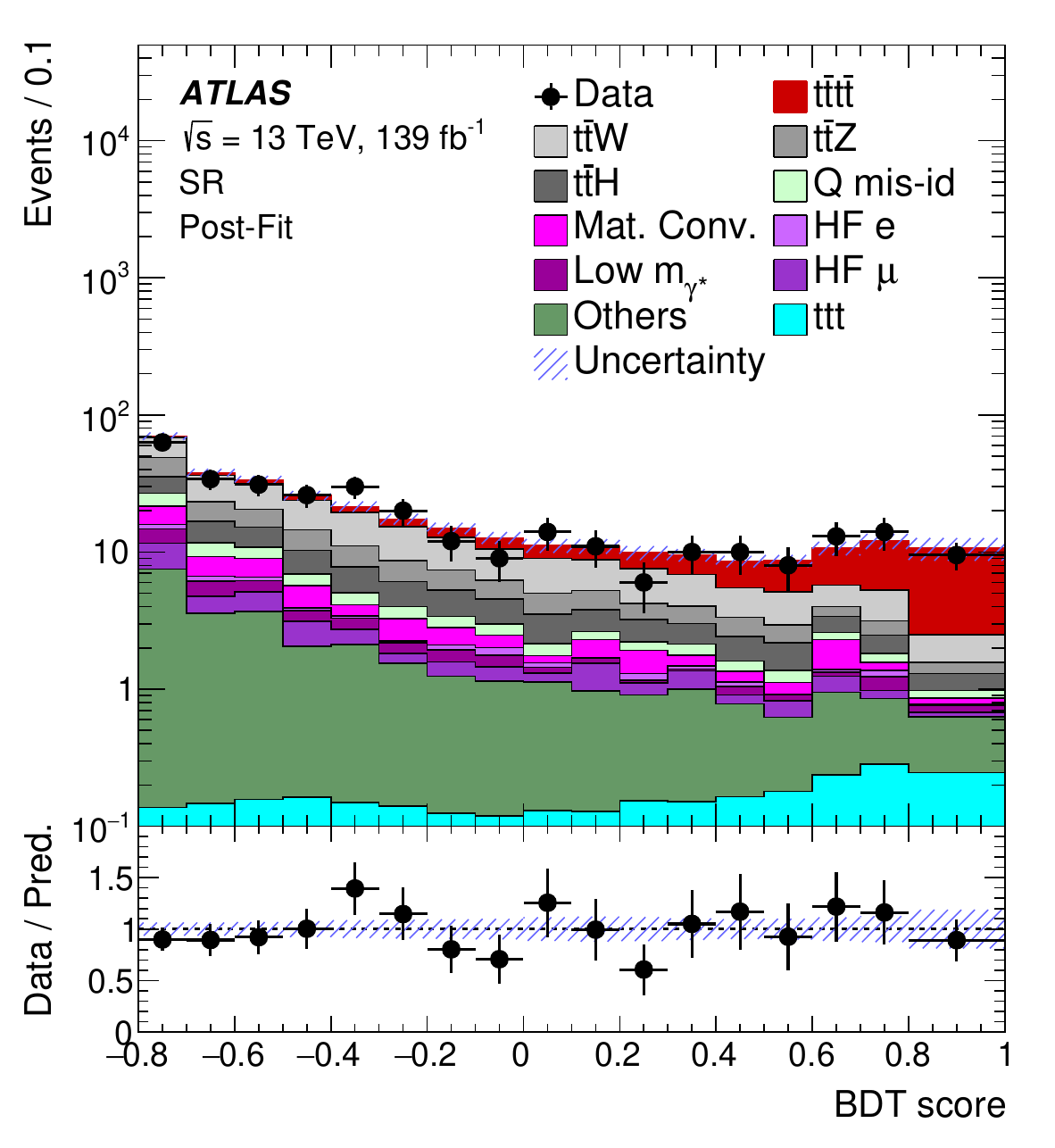}
\end{minipage}
\caption{
(Top left) Examples of Feynman diagrams of four top quark production. (Top right) Distribution over the BDT discriminant for multilepton events observed by ATLAS based on the full Run II data set.  
From Ref~\cite{ATLAS:2021kqb,ATLAS:2020hpj}.}
\label{fig:Atlas_4t}
\end{figure}

The analysis of 4-top events necessitates several analysis channels due to the four $W$~bosons in the final state.
The data set is typically divided into the same-sign channel (which includes same-sign dilepton and multilepton events) and the lepton+jets channel (which includes lepton+jets and opposite sign dilepton events), with the former being the most sensitive. The same-sign events have smaller expected event counts but also smaller backgrounds, while the lepton+jets events have larger expected event counts and larger backgrounds. Multivariate discriminants are constructed to separate signal from backgrounds in each channel. In Figure~\ref{fig:Atlas_4t} (right), a distribution over the multivariate discriminant is shown for multilepton events observed by the ATLAS experiment~\cite{ATLAS:2021kqb,ATLAS:2020hpj}. The signal shown in red is clearly visible at the high values of the discriminant. 
Based on the full Run~2 data set, ATLAS reported an observation of the four-top signal with a significance of 4.7 standard deviations for a combination of both channels, with the expected significance of 2.6. 
Overall, the measured value is in agreement with the SM within 2 standard deviations. 

Using the multilepton channel, CMS reported observed and expected significances of 2.6 and 2.7 standard deviations, respectively~\cite{CMS:2019rvj}. This measurement allowed to constrain the ratio of the top Yukawa coupling to its SM value to be less that 1.7 at 95\% CL. 

\begin{figure}[htb]
\centering
\includegraphics[width=0.6\textwidth,angle=-90]{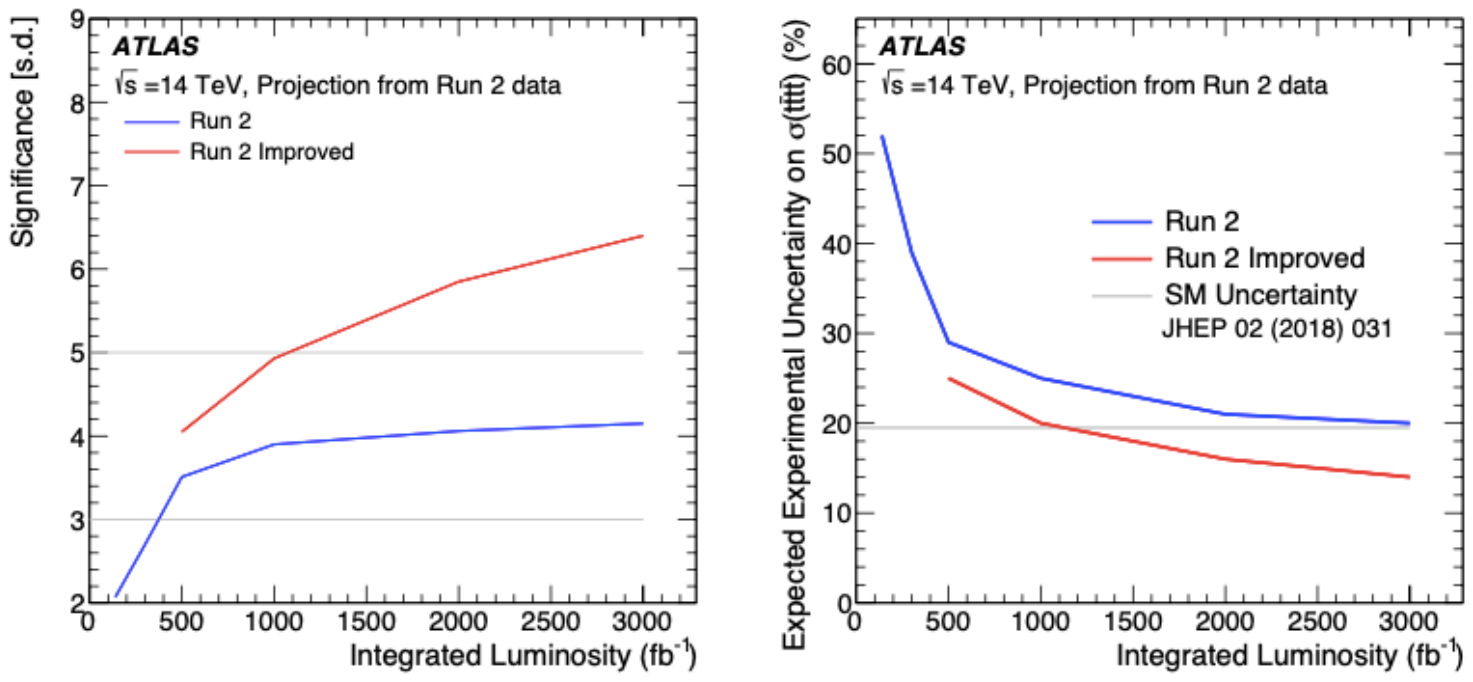}
\caption{Expected significance for the measured $t\bar t t \bar t$ cross section (left) and expected experimental uncertainty (right), assuming two different scaling scenarios for the systematic uncertainties. Taken from Ref.~\cite{ATLAS:2022kld}.}
\label{fig:4Top_ATLAS}
\end{figure}

The CMS searches for the production of four top quarks in the same-sign~\cite{CMS:2019rvj} and lepton+jets channels~\cite{CMS:2019jsc} were used to provide
projections for the High-Luminosity LHC and High-Energy LHC~\cite{CMS:2018nqq}. 
Several different scenarios for the systematic uncertainties are
considered. For proton-proton collisions at $\sqrt{s} = 14$ TeV, the existing analysis strategies are expected to become dominated by systematic uncertainties. Evidence for $t\bar{t}t\bar{t}$
in a single analysis will become 
possible with around 300~fb$^{-1}$ of High-Luminosity
LHC data at 14~TeV. With these data sets, the uncertainty on the
measured cross section will be of the order of 40\%. With 3~ab$^{-1}$ of High-Luminosity LHC data, the cross section can be constrained to 9\% statistical uncertainty and 18 to 28\% total uncertainty. 
A study done by ATLAS~\cite{ATLAS:2022kld} demonstrates a similar increase in the signal significant and decrease in the uncertainties as a function of the integrated luminosity (see Figure~\ref{fig:4Top_ATLAS}).

Comparison between different machines in terms of their ability to constrain the four-top operator~\cite{Banelli:2020iau} is summarized in Table~\ref{tab:TOPHF-bound-4top}. The 4-top operator limits from the LHC can be improved upon at a lepton collider, though a significant improvement requires high-energy collisions of at least a TeV. 

 \begin{table}[!h!t]
\begin{center}
\begin{tabular}{l|c|c|c|c|c}
 & HL-LHC($pp\to tttt$) & FCC-ee & ILC & CLIC  & FCC-hh \\ 
 & ($pp\to tttt$) & ($e^+e^-\to tt$) & ($e^+e^-\to tt$) & ($e^+e^-\to tt$) & ($pp\to tttt$)\\ \hline
$\sqrt{s}$ [TeV] & 14 & 0.365 & 1  & 3 & 100 \\
$\L [\abinv]$       & 3 & 1.5    & 1    &   3 &  30 \\ \hline
 $\Lambda/\sqrt{|c_{tt}|}$ [TeV]   & 1.3 & 1.6 & 4.1 & 7.7  & 6.5  \\ \hline
\end{tabular}
\caption{Bounds on the four-top operator. Taken from \cite{Banelli:2020iau}}, which is a phenomenological study and the HL-LHC numbers are not based on the ATLAS~\cite{ATLAS:2022kld} and CMS~\cite{CMS:2018nqq} studies.
\label{tab:TOPHF-bound-4top}
\end{center}
\end{table}

\begin{figure}[htb]
\centering
\includegraphics[width=0.7\textwidth]{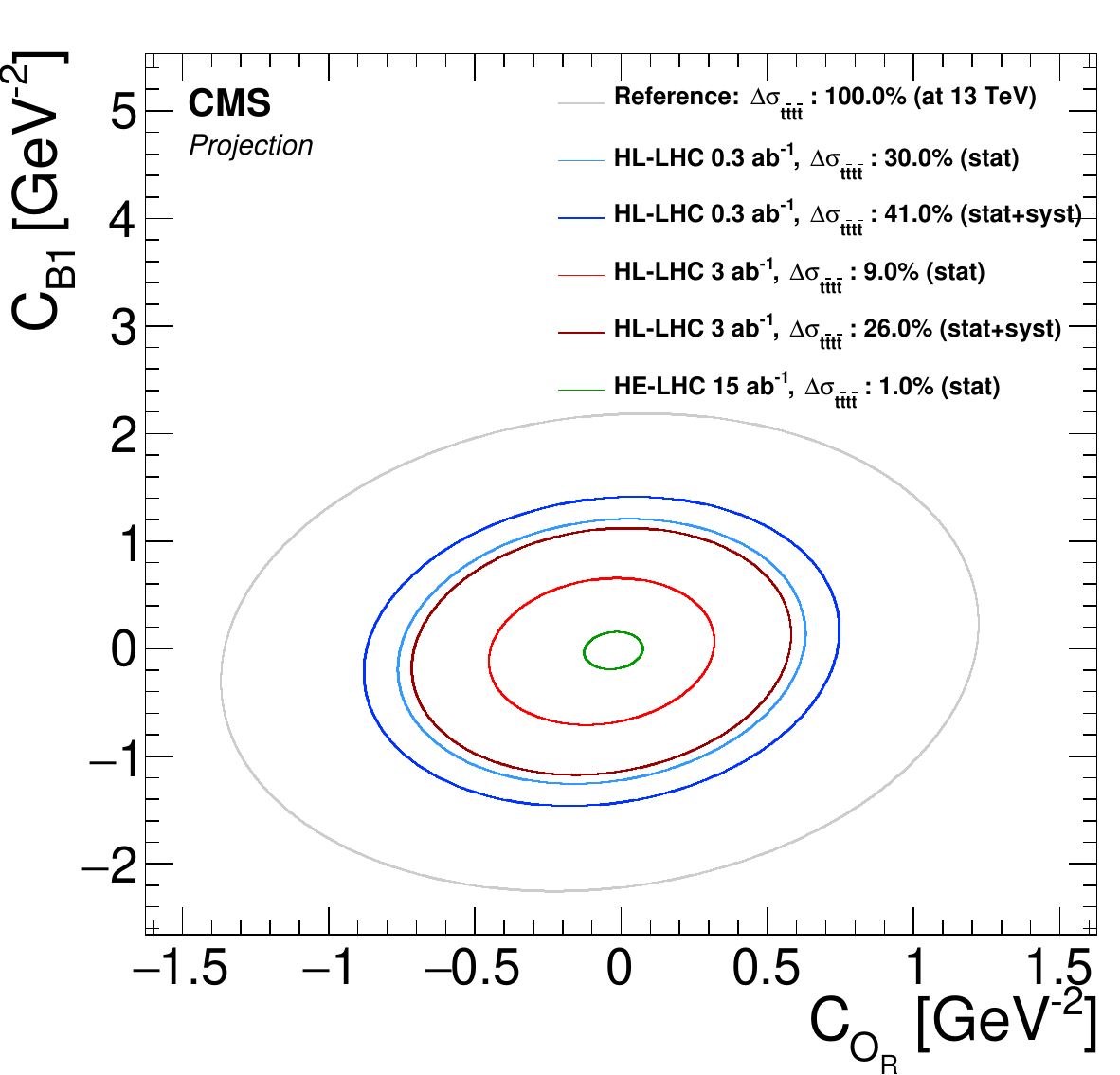}
\caption{Expected limits on EFT contact interaction operators for the $t\bar t t \bar t$ process. Taken from Ref~\cite{CMS:2018nqq}.}
\label{fig:4topcontact_CMS}
\end{figure}

With such increase in precision it will be possible to constrain the EFT contact interaction operator even further. The corresponding  projections for the HL-LHC are shown in  Figure~\ref{fig:4topcontact_CMS}. It is clear that the increase in energy provides with the most significant improvement in precision (27~TeV HE-LHC shown in green). 

\FloatBarrier
\section{Top-quark coupling measurements from EFT fits}
\label{sec:TOPHF-EW-couplings}

\begin{figure}[!h!t]
\hspace*{1.cm}%
\includegraphics[width=0.65\columnwidth]{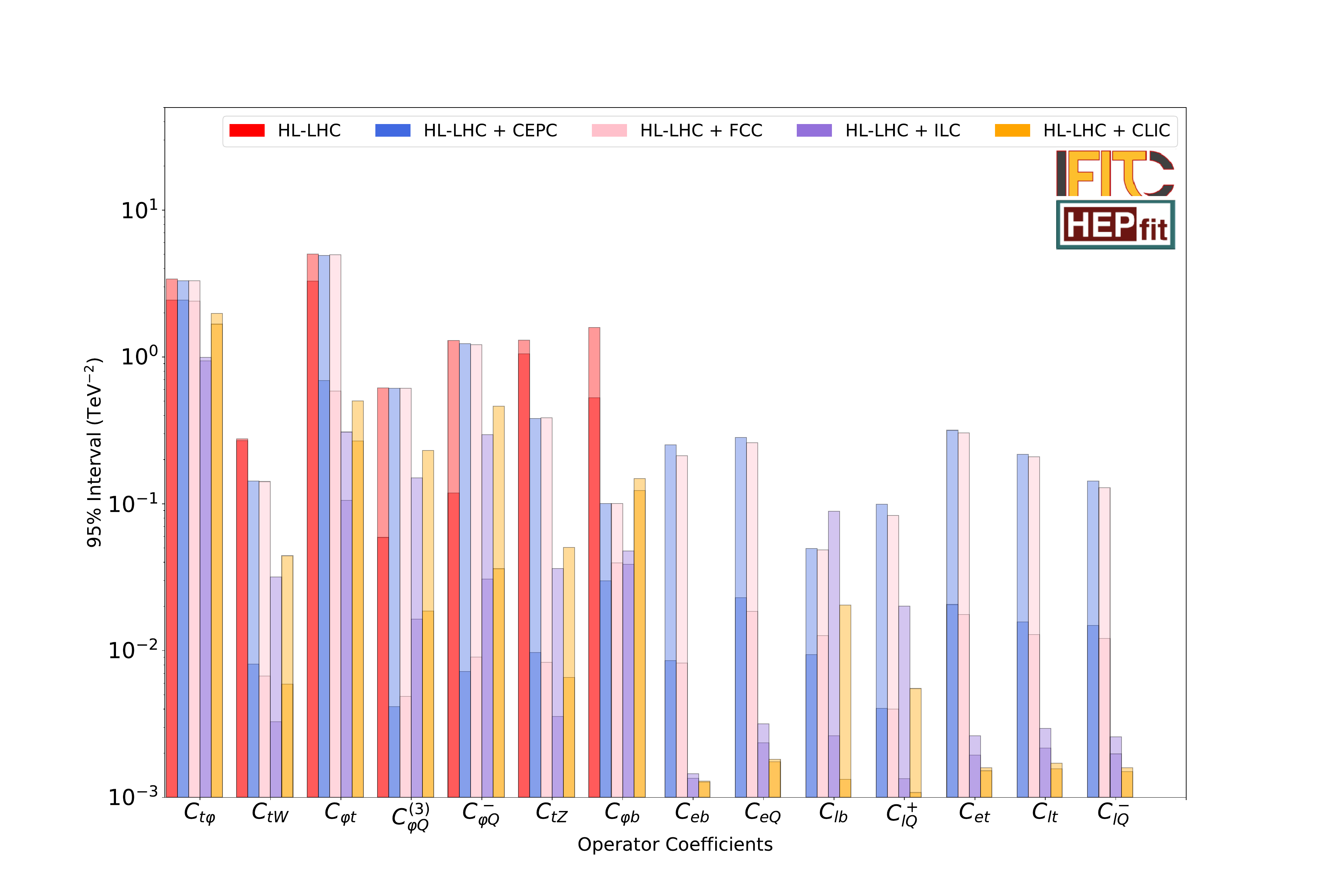}
\caption{Comparison of the constraints expected from a combination of HL-LHC and lepton collider data on Wilson coefficients for EFT operators relevant to top-quark couplings.
The solid bars provide the individual limits of the single-parameter fit and the shaded ones the marginalised limits of the global fit.
}
\label{fig:TOPHF-top-EFT}
\end{figure}

Measurements of (differential) cross-sections of top-quark production processes provide important inputs to global SMEFT fits~\cite{Brivio:2019ius}. The studies of top-quark decay and top-quark final-state correlations and associated production processes provide further constraints~\cite{Durieux:2018tev,deBlas:2022ofj}. In this section, the LHC Run 2 results and projections for future colliders as well as for theoretical uncertainties are listed which have been included in global EFT fits performed in the Topical Group EF04. The resulting limits on top-quark-related EFT operators are found in the EF04 report~\cite{EWreport} and are also shown here in Figure~\ref{fig:TOPHF-top-EFT}.

\subsection{Prospects at the HL-LHC}

Precision measurements of top-pair, single top and top-associated processes are all used as inputs to the global EFT fits. The $\ttbar$ and $t\bar{t}X$ observables are described below. For single top-quark production, the inclusive cross-sections are used for $t$-channel, $s$-channel and $tW$ production. The $s$-channel single top production at the Tevatron is also included. At the LHC, associated production of a single-top quark in the $t$-channel with either a photon or a $Z$~boson provides additional information.

\subsubsection{Observables in \texorpdfstring{$t\bar{t}$}{tt} production}

The projections for the HL-LHC fit shown in Figure~\ref{fig:TOPHF-top-EFT} are based  on an extrapolation from current (Run~2) measurements. The measurements that form the basis for the HL-LHC projection are listed in Table~\ref{tab:measurements}. This includes the production processes discussed in more detail in Sections~\ref{sec:TOPHF-XS-LHC-tt-theo}, \ref{sec:TOPHF-XS-LHC-ttexp}, \ref{sec:TOPHF-XS-LHC-stexp} and~\ref{sec:TOPHF-XS-LHC-ttX}.

\begin{table*}[tb]
\centering
\resizebox{\textwidth}{!}{%
\begin{tabular}{|l|c|c|c|c|c|c|}
\hline
Process & Observable & $\sqrt{s}$  & $L_\text{int}$  & Experiment & SM  & Ref.\\ \hline
$pp \rightarrow \ttbar $ & $d\sigma/dm_\ttbar$ (15+3 bins) & 13 TeV & 140~\ifb & CMS & \cite{Czakon:2013goa} & \cite{CMS:2021vhb} \\
$pp \rightarrow \ttbar $ & $dA_C/dm_\ttbar$ (4+2 bins) & 13 TeV & 140~\ifb & ATLAS & \cite{Czakon:2013goa} & \cite{ATLAS-CONF-2019-026} \\
$p p \rightarrow t \bar{t} H+ tHq$ & $\sigma$ & 13 TeV & 140~\ifb & ATLAS & 
\cite{deFlorian:2016spz} &  \cite{ATLAS:2020qdt} \\ 
$p p \rightarrow t \bar{t} Z$ &  $d\sigma/dp_T^Z$ (7 bins) & 13 TeV &  140~\ifb  & ATLAS &
\cite{Broggio:2019ewu} & \cite{ATLAS:2020cxf} \\ 
$p p \rightarrow t \bar{t} \gamma$ & $d\sigma/dp_T^\gamma$ (11 bins) & 13 TeV & 140~\ifb & ATLAS &
\cite{Bevilacqua:2018woc,Bevilacqua:2018dny} & \cite{Aad:2020axn}  \\ 
$p p \rightarrow tZq$ & $\sigma$ & 13 TeV & 77.4~\ifb{}  & CMS &
 \cite{Sirunyan:2017nbr} & \cite{Sirunyan:2018zgs} \\ 
$p p \rightarrow t\gamma q$ & $\sigma$ & 13 TeV & 36~\ifb{}  & CMS &
 \cite{Sirunyan:2018bsr} & \cite{Sirunyan:2018bsr} \\
 $p p \rightarrow t \bar{t} W$ & $\sigma$ & 13 TeV & 36~\ifb  & CMS &
\cite{deFlorian:2016spz,Frederix:2017wme} &  \cite{Sirunyan:2017uzs} \\
 $p p \rightarrow t\bar{b}$ (s-ch) & $\sigma$ & 8~\tev{} & 20~\ifb{} & LHC &
\cite{Aliev:2010zk,Kant:2014oha} & \cite{Aaboud:2019pkc} \\ 
$p p \rightarrow tW$ & $\sigma$ & 8~\TeV & 20~\ifb{}  & LHC &
 \cite{Kidonakis:2010ux} &  \cite{Aaboud:2019pkc}   \\ 
$p p \rightarrow tq$ (t-ch) & $\sigma$ & 8~\tev{} & 20~\ifb{} & LHC &
\cite{Aliev:2010zk,Kant:2014oha} & \cite{Aaboud:2019pkc} \\ 
$t \rightarrow Wb $ & $F_0$, $F_L$  & 8~\tev{} & 20~\ifb{} & LHC &
\cite{PhysRevD.81.111503}  & \cite{Aad:2020jvx} \\
$p\bar{p} \rightarrow t\bar{b}$ (s-ch) & $\sigma$ & 1.96~\tev & 9.7~\ifb & Tevatron & \cite{Kidonakis:2010tc} & \cite{CDF:2014uma} \\
$e^{-} e^{+} \rightarrow b \bar{b} $ & $R_{b}$ ,  $A_{FBLR}^{bb}$ & $\sim$ 91~\GeV & 202.1~\ipb  & 
LEP/SLD &
 - & \cite{ALEPH:2005ab}  \\ \hline
\end{tabular}%
}
\caption{Measurements included in the EFT fit of the top-quark electroweak sector. For each measurement, the process, the observable, the centre-of-mass energy, the integrated luminosity and the experiment/collider are given. The last two columns list the references for the predictions and measurements that are included in the fit. LHC refers to the combination of ATLAS and CMS measurements. In a similar way, Tevatron refers to the combination of CDF and D0 results, and LEP/SLD to different experiments from those two accelerators.}
\label{tab:measurements}
\end{table*}

For the top-quark pair production process, statistics is abundant and measurements in the bulk already reach a precision of a few \%. 

Experimental uncertainties on the inclusive cross-section as well as many differential distributions are expected to be reduced to approximately 1\% at the HL-LHC~\cite{ATLAS:2016zet,Azzi:2019yne}, see Section~\ref{sec:TOPHF-XS-LHC-ttexp}. Currently, theory uncertainties of the N$^2$LO calculation are at the level of 3--4\% for the inclusive cross section~\cite{Czakon:2013goa}. These might be reduced to roughly half with the calculation of the N$^3$LO corrections (see Section~\ref{sec:TOPHF-XS-LHC-theo}) and the improvement of the proton PDFs (see Section~\ref{sec:TOPHF-mtop-pdf} and Ref.~\cite{QCDreport}). Even in that case, theory uncertainties are likely to remain the limiting factor.

The $t\bar{t}$ charge asymmetry is a subtle effect at the LHC, but it  brings important information to EFT fits~\cite{Zhang:2012cd}. As a ratio, it can be precisely predicted~\cite{Czakon:2017lgo}. Modelling uncertainties play an important role~\cite{ATLAS-CONF-2019-026} and are likely to limit future progress in the inclusive measurement. Therefore, a less aggressive scenario is adopted, where all experimental systematic uncertainties are improved by a factor 1/2 and only the statistical uncertainty scales with $1/\sqrt{L_\text{int}}$.

For top-quark pair production, differential measurements of the cross section~\cite{CMS:2021vhb} and the charge asymmetry~\cite{ATLAS-CONF-2019-026} as a function of the invariant mass of the $t\bar{t}$ system are considered. A promising avenue for progress is the boosted regime, where the sensitivity to four-fermion operators increases considerably~\cite{Rosello:2015sck}. Measurements of the cross section and charge asymmetry for $t\bar{t}$ systems produced at large invariant mass already play an important role in the constraints on the four-fermion operators and their weight will increase if measurements on bulk $t\bar{t}$ are limited by experimental or theoretical systematic uncertainties. To take maximal advantage of this potential, the range of the projections is extended further into the high-$m_\ttbar$ tail than the current Run 2 measurements. 


\subsubsection{Observables in associated production processes}
\label{sec:TOPHF-xs-assoc}


The theoretical challenges for the production of a top-quark pair with another particle are described in Section~\ref{sec:TOPHF-XS-LHC-ttX}. Here, the experimental measurements are given that are projected to the HL-LHC.
The projections of the measurements of rare top-quark production processes are modelled on the S2 scenario used to predict the precision of Higgs coupling measurements in Ref.~\cite{Cepeda:2019klc}. This scenario envisages that many statistical and experimental uncertainties scale as $1/\sqrt{L_\text{int}}$, where $L_\text{int}$ is the integrated luminosity. For the complete HL-LHC program, experimental uncertainties are expected to be reduced by a factor 5 with respect to the current Run~2 results. Theory and modelling uncertainties are divided by two, with respect to today's state of the art. In all measurements of inclusive $pp \rightarrow t\bar{t}X$ and $pp \rightarrow tqX$ rates, the modelling uncertainties become the dominant source of uncertainty in this scenario, and the theory uncertainty on the SM prediction is also more sizable than statistical and experimental uncertainties.
Even if N$^2$LO calculations\cite{Catani:2021cbl} including EW corrections are achieved for associated production processes (see Section~\ref{sec:TOPHF-XS-LHC-ttX}), theory and modelling uncertainties are still expected to limit the precision of the comparison.

To gain the maximal sensitivity to the EFT coefficients, differential measurements are included in the global analysis. As in Ref.~\cite{Miralles:2021dyw}, for the $pp \rightarrow t\bar{t}Z$ and $pp \rightarrow t\bar{t}\gamma$ processes, differential measurements as a function of the $Z$-boson and photon $p_T$ are included, enhancing the sensitivity to $C_{tZ}$, in particular~\cite{Bylund:2016phk}.

\subsubsection{Discussion}

Generally, the progress envisaged in the S2 uncertainty scenario is limited by the theory and modelling uncertainties, while statistical and experimental uncertainties  are expected to be sub-dominant in nearly all measurements. Improvements are only possible experimentally through studying rare processes and more extreme phase space regions~\cite{Maltoni:2019aot}. Theoretically, improving the accuracy of fixed-order predictions beyond a factor two will lead to a direct improvement of the sensitivity of EFT fits, even if these calculations become available after the HL-LHC program is complete. This will, however, likely require N$^3$LO precision for $2 \rightarrow 3$ processes with top quarks in the final state. 

The boosted regime is one of the keys to improving bounds on the operators that affect the top-quark pair production process. In particular, the high-$m_\ttbar$ tail of the top-quark pair production measurements provides a significant reduction in the allowed regions of the four-quark operators, which shrink by a factor between two and five (depending on the operator) thanks to the enhanced sensitivity in this regime and the more pronounced improvement in the measurement. This effect is present even in a fit that only includes the linear (${\cal O} (\Lambda^{-2}$)) terms in the parameterization of the EFT dependence~\cite{EWreport}. 

The marginalised bounds on the four-fermion operators remain an order of magnitude worse than the individual bounds after the HL-LHC, even if both individual and global bounds improve considerably. This is due to unresolved correlations between the coefficients. The same feature is observed in recent fits to the top sector of the SMEFT~\cite{Brivio:2019ius,Hartland:2019bjb} and in global Higgs/EW/top fits~\cite{Ethier:2021bye,Ellis:2020unq}. Stricter limits can be obtained if the dimension-six-squared terms proportional to $\Lambda^{-4}$ are included in the fit~\cite{Ethier:2021bye}.

Two-quark two-lepton operators, omitted in this section, can also be probed at the LHC.
Dedicated signal regions, for instance with off-$Z$-peak dilepton invariant masses in $pp\to t\bar{t}\ell^+\ell^-$~\cite{Durieux:2014xla, Chala:2018agk,Sirunyan:2020tqm}, would increase their sensitivity.

\subsection{Prospects for \texorpdfstring{$e^+e^-$}{e+e-} colliders}\label{sec:TOPHF-EW-eett}

The $e^+e^- \rightarrow \gamma,Z \rightarrow t\bar{t}$ process opens up for centre-of-mass energies that exceed twice the top mass (i.e. $\sqrt{s} \gtrsim$ 350~\GeV) and directly probes the electroweak couplings of the top quark.
Data taken with different beam polarisations at linear colliders can be used to distinguish the photon and $Z$-boson couplings~\cite{Amjad:2013tlv,Amjad:2015mma,Durieux:2018tev,CLICdp:2018esa} to the top quark.
At circular colliders, a measurement of the final state polarisation using the semi-leptonically decaying top quarks can also be used to separate the two contributions~\cite{Janot:2015yza}.

Realistic estimates of the impact of acceptance, identification and reconstruction efficiencies are taken from full-simulation studies for the ILC and CLIC in Ref.~\cite{Amjad:2013tlv,Abramowicz:2016zbo}.
In runs at the \ttbar{} threshold or slightly above, the cross section can be measured to a few per mille precision~\cite{Amjad:2013tlv,Abramowicz:2016zbo}. The precision drops to several \% for the highest center-of-mass energy at CLIC, due to the $1/s$ decrease of the cross section, that is only partially compensated by the luminosity, in combination with a degradation of the top-selection and flavour-tagging capabilities and the broad CLIC luminosity spectrum~\cite{Abramowicz:2016zbo}. Experimental systematic uncertainties and the precision of theory calculations are expected to be sub-dominant in the comparison of data with the Standard Model predictions.  For an overview of the current status of theory calculations for continuum $t\bar t$ and $t\bar tH$ production see, e.g., Ref.~\cite{ChokoufeNejad:2016qux}, where a study at the differential level for the full processes $e^+e^-\to \mu^+ \nu_\mu e^-\bar \nu_e b\bar b$ and $e^+e^-\to \mu^+ \nu_\mu e^-\bar \nu_e b\bar bH$ has been performed at NLO QCD including non-resonant contributions, off-shell effects and interferences. 

Numerical prospects for the broader $e^+e^-$ top-physics program are based on the study of statistically optimal observables defined at leading order on the $e^+e^- \to t\bar{t} \rightarrow WbWb$ differential distribution~\cite{Durieux:2018tev}. The optimal observable capture not only the information in classical observables such as the cross section and the forward-backward asymmetry, but also take advantage of the top polarization to find the optimal constraint on the complete set of EFT operator coefficients. We note that the $WbWb$ final state also receives contribution from single top production which become sizable at high centre-of-mass energies, but the analysis does not include the associated $e^+e^- \rightarrow t\bar{t}H$ production process, that provides a direct constraint on the top quark Yukawa coupling~\cite{Price:2014oca}, or top-quark pair production in vector-boson-fusion. 
The $e^+e^- \rightarrow t\bar{t}H$ process is included directly~\cite{EWreport}. 

The results of these full-simulation studies are extrapolated to realistic operating scenarios of the electron-positron collider projects listed in Table~\ref{tab:epem_setup} using a parameterization of the acceptance and efficiency as a function of the centre-of-mass energy. 
\begin{table}[tb]
    \centering
    \renewcommand{\arraystretch}{1.1}
    \begin{tabular*}{\textwidth}{@{\extracolsep{\fill}}|c|c@{\quad}|c@{\quad}|c@{\quad}|c@{\quad}|}
    \hline
       Machine & Polarisation & Energy & Luminosity & Reference\\\hline
        \multirow{3}{*}{ILC} & \multirow{3}{*}{P($e^+$, $e^-$):$(\pm30\%,\,\mp80\%)$}  & 250 GeV & 2 \iab{} & \multirow{3}{*}{\cite{AlexanderAryshev:2022pkx}} \\
         &  & 500 GeV & 4 \iab{} & \\
         &  & 1 TeV & 8 \iab{} & \\\hline
        \multirow{3}{*}{CLIC} & \multirow{3}{*}{P($e^+$, $e^-$):$(0\%,\,\pm80\%)$} &  380 GeV & 1 \iab{} & \multirow{3}{*}{\cite{Robson:2018zje}}\\
         &  & 1.4 TeV & 2.5 \iab{} & \\
         &  & 3 TeV & 5 \iab{} & \\\hline
         \multirow{4}{*}{FCC-$ee$} & \multirow{4}{*}{Unpolarised} &  Z-pole & 150 \iab{} & \multirow{4}{*}{\cite{Bernardi:2022hny}}\\
         &  & 240 GeV & 5 \iab{} & \\
         &  & 350 GeV & 0.2 \iab{} & \\
         &  & 365 GeV & 1.5 \iab{} & \\\hline
          \multirow{4}{*}{CEPC} & \multirow{4}{*}{Unpolarised} &  Z-pole & 57.5 \iab{} & \multirow{4}{*}{\cite{Bernardi:2022hny}}\\
         &  & 240 GeV & 20 \iab{} & \\
         &  & 350 GeV & 0.2 \iab{} & \\
         &  & 360 GeV & 1 \iab{} & \\\hline
    \end{tabular*}
    \caption{Configurations for future $e^+e^-$ colliders.}
    \label{tab:epem_setup}
\end{table}

Runs at two different centre-of-mass energies above the top-quark pair production threshold are important to disentangle the $e^+e^-t\bar{t}$ operator coefficients from the two-fermion EFT operator coefficients~\cite{Durieux:2018tev}.
The two sets of operators have very different scaling with energy: the sensitivity to four-fermion operators grows quadratically, while it is constant or grows only linearly for two-fermion operators.
In a fit to data taken at a single center of mass, linear combinations of their coefficients remain degenerate and form blind directions.
The combination of runs at two different centre-of-mass energies effectively disentangles them and provides global fit constraints close to the individual bounds.

\FloatBarrier
\section{BSM physics from Top-quark physics}
\label{sec:TOPHF-BSM}

The top quark is a sensitive probe in direct searches for BSM physics~\cite{Bose:2022obr} and indirectly in EFT fits~\cite{EWreport}, see also Section~\ref{sec:TOPHF-EW-couplings}. At hadron colliders, precision measurements of top-pair production are sensitive to the supersymmetric partners of the top quark, the top squarks, with masses close to the top-quark mass (see, e.g., Section~\ref{sec:TOPHF-spin-corr}). At all colliders, flavor-changing neutral currents can be probed in the production and in the decay of top quarks. 

\begin{figure}[ht]
\centering
\includegraphics[width=0.60 \linewidth
]{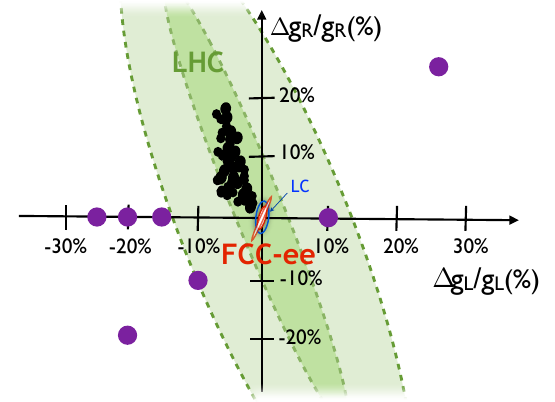}
\caption{Expected relative
  precision on the $Z t_Lt_L$ and $ Zt_Rt_R$ couplings at
  the LHC (lighter green), the HL-LHC (darker  green), the ILC (blue)
  and the FCC-ee (orange, red).  The black dots indicate the deviations expected for different
  parameter choices of 4D composite Higgs models, with $f<2$\,TeV (purple dots: examples for typical deviations in various BSM models).
From~\cite{Janot:2015mqv}. 
}
\label{fig:gLgR_Ztt}
\end{figure}

As mentioned earlier, the top quark with its large Yukawa coupling is intimately connected to the Higgs sector. For instance, in the SM the largest loop-induced corrections to the Higgs boson mass originate from the top quark, requiring an unnatural amount of fine-tuning.  BSM models which aim to protect the Higgs boson mass from these large quadratic corrections beyond the EW energy scale are therefore often closely connected to the top quark. An important example where a symmetry protects the Higgs boson mass is low-energy supersymmetry (SUSY). In unbroken SUSY the corrections induced by the top quark and its bosonic SUSY partners exactly cancel. Another option to obtain a natural, less fine-tuned, Higgs boson mass are composite Higgs models 
(for a review see, e.g., Refs.~\cite{Bellazzini:2014yua,Panico:2015jxa}), where the Higgs boson emerges from new strong dynamics after spontaneous global symmetry breaking as a composite Pseudo-Nambu-Goldstone boson. The large top Yukawa coupling (and large mass) can be achieved through partial compositeness~\cite{Kaplan:1991dc} where the physical top quark is considered to be a combination of elementary and composite degrees of freedom (for a review see, e.g.,\cite{Panico:2015jxa}). Examples of the phenomenology of these type of models in the top sector are the occurrence of new fermionic resonances (top partners), anomalous 4-top quark production, and modified top Yukawa and top-EW couplings. The latter is illustrated in Fig.~\ref{fig:gLgR_Ztt} where the deviations in the $Zt_L\bar t_L$ and $Zt_R\bar t_R$ couplings from the SM are shown in 4D composite Higgs models for various choices of the model parameters~\cite{Janot:2015mqv}. As can be seen, precision measurements of EW top-quark couplings at future lepton colliders can considerably constrain these models. 

In the following we provide some more examples for the power of top-quark observables to constrain a wide variety of BSM scenarios. A detailed study of achievable constraints on BSM physics in the top-quark sector in form of higher-dimensional operators in SMEFT (including those inducing $eett$ contact interactions) from global fits can be found in the EF04 report~\cite{EWreport} (see also Section~\ref{sec:TOPHF-EW-couplings}).  

\FloatBarrier
\subsection{Top-quark spin correlations}
\label{sec:TOPHF-spin-corr}
Summary of white paper contribution~\cite{CMS:spincor}

\begin{figure}[!h!t]
    \centering
    \includegraphics[width=0.6\textwidth]{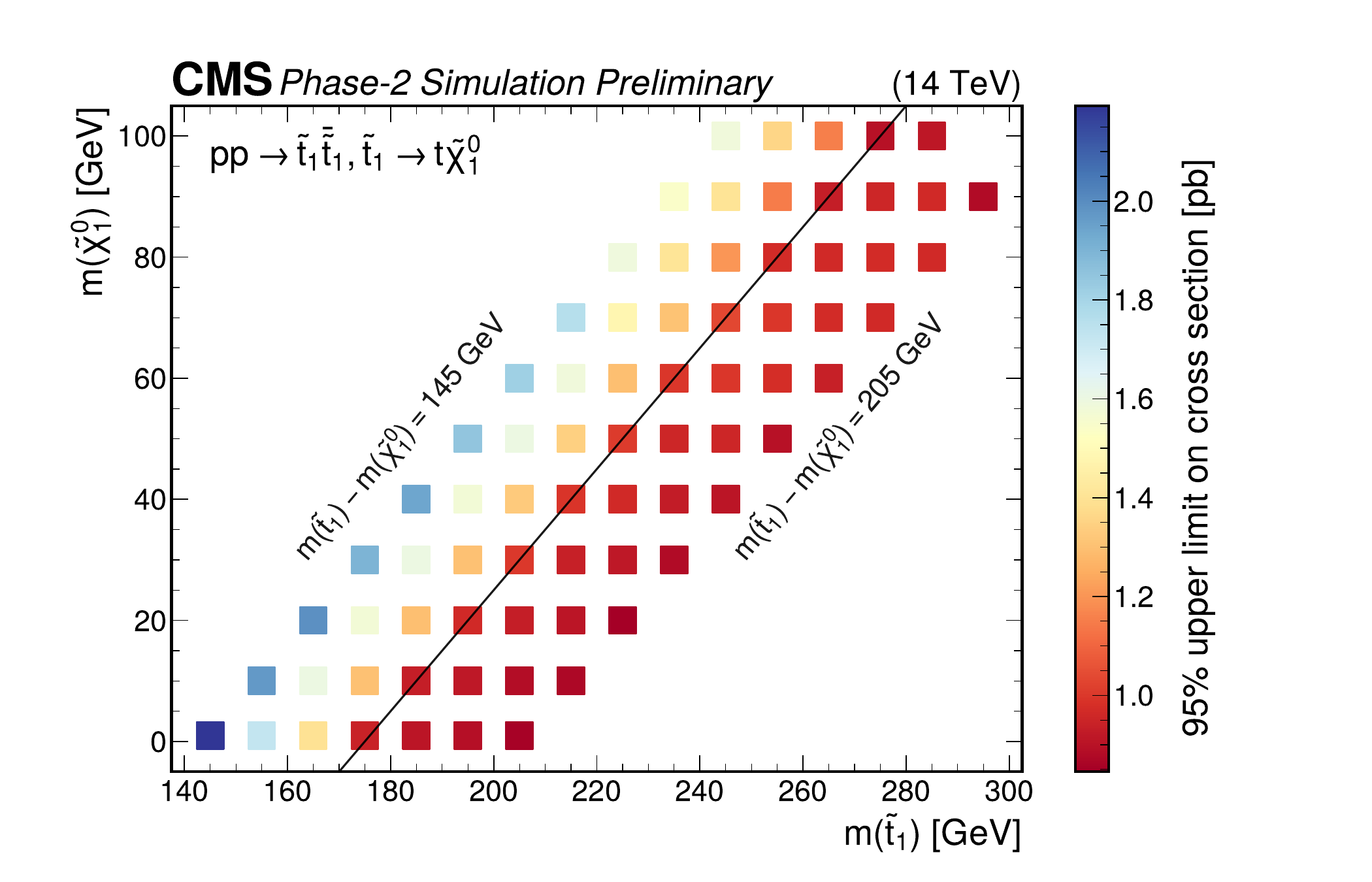}
\caption{Limit on the cross-section for SUSY stop production in the compressed region where the stop mass ($m(\tilde{t})$) is close to the neutralino mass ($m(\chi_0)$), $m(\tilde{t}) - m(\chi_0) = 175$~GeV. Every point in this plot is excluded. From Ref.~\cite{CMS:spincor}. }
    \label{fig:TOPHF-spincor}
\end{figure}

The correlation of the spins of the two top quarks in \ttbar production can be measured precisely from the correlations of the two leptons in dilepton \ttbar events. This is an important measurement to understand the EW interactions of the top quark and can be used in EFT fits~\cite{EWreport}. It is also a sensitive probe of BSM physics with new particles that are close in mass to the top quark and have the same final state (leptons and bottom quark jets). One example is SUSY stop quarks in the compressed region (stop mass close to top mass and small neutralino mass). Figure~\ref{fig:TOPHF-spincor} shows the projected limit for a 30~GeV-wide corridor in stop mass ($m(\tilde{t})$) and neutralino mass ($m(\chi_0)$) around the top quark mass ($m(\tilde{t})-m(\chi_0)-m(t)|<30$~GeV)~\cite{CMS:spincor}. The width of the corridor corresponds to the experimental resolution and the region where direct stop searches are not sensitive because of the large \ttbar background. The limits expected for the HL-LHC are a factor two (at low $m(\tilde{t})$) to ten (at high $m(\tilde{t})$) better than the Run~2 limits in this region. The predicted cross-section for SUSY stop pair production in this region is 10~pb to 100~pb, meaning all of the points shown in Figure~\ref{fig:TOPHF-spincor} will be excluded at the HL-LHC. The exclusion reaches all the way up to stop masses of 600~GeV.

\FloatBarrier
\subsection{Azimuthal angular correlation as a new boosted top-jet substructure}
\label{sec:TOPHF-EW-topjet-substructure}
Summary of white paper contribution~\cite{Yu:2203.02760}

\begin{figure}[htb]
\centering
\includegraphics[width=0.5\textwidth]{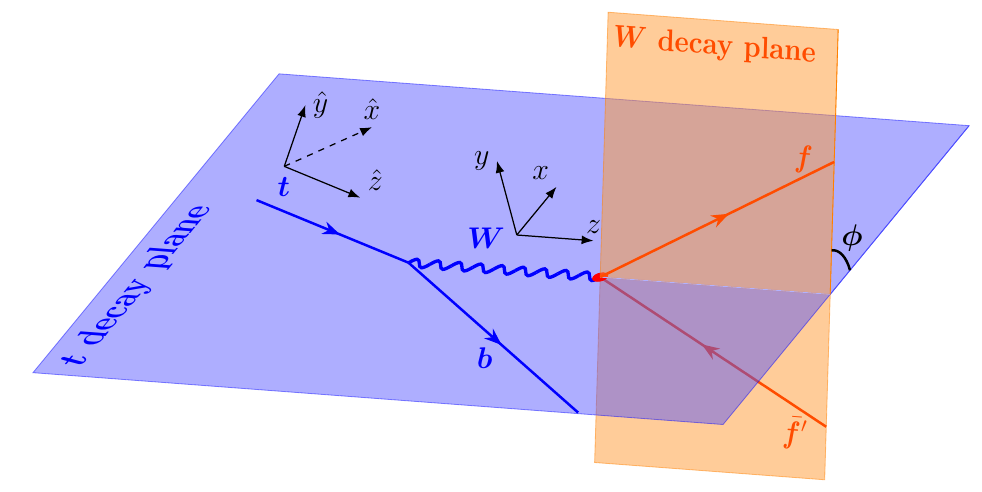}
\includegraphics[width=0.4\textwidth]{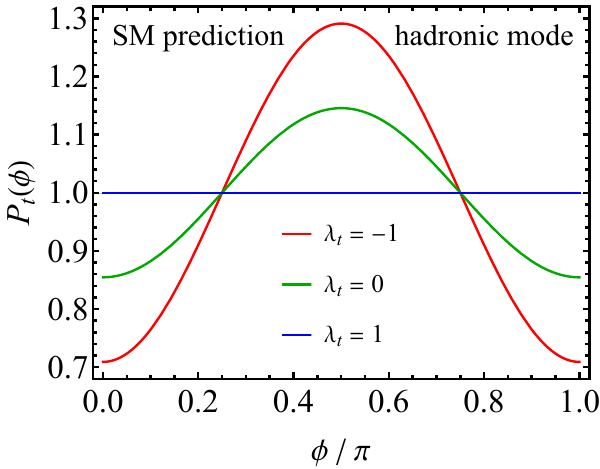}
\caption{Left: Sketch of the top-quark decay products defining the azimuthal angle $\phi$. Right: Azimuthal angular correlation in the decay of boosted top quarks for different values of the top-quark longitudinal polarization $\lambda_t$. Taken from Ref~\cite{Yu:2203.02760}.}
\label{fig:topjet_sub}
\end{figure}

When a top quark is highly boosted, the $W$ boson from its decay has a substantial linear polarization that results in a $\cos2\phi$ azimuthal angular correlation among the top-decay products~\cite{Yu:2203.02760}. The angle $\phi$ is shown in the sketch in Figure~\ref{fig:topjet_sub} (left). This correlation can be measured for hadronically decayed boosted top quarks, and in Ref.~\cite{Yu:2203.02760} an experimental method has been proposed to measure the degree of such azimuthal correlation that only requires $b$-tagging in the top-quark jet. The magnitude of the azimuthal correlation provides a way to measure the longitudinal polarization of a boosted top quark ($\lambda_t$), which differs from the methods proposed in the literature that exploit the energy fractions of the subjets. $\lambda_t$ is an important probe of new physics that couples to the top-quark sector. A simple example is the $W'$ model in which a heavy vector boson $W'$ can couple to top and bottom quarks in any arbitrary chiral combination so that the top quark can be left-handed ($\lambda_t=-1$), right-handed  ($\lambda_t=+1$), or unpolarized ($\lambda_t=0$) as shown in 
Figure~\ref{fig:topjet_sub} (right).

\FloatBarrier
\subsection{Top-quark flavor-changing neutral currents}
\label{sec:TOPHF-FCNC}

\begin{figure}[ht]
\centering
\includegraphics[width=0.88 \linewidth
]{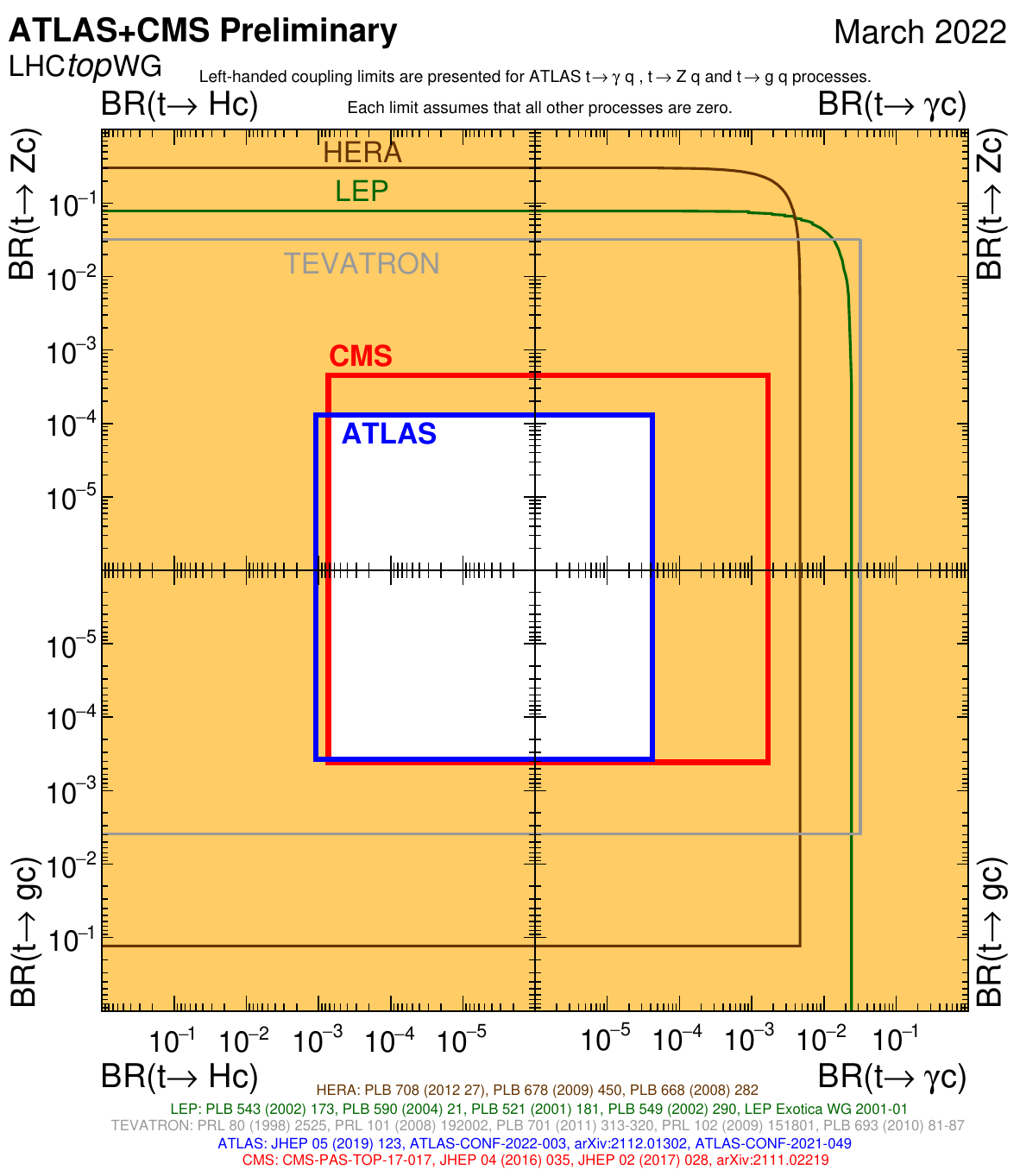}
\caption{Summary of flavor-changing neutral current limits between the top quark and the charm quark from the LHC, Tevatron, HERA and LEP. Each limit assumes that all other FCNC processes vanish. Top-quark decay branching ratios values outside the boxes are excluded at 95\% confidence level.  From~\cite{LHCtopWG}.
}
\label{fig:fcncsum1}
\end{figure}

\begin{figure}[ht]
\centering
\includegraphics[width=0.88 \linewidth
]{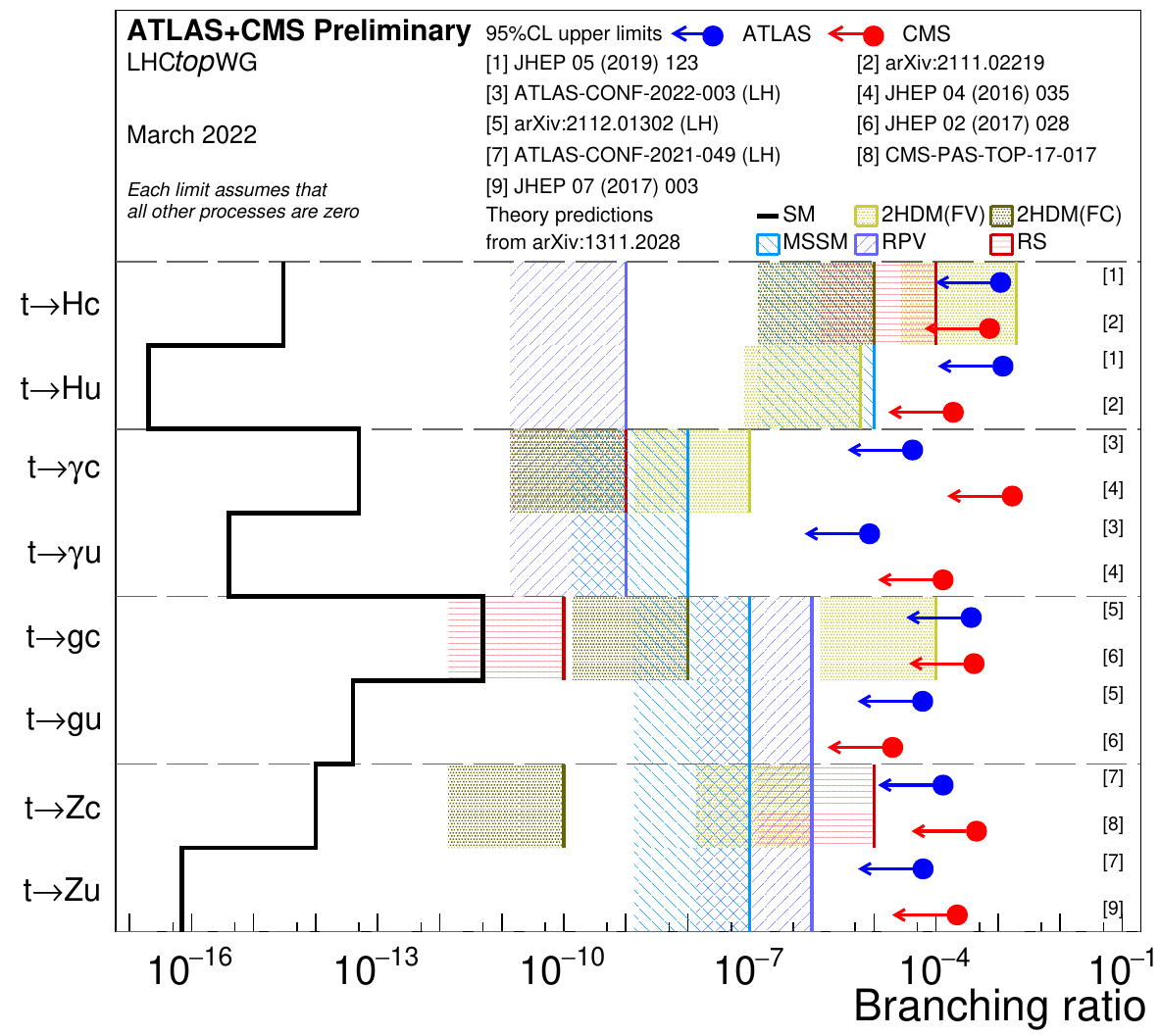}
\caption{Summary of flavor-changing neutral current limits by ATLAS and CMS compared to predictions from BSM scenarios. From~\cite{LHCtopWG}.
}
\label{fig:fcncsum2}
\end{figure}

Top-quark interactions are an excellent probe of flavor-changing neutral currents (FCNC) involving the third generation of leptons. The theoretical models used for many of the FCNC searches at the LHC were summarized for the Snowmass 2013 Top Group report~\cite{Agashe:2013hma}. The searches by ATLAS and CMS, as well as limits previously obtained at the Tevatron, LEP and HERA are shown in the summary plots maintained by the LHC top working group~\cite{LHCtopWG}. Figure~\ref{fig:fcncsum1} shows the summary of searches for FCNC couplings between the top quark and the charm quark, separately for the couplings to photons, gluons, $Z$ and Higgs bosons. No dedicated studies were performed for this report, here we summarize current results and previous studies~\cite{Cerri:2018ypt,Azzi:2019yne}. The theoretical description of FCNC interactions in the effective operator framework is possible~\cite{Durieux:2014xla}. Here we explore sensitivity in terms of the branching ratios of top quark decays, which is easy to understand and interpret and compare. Figure~\ref{fig:fcncsum2} shows the same comparison, and in addition also BSM models that predict sizable FCNC interactions.

The top-photon FCNC coupling to an up and a charm quark was studied by CMS~\cite{Azzi:2019yne}. For 3000~fb$^{-1}$, the expected limits (including systematic uncertainties) on the branching ratio are $8.6\times 10^{-6}$ for $t\gamma u$ and $74\times 10^{-6}$ for $t\gamma c$. These correspond to improvements by almost a factor ten over the current limits from ATLAS~\cite{ATLAS:2022skc} and CMS~\cite{CMS:2015kek}.
The top-gluon FCNC coupling to an up or a charm quark was studied by CMS~\cite{Azzi:2019yne,CMS:2018kwi}. For 3000~fb$^{-1}$, the expected limits (including systematic uncertainties) on the branching ratio are $3.8\times 10^{-6}$ for $tgu$ and $32\times 10^{-6}$ for $tgc$. These correspond to improvements by about a factor ten over the current limits from ATLAS~\cite{ATLAS:2021amo} and CMS~\cite{Khachatryan:2016sib}. The projections for both top-photon and top-gluon FCNC are based on single-top quark production. This leads to the large improvement compared to the current limits: larger dataset allow for more strict selection cuts to isolate the signal. This also leads to the difference between up and charm quark limits, since the coupling of the top to the parton in the initial state is probed. Dedicated studies for FCC-hh were not performed, nevertheless, it can be expected that the larger CM energy and larger integrated luminosities will provide sensitivity to significantly smaller branching ratios. 

The top-$Z$ boson FCNC coupling to an up and a charm quark was studied by ATLAS~\cite{ATL-PHYS-PUB-2019-001, ATL-PHYS-PUB-2016-019}. For 3000~fb$^{-1}$, the expected branching ratio limits are at the level of 4~to~$5\times 10^{-5}$ depending on the considered scenarios for the systematic uncertainties. These correspond to an improvement by a factor five over the current limits from ATLAS~\cite{ATLAS:2021stq} and CMS~\cite{CMS:2017twu,CMS:2017wcz}. There is not much difference between the sensitivities to up and charm quarks here because the FCNC coupling is probed in the top quark decay. Moreover, it is not clear how much improvement can be achieved at FCC-hh since the systematic uncertainties dominate. The CMS study at 8~TeV~\cite{CMS:2017wcz} is based on single top-quark production, and extrapolating that to the FCC-hh should lead to significant improvements.

The top-Higgs boson FCNC coupling to an up and a charm quark was studied by ATLAS~\cite{ATL-PHYS-PUB-2016-019}. For 3000~fb$^{-1}$, the expected branching ratio limits are at the level of $10^{-4}$ for both up and charm quarks. These limits are about a factor ten better than the current limits by ATLAS~\cite{ATLAS:2018jqi} and CMS~\cite{CMS:2021cqc}. The top-Higgs FCNC searches have less sensitivity than those for the other bosons because of the difficulty to identify Higgs boson decays cleanly (for example in the $b\bar{b}$ final sate) or because of the low branching ratio (for example in the $\gamma \gamma$ final state).

Lepton colliders are sensitive to FCNC couplings of the top quark to the photon and the $Z$~boson, especially at energies below the $\ttbar$ production threshold~\cite{Aryshev:2203.07622,Bernardi:2022hny,Shi:2019epw}. The production of a single top quark together with an up or charm quark provides a unique final state signature.
At the ILC or CLIC, the sensitivity to the branching ratio is expected to be one or two orders of magnitude better than for HL-LHC~\cite{Aryshev:2203.07622,deBlas:2018mhx}. Studies at FCC-ee show a sensitivity to a branching ratio for $t\gamma q$ and $tZq$ of about $10^{-5}$. This is slightly less sensitive that the HL-LHC for $t\gamma u$, but more sensitive than the HL-LHC for $t\gamma c$ and $tZq$. However, the description in terms of decay branching ratio is more suitable for hadron colliders, where the main challenge is suppressing backgrounds. The power of lepton colliders in top-quark FCNC searches becomes clear when looking at all possible operators that can contribute to the $tq$ final state in an EFT framework~\cite{deBlas:2018mhx,Durieux:2014xla}. The clean final states make it easier to search for FCNC operators individually, including top-photon, top-$Z$, and four-fermion operators. Limits that are an order to two orders of magnitude better than the expected HL-LHC sensitivity can be reached. Combining runs at multiple CM energies provides additional sensitivity, especially at the highest energies reached in $e^+ e-$ only by CLIC~\cite{deBlas:2018mhx}. This is an area where a muon collider might also provide additional sensitivity.

At higher energies (at or above the $\ttbar$ production threshold), lepton colliders are sensitive to FCNC couplings also in the top-quark decay~\cite{Aryshev:2203.07622,Bernardi:2022hny}. The number of $\ttbar$ events at a lepton collider will always be much less than those produced at HL-LHC, but the final states will be much cleaner, leading to much larger identification efficiencies and competitive FCNC limits also for the other couplings (gluon and Higgs boson). 

\FloatBarrier
\subsection{Top-quark compositeness}
\label{sec:TOPHF-compositeness}

\begin{figure}[ht]
\centering
\includegraphics[width=0.70 \linewidth
]{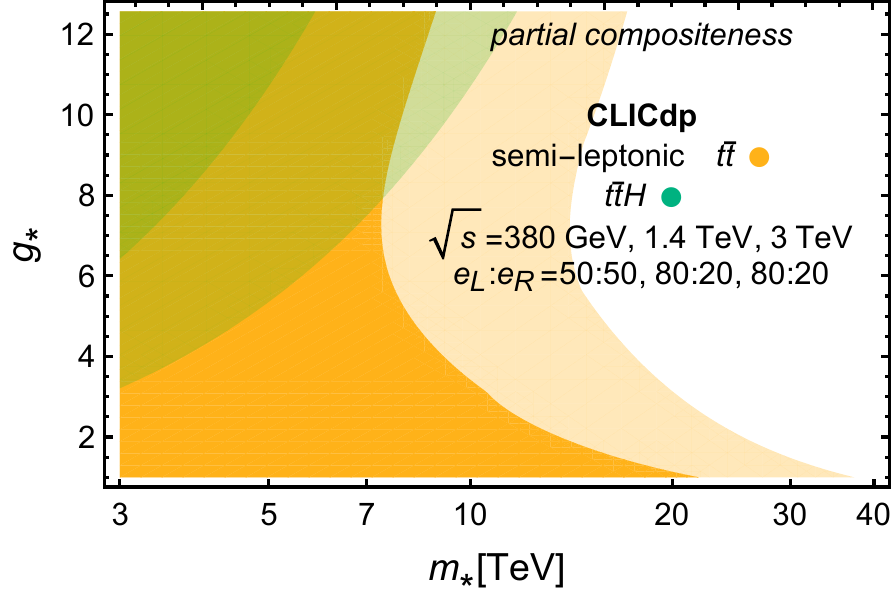}
\caption{`Optimistic'' (light color) and ``pessimistic'' (dark color) $5\sigma$ discovery regions for a partial right-handed top compositeness scenario at a high-energy $e^+ e^-$ collider. The orange contours are derived from a $t \bar t$ global fit, and the green contour is derived from a top Yukawa analysis. From~\cite{CLICdp:2018esa}. 
}
\label{fig:top_pc1}
\end{figure}

High-energy lepton colliders are sensitive probes of top-quark compositeness. For example, Fig.~\ref{fig:top_pc1} shows the reach in the composite sector confinement scale $m_*$ and the composite coupling strength parameter $g_*$ of a partial top compositeness scenario at a multi-TeV $e^+e^-$ collider~\cite{CLICdp:2018esa} (see also~\cite{Aryshev:2203.07622}).  

\begin{figure}[ht]
\centering
\includegraphics[width=0.45 \linewidth]{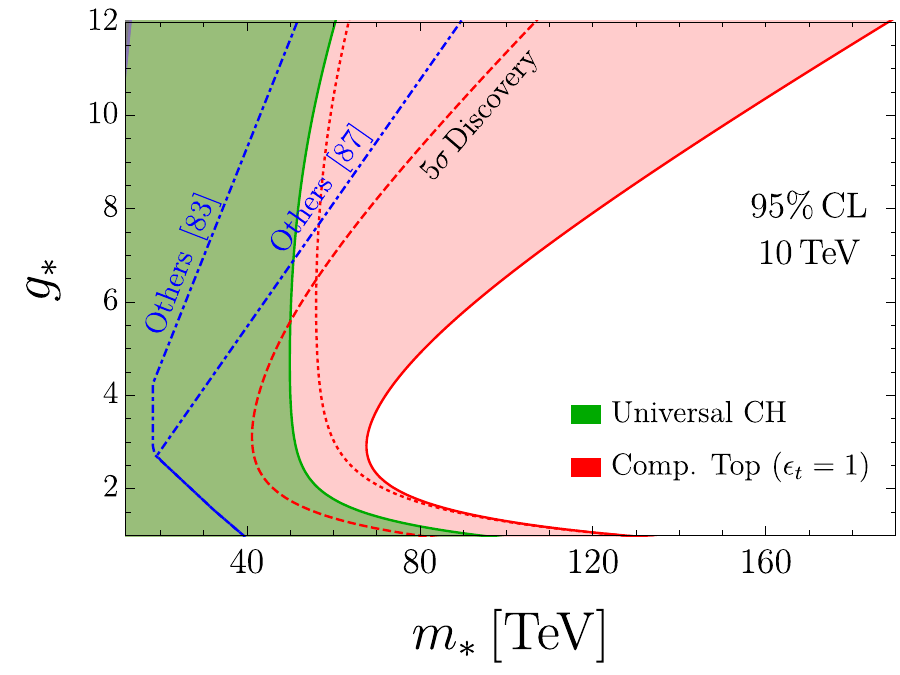}
\includegraphics[width=0.45 \linewidth
]{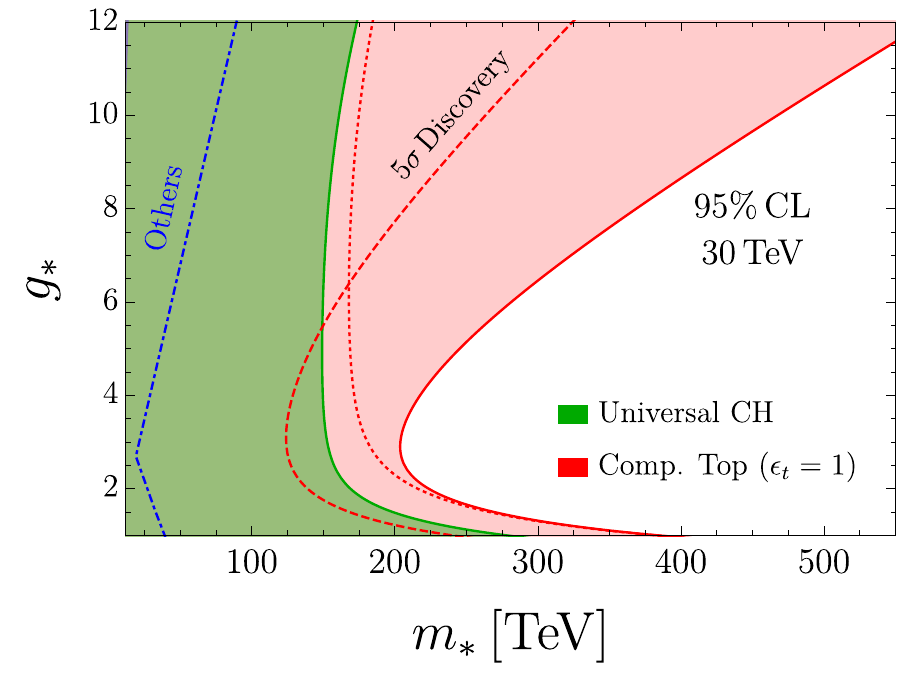}
\caption{Projected 95\% exclusion reach at a 10 TeV (left) and 30 TeV (right) muon collider when including $t\bar t$ (and $b\bar b$) production (red shaded area) in a top compositeness model where the right-handed top quark is considered fully composite. Also shown are projections obtained for the HL-LHC (labeled as 'Others [83]')~\cite{EuropeanStrategyforParticlePhysicsPreparatoryGroup:2019qin} and for CLIC (labeled as 'Others [87]')~\cite{Banelli:2020iau}.  From~\cite{Chen:2022msz}.
}
\label{fig:top_pc2}
\end{figure}

\begin{figure}[ht]
\centering
\includegraphics[width=0.6 \linewidth]{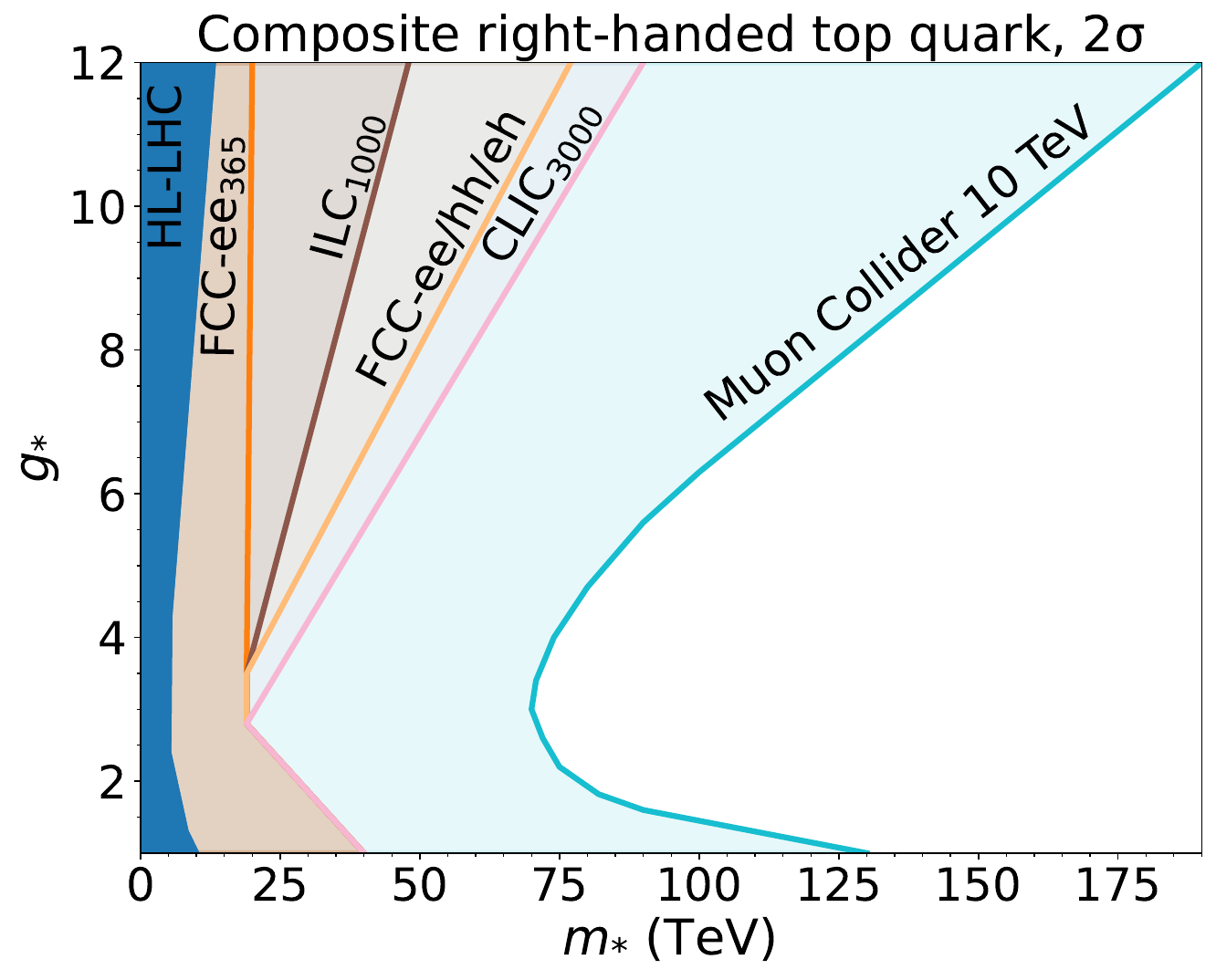} 
\caption{Exclusion (2-$\sigma$) sensitivity projections for compositeness models for future colliders as labeled, for models where both the Higgs boson and the top quark with right-handed couplings are composite. Plot based on Refs.~\cite{Chen:2022msz,Banelli:2020iau}.}\label{fig:TOPHF-composite}
\end{figure}

In Figure~\ref{fig:top_pc2}~\cite{Chen:2022msz} the red shaded area shows the impact of including $t\bar t$ (and $b\bar b$) production on the 95\% exclusion reach in the $(g_*,m_*)$ plane in a top compositeness scenario where the right-handed top quark is assumed to be fully composite ($\epsilon_t=1$) at a 10 TeV (left) and 30 TeV (right) muon collider. The increase in sensitivity with the center-of-mass energy ($E_{cm}$) is due to the fact that the relevant dimension-6 operators considered in this study give rise to contributions which grow with $E_{cm}$~\cite{Chen:2022msz}. The green shaded area indicates the sensitivity related to Higgs compositeness. Also shown are projections obtained for the HL-LHC (labeled as 'Others [83]')~\cite{EuropeanStrategyforParticlePhysicsPreparatoryGroup:2019qin} and for CLIC (labeled as 'Others [87]')~\cite{Banelli:2020iau}. An overview of projections for a wider range of future colliders is shown in Figure~\ref{fig:TOPHF-composite}. Note that the projected reaches in the mass scale $m_*$ are an order of magnitude larger than $E_{cm}$ of the colliders.

\FloatBarrier
\section{Heavy flavor, bottom quark studies}

\subsection{Running bottom-quark mass}
\label{sec:TOPHF-bmass}
Summary of white paper contribution~\cite{Aparisi:2022yfn}

The prospects for measurements of the bottom-quark mass and tests of the scale evolution or "running" of the mass value predicted by the Standard Model are discussed in detail in Ref.~\cite{Aparisi:2022yfn}.

The bottom-quark mass is currently measured most precisely using low-energy experimental inputs. The most precise determinations~\cite{Narison:2019tym,Peset:2018ria,Kiyo:2015ufa,Penin:2014zaa,Alberti:2014yda,Beneke:2014pta,Dehnadi:2015fra,Lucha:2013gta,Bodenstein:2011fv,Laschka:2011zr,Chetyrkin:2009fv} rely on the measurement of the mass of bottomonium bound states and the $e^+e^- \rightarrow $ hadrons cross section as experimental input, in combination with QCD sum rules and perturbative QCD calculations. Several lattice QCD groups have also published results, the most recent of which reaches a precision of approximately 0.3\%~\cite{Bazavov:2018omf,Colquhoun:2014ica,Bernardoni:2013xba,Lee:2013mla,Dimopoulos:2011gx} (see also the FLAG report~\cite{Aoki:2019cca}).The combination of all these measurements yields a world average with a sub-\% precision~\cite{Zyla:2020zbs}. Expressed in the \MSbar{} scheme, the value of the bottom-quark mass at the scale given by the mass itself is: $m_b(m_b) =  4.18^{+0.03}_{-0.02}~\gev{}$.

Measurements of the bottom quark mass at higher scale were performed at the $Z$-pole, where jet rates and event shapes are sensitive to subleading mass effects~\cite{Bilenky:1994ad,Rodrigo:1997gy,Bilenky:1998nk,Rodrigo:1999qg,Bernreuther:1997jn,Brandenburg:1997pu,Nason:1997tz,Nason:1997nw}. Measurements were performed by the LEP and SLC experiments~\cite{Abreu:1997ey,Brandenburg:1999nb,Abe:1998kr, Barate:2000ab,Abbiendi:2001tw,Abdallah:2005cv,Abdallah:2008ac}. The combination of the most precise determinations from three-jet rates of each experiment yields $m_b(m_Z) = 2.82 \pm 0.28~\GeV{}$.

Future $e^+e^-$ colliders can improve the precision of the $m_b(m_Z)$ measurement. A dedicated high-luminosity run at the $Z$-pole, i.e. the ``GigaZ" program of a linear
 collider or the ``TeraZ'' run at the circular colliders, yields a sample of $Z$-bosons that exceeds that of the LEP experiments and SLD by orders of magnitude. We adopt the extrapolation of LEP/SLD results in Ref.~\cite{ILDnote2020} that assumes that the extraction of $m_b(m_Z)$ from the three-jet rates will be limited by the theory uncertainty and hadronization uncertainties. Both sources of uncertainty are assumed to be reduced by a factor 2. This requires fixed-order calculations at NNLO accuracy, with full consideration of mass effects, which is available for Higgs decays~\cite{Bernreuther:2018ynm}.  

The Higgs factory program itself, with several inverse attobarn at a center-of-mass energy of 240-250~\GeV{}, can take advantage of radiative-return events. The Lorentz-boost of the $Z$-bosons complicates the selection, reconstruction and interpretation. A dedicated full-simulation study is therefore required to provide a reliable, quantitative projection. However, it is clear that the radiative-return data has the potential to significantly improve the precision of existing LEP/SLC analysis.

Finally, a high-energy electron-positron collider operated at a center-of-mass energy of 250~\GeV{} or above can extend the analysis to higher energies and thus probe the 
effect of coloured states with masses heavier than that the Higgs boson on the running of the bottom-quark mass. The potential of the three-jet rate measurement to determine $m_b(\mu)$ for $\mu=$ 250~\GeV{} has been studied in Ref.~\cite{ILDnote2020}. The mass dependence of the observable is found to drop rapidly with increasing $\mu$, since the bottom-quark mass dependence is a power-suppressed correction. A measurement with a precision of 1~\GeV{} is feasible for $\mu=$ 250~\GeV{}.

The bottom-quark mass at the scale of the $Z$-boson mass can also be inferred from the $Z \rightarrow b\bar{b}$ decay width. Currently, this method does not offer a competitive precision. Using $R_{0,b} = \Gamma(Z \rightarrow b\bar{b})/\Gamma_{\rm total} = 0.21582 \pm 0.00066$, as reported by the LEP/SLC Electro-weak Working Group~\cite{ALEPH:2005ab}, Ref.~\cite{Kluth:2022ucw} finds an uncertainty greater than 1~\GeV.

A future high-statistics $Z$-pole run, together with theory improvements, can significantly enhance the potential of this approach. Following the FCCee Conceptual Design Report~\cite{FCC:2018byv,FCC:2018evy}, that predicts a ten-fold increase of the precision of $R_{0,b}$,  one can expect a precision of 140~MeV (5\%) on $m_b(m_Z)$ after the
 ``TeraZ'' program.
This requires considerable improvements in the modelling of B- and D-hadron decays, compared to the reference analysis performed by SLC that forms the basis for the extrapolation by the FCCee study.

Recently, the bottom-quark mass at the scale of the Higgs boson mass was extracted from Higgs boson decay rates measured by ATLAS\cite{ATLAS:2020qdt} and CMS~\cite{Sirunyan:2018koj} at the LHC~\cite{Aparisi:2021tym}, yielding a value: $m_b(m_H) = 2.60 ^{+0.36}_{-0.30}~\GeV{}$.
The precision of this determination is still statistics-limited, but already quite comparable with that of the LEP/SLC measurement of $m_b(m_Z)$.
The measurement of $m_b(m_H)$ from the Higgs decay width to a bottom-antibottom quark pair is expected to increase rapidly in precision as the precision of Higgs coupling 
measurement improves.
The method of Ref.~\cite{Aparisi:2021tym} provides a very clean theoretical basis that allows for steady progress as the experimental precision improves. The key aspect of
 this method is that the Higgs boson is a color-less spin-0 state with a relatively small decay width, such that the analysis is essentially insensitive to the theoretical
 knowledge of the Higgs production rate. For the same reason very precise theoretical predictions can be made for the Higgs partial width into a bottom-antibottom quark pair.

\begin{figure}[!h!tbp]
\includegraphics[width=0.7\columnwidth]{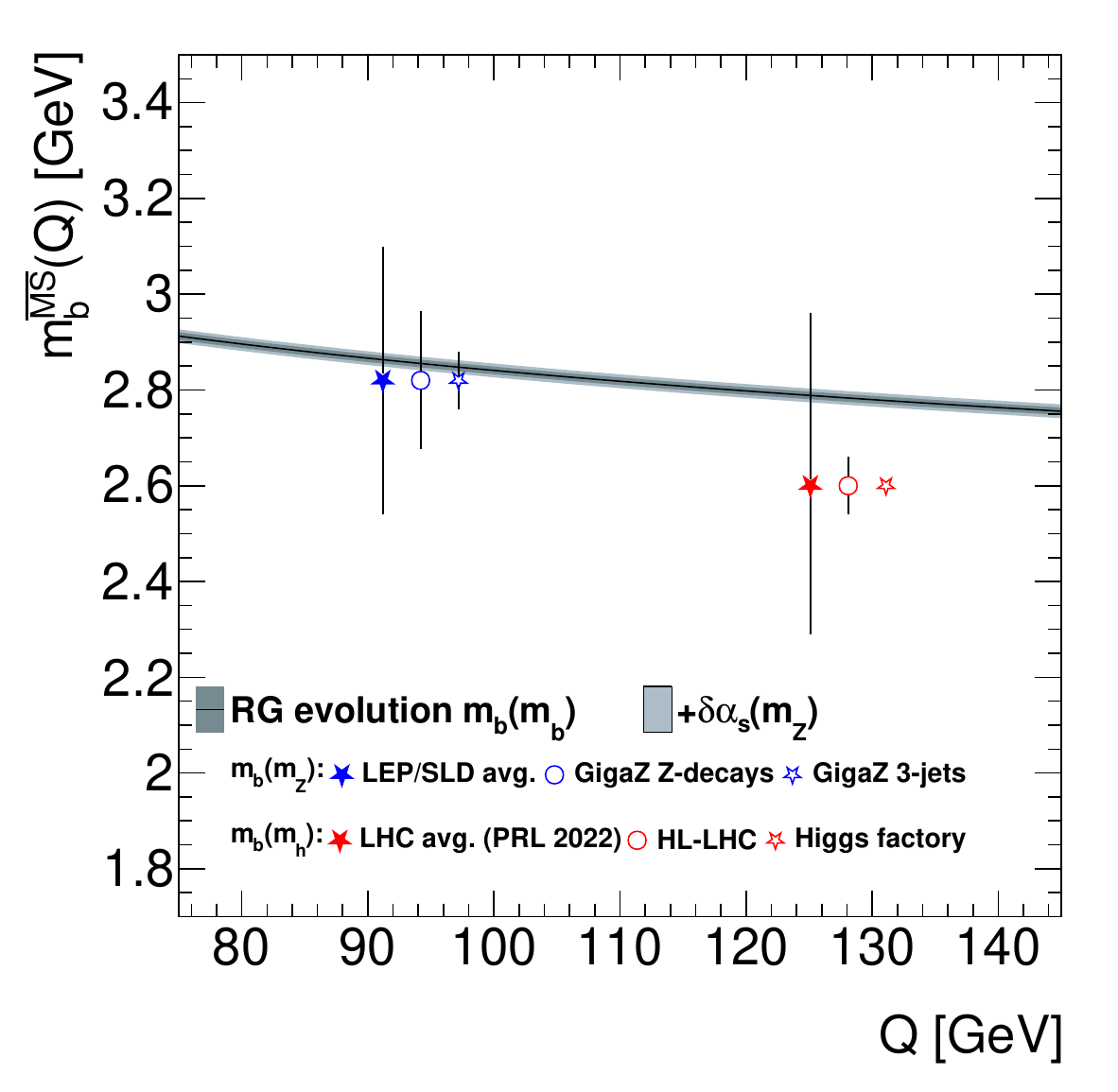}%
\caption{\label{fig:running_mass_projection} Prospects for measurements of the scale evolution of the bottom-quark \MSbar{} mass at future colliders. The markers are projections for $m_b(m_Z)$ from three-jet rates at the $Z$-pole and for $m_b(m_H)$ from Higgs boson branching fractions. The RGE evolution of the mass is calculated at five-loop precision with REvolver~\cite{Hoang:2021fhn}. }
\end{figure}

The projections and extrapolations discussed above have been included in Fig.~\ref{fig:running_mass_projection}. The markers are centered on the current
 central values for $m_b(m_Z)$ and $m_b(m_H)$ and the error bars indicate the projected precision. The solid line indicates the evolution of the PDG world average from $m_b(m_b)$ to a higher scale using the RGE calculation included in the REvolver code~\cite{Hoang:2021fhn} at five-loop precision. The uncertainty band includes the projected
uncertainty of 10~\MeV{} on $m_b(m_b)$ (dark grey) and an 0.5\% uncertainty on $\alpha_s(m_Z)$. 
 
The independent determinations of the bottom-quark mass at different energies yield a precision test of the scale evolution of the bottom quark mass. High-scale determinations can be used to search for the impact of new massive coloured states on the scale evolution, using a similar strategy to studies of $\alpha_s$~\cite{Llorente:2018wup,Jezabek:1992sq}, and possibly incorporating the analysis of $\alpha_s$ and $m_b$ in a combined fit. The implementation of this program, and a precise estimate of its sensitivity, is left for future work.

In the next decades, with the completion of the high luminosity program of the LHC and the construction of a new ``Higgs factory" electron-positron collider, rapid progress is envisaged in the measurement of Higgs coupling measurement. These precise measurements will enable an extraction of the \MSbar{} bottom-quark mass $m_b(\mu)$ at the
 scale given by the Higgs boson mass, $m_b(m_H)$, with a precision of the order of 10~\MeV. With a relative precision of 2 per mille, the high-scale measurement can reach 
a similar precision as $m_b(m_b)$ based on low-energy measurements.

Together with improved measurements of $m_b(m_b)$ from low-energy data, $m_b(m_Z)$ from three-jet rates in $e^+e^-$ collisions (and possibly new measurements at scales smaller than $m_Z$ and larger than $m_H$), one can expect to map out the scale evolution of the bottom-quark mass from $m_b$ to $m_H$ with a precision at the few per mille level. At the same time, improved measurements of the strong coupling at each of these scales reduce the uncertainty in the evolution between the two energies. When all these elements are brought together, they form a powerful test of the ``running" of quark masses predicted by the Standard Model and allow for stringent limits on coloured states with mass below the electroweak scale.

\FloatBarrier
\subsection{Probing heavy-flavor parton distribution functions at hadron colliders}
\label{sec:TOPHF-HF-PDF}
Summary of white paper contribution~\cite{Xie:2203.06207}

Modern global QCD analyses~\cite{Hou:2019efy,NNPDF:2017mvq,Bailey:2020ooq,Alekhin:2017kpj} extract collinear parton distribution functions (PDFs) and their combinations using deep-inelastic scattering (DIS) and fixed-target cross section measurements together with a large variety of high-precision data from the LHC, e.g., single-inclusive jet production, production of Drell-Yan pairs, top-quark pairs, and high-$p_T$ $Z$ bosons. Despite all efforts, PDFs and fragmentation functions (FFs) still represent one of the major sources of uncertainty in precision calculations of standard candle observables at the LHC. In addition, heavy-flavor (HF) PDFs deserve particular attention as they are currently poorly constrained as compared to the other PDFs. Constraining HF PDFs is a twofold task. First, it requires a specific QCD framework (i.e., a general mass variable flavor number (GMVFN) scheme) to give correct theory predictions for observables involving heavy quarks (HQ) when the number of quark flavors changes with energy. Second, it must allow one to directly access HQ PDFs parametrized at the initial scale. This motivates the development of a general-mass (GM) factorization scheme for proton-proton ($pp$) collisions, which in Ref.~\cite{Xie:2019eoe,Xie:2021ycd} is named S-ACOT-MPS~\cite{Xie:2019eoe,Xie:2021ycd}, to probe and constrain HF PDFs using high-precision data from the LHC in future global QCD analyses.
S-ACOT-MPS is based on an amended version of S-ACOT that was developed for DIS~\cite{Aivazis:1993kh,Aivazis:1993pi,Tung:2001mv,Nadolsky:2009ge,Guzzi:2011ew} and is applied to proton-proton collisions. It differs from other available GMVFN schemes~\cite{Kniehl:2004fy,Kniehl:2005mk,Helenius:2018uul} in the treatment of the phase space.
More details about the S-ACOT-MPS scheme can be found in~\cite{Xie:2019eoe,Xie:2021ycd,Xie:2022sqa} and will also be presented in a forthcoming study~\cite{S-ACOT-MPS}.

\begin{figure}
    \centering
    \includegraphics[width=0.49\textwidth]{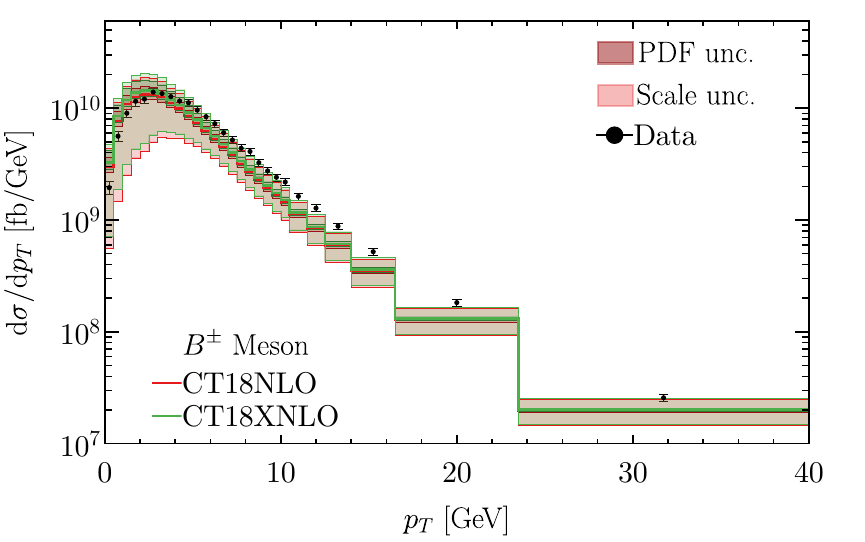}
    \includegraphics[width=0.49\textwidth]{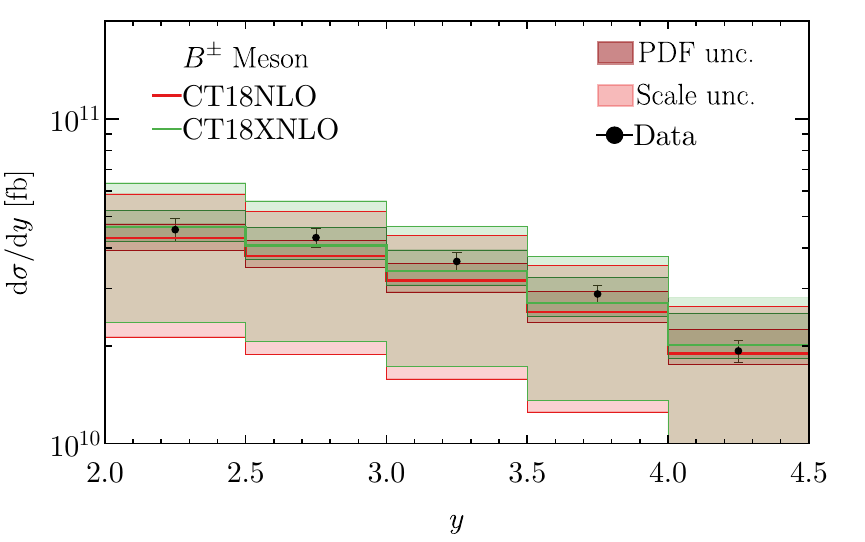}
    \caption{NLO theory predictions for the $p_T$ and $y$ distributions obtained with CT18NLO and CT18XNLO PDFs compared to the LHCb data for $B^\pm$ production at 13 TeV~\cite{LHCb:2017vec}.}
    \label{fig:TOPHF_hadron}
\end{figure}

In~\cite{Xie:2203.06207} S-ACOT-MPS is applied to the case of $b$-quark hadroproduction at next-to-leading order (NLO) in QCD in $pp$ collisions. We calculate theory predictions
for $b$-quark hadroproduction cross sections including $b$-quark fragmentation contributions and compare the $p_T$ and rapidity $y$ spectra at particle level to precision measurements~\cite{LHCb:2017vec} of $B^\pm$ meson production at LHCb at a center of mass energy of 13 TeV.
A data vs theory comparison is illustrated in Fig.~\ref{fig:TOPHF_hadron} where theory predictions are obtained with the CT18 and CT18X PDFs~\cite{Hou:2019efy}. 
The agreement between theory and data is overall good and within the quoted uncertainties.
In particular, the predictions for the $p_T$ spectrum are in good agreement with data at low $p_T$. However, at $p_T > 10$ GeV the data lie on the upper edge of the theoretical error bands. This suggests that higher-order corrections at next-to-next-to-leading order (NNLO) may improve the agreement between data and theory. The predictions for the rapidity distribution agree well with the data, although theory uncertainties are large. The rapidity central value, obtained with the CT18X PDFs, is in slightly better agreement with respect to that obtained with CT18. This may be ascribed to the enhanced CT18X gluon and $b$-quark PDFs as compared to CT18, which reflect small-$x$ dynamics effects captured by the CT18X fit.
The inclusion of $b$-quark hadroproduction processes in future global QCD analyses will be important to improve PDF determinations that aim at reducing uncertainties of heavy-flavor PDFs, provided that a consistent general mass treatment for $pp$ reactions is utilized.
In addition, a recent study based on cross section measurements at forward rapidity for $Z+c$ production at LHCb~\cite{LHCb:2021stx} has suggested a valence-like intrinsic-charm component in the proton wave function. This needs to be further explored in new global PDF analyses using precision measurements at the LHC that are sensitive to heavy-flavor PDFs, and a consistent general mass treatment to correctly account for mass effects.

\FloatBarrier
\subsection{Bottom-quark production measurements for EFT fits}
\label{sec:TOPHF-HF-EFT}

Bottom- and charm-quark production measurements at e+e- colliders provide inputs to EW precision fits and for EFT fits, constraining operators related to second- and third-generation quarks. These are otherwise only accessed through low-energy decay-based measurements.
The impact of precision measurements in $e^+e^- \rightarrow  b\bar{b} $ and $e^+e^- \rightarrow  c\bar{c} $ production on precision EW fits is explored in Section~2.2 of the EW report~\cite{EWreport}. 
The measurements of the $Z$~boson decay branching ratio to bottom quarks ($R_b$) and the forward-backward asymmetry in $b\bar{b}$ production ($A^{bb}_{FBLR}$) at the $Z$~peak at LEP and SLD are included in the top-quark fit presented in Section~\ref{sec:TOPHF-EW-couplings}~\cite{ALEPH:2005ab}. 

Prospects for the $e^+e^- \rightarrow b\bar{b}$ and $e^+e^- \rightarrow c\bar{c}$ processes are based on the full-simulation studies for $b\bar{b}$ of the ILD concept~\cite{Okugawa:2019ycm} at $\sqrt{s} = $ 250~\GeV. The studies are based on realistic estimates of efficiency and acceptance, including the signal losses required to ensure a robust calibration of the flavor tagging efficiency. The statistical uncertainties on the measurements of the cross section and forward-backward asymmetry are complemented by polarisation and flavour-tagging systematics. Section~2.2 in the EW report~\cite{EWreport} gives more details on the variables, configurations and uncertainties considered. Projections for the $Z$-pole measurements of $R_b$ and $A_{FB}$ are provided by the FCCee and CEPC projects for their ``TeraZ'' runs at the $Z$-pole~\cite{Bernardi:2022hny,CEPCStudyGroup:2018ghi}. These studies show that improvements in uncertainties by more than an order of magnitude over LEP can be expected. The largest systematic uncertainties are from the modeling of QCD interactions; improvements in MC generators, QCD radiation and hadronization modeling are required to reach this precision.

\FloatBarrier
\section{Conclusions}
\label{sec:TOPHF-conclusions}
\subsection{Summary}
The broad program of top-quark measurements at the LHC and related advanced theoretical calculations will continue into the future, at the HL-LHC and possibly at future lepton and hadron colliders. At lepton colliders, the top-quark program will only start in earnest once the CM energy reaches the top-pair production threshold. 

Studies of the top quark are directly connected to the important questions at the energy frontier, see Figure~\ref{fig:topoverview}. Of particular importance is the top-quark mass, which is a key ingredient in EW precision fits~\cite{EWreport} and QCD calculations. Mass measurements at hadron colliders are limited by theory modeling uncertainties, the best precision will only be achieved at future lepton colliders running at the top threshold. Top-quark production processes probe all aspects of the top-quark couplings to the SM bosons and top-quark final states are sensitive to BSM particles such as SUSY top squarks. The HL-LHC will provide the event samples required to study many of these processes with percent-level precision, which necessitates (N)NNLO and higher-order QCD calculations, and the inclusion of (N)NLO EW corrections. 

The top-quark program at linear colliders greatly benefits from CM energies above 500~GeV. The FCC-ee and CEPC are able to scan the top-production threshold at 340~GeV with small statistical uncertainties. However, they are not able to reach energies required to produce top-quarks in association with bosons in sufficient numbers to study the top-Higgs coupling directly. The ILC running at 500~GeV will be able to do that, and CLIC, with even higher accessible energies, will be able to probe additional processes. 

Searches for BSM physics in top-quark final states focus on the third-generation coupling of BSM particles, indirectly through EFT fits~\cite{EWreport} or searches for flavor-changing neutral currents, and directly through searches for SUSY and other new particles~\cite{Bose:2022obr}. Linear colliders will expand the reach of FCNC searches of the top-$Z$ interaction.

Contact interaction and searches for compositeness are examples of BSM physics that top-quark production is sensitive to at TeV energies and above, these are probed at CLIC, FCC-hh as well as muon colliders.

Table~\ref{tab:TOPHF-couplings} compares a few top-quark measurements between different future collider options. Each of the measurements can be improved at future colliders beyond the precision at the HL-LHC. Significantly improving the precision of the top-quark Yukawa coupling beyond the 2-4\% uncertainty expected at the HL-LHC~\cite{ATLAS:2022hsp} requires a high-energy lepton collider at a CM energy of 500~GeV or the FCC-hh. The precision of the coupling measurements to the SM bosons will all be significantly improved at a lepton collider running at or above the top-production threshold. The four-top coupling can be probed at hadron colliders, or at lepton colliders running at sufficiently high energies. 
Searches for flavor-changing neutral currents via the $Z$ boson or photon are done in top-quark decays at hadron colliders, the sensitivity is significantly extended at lepton colliders running as a Higgs factory.

\begin{table}[!h!t]
\begin{center}
\begin{tabular}{l|c|c|c|c}
Parameter & HL-LHC  & ILC 500 &  FCC-ee & FCC-hh \\ \hline
$\sqrt{s}$ [TeV] & 14 & 0.5 & 0.36 & 100  \\ \hline
Yukawa coupling $y_t$ (\%) & 3.4  & 2.8 & 3.1 & 1.0 \\
Top mass $m_t$ (MeV/\%) & 170/0.10 & 50/0.031 & 40/0.025 & -- \\
Left-handed top-$W$ coupling $C_{\phi Q}^3$  (TeV$^{-2}$) & 0.08 & 0.02 & 0.006 & -- \\
Right-handed top-$W$ coupling $C_{tW}$ (TeV$^{-2}$) & 0.3 & 0.003 & 0.007 & -- \\
Right-handed top-$Z$ coupling $C_{tZ}$ (TeV$^{-2}$) & 1 & 0.004 & 0.008 & -- \\
Top-Higgs coupling $C_{\phi t}$ (TeV$^{-2}$) & 3 & 0.1 & 0.6 & \\
Four-top coupling $c_{tt}$ (TeV$^{-2}$) & 0.6 & 0.06 & -- & 0.024 \\
FCNC $t\gamma u$, $tZu$ BR & $10^{-5}$ & $10^{-6}$ & $10^{-5}$ & -- \\
\end{tabular}
\caption{Anticipated precision of top-quark Yukawa coupling and mass measurements, and of example EFT Wilson coefficient for the top-quark coupling to $W$, $Z$ and Higgs bosons, as well as a four-top Wilson coefficient. The expected reach of the CEPC should mirror that of the FCC-ee.}
\label{tab:TOPHF-couplings}
\end{center}
\end{table}

\subsection{Theory challenges}
Significant theoretical effort is required to exploit the full potential of future colliders, as pointed out throughout this document. Some of the biggest challenges are:
\begin{itemize}
    \item Calibration of the top quark MC mass to a well-defined scheme in perturbation theory with a precision comparable to the experimental uncertainty.
    \item Computing cross-sections, inclusively and differentially at higher orders in perturbation theory, going to N$^{3}$LO in QCD for top pair production plus resummation, going to NNLO in QCD for associated production processes, and including EW higher order corrections, see also the Les Houches wishlist~\cite{Huss:2022ful}.
    \item Reducing the PDF uncertainties, which are already now the largest theory uncertainties for several processes, most importantly top-pair production. This requires close interconnections between theory and experiment and new differential measurements of top production processes.
    \item Improving the modeling of the full event at the LHC and future hadron and lepton colliders and reducing parton shower uncertainties.
\end{itemize}
For more details about the status and necessary advances in high-precision theory see also the EF06 report~\cite{QCDreport}, and the Theory Frontier Topical Group reports on {\it Theory Techniques for Precision Physics} (TF06) and {\it Theory of Collider Phenomena} (TF07).

\FloatBarrier
\Acknowledgements


We acknowledge the support of the U.S. Department of Energy and the National Science Foundation. 
We thank Andre Hoang for writing the section on {\it Top-quark mass: Theory aspects and challenges}~\ref{sec:TOPHF-mtop-theory}, Tony Liss for writing the section on {\it Top-quark mass measurements from top-quark decays at hadron colliders}~\ref{sec:TOPHF-mtop-direct}, Maria Vittoria Garzelli for writing the section on {\it Top-quark mass measurements in well-defined schemes at hadron colliders}~\ref{sec:TOPHF-mtop-pole}, Frank Simon for writing the section on {\it Top-quark mass measurements at $e^+e^-$ colliders}~\ref{sec:TOPHF-mtop-ee}, Nikolaos Kidonakis for writing parts of the section on {\it Top-quark pair and single top-quark production at the LHC: a brief review of theory calculations}~\ref{sec:TOPHF-XS-LHC-theo}, Regina Demina for writing the sections on {\it Experimental aspects of $pp \to t\bar t$} ~\ref{sec:TOPHF-XS-LHC-ttexp} and {\it Four-top production}~\ref{sec:TOPHF-4top-production}:, Manfred Kraus and Laura Reina for writing the section on {\it $ttX,  X=j,\gamma,Z,W^\pm,H,b\bar{b}$: review of theory predictions for LHC and HL-LHC}~\ref{sec:TOPHF-XS-LHC-ttX}, Victor Miralles and Marcel Vos for writing the section on {\it Top-quark coupling measurements from EFT fits}~\ref{sec:TOPHF-EW-couplings}. We thank Chip Brock, Marcel Vos, and Andre Hoang for reviewing early versions of this manuscript, and the many people who provided feedback. We are grateful to the authors of the EF03 white papers and of selected publications for summarizing their contribution briefly for this report.

\bibliographystyle{utphys.bst}
\bibliography{TOPHF/bib/Snowmass-EF,TOPHF/bib/References,TOPHF/bib/EFTinputs,TOPHF/bib/ttX}

\end{document}